\documentclass[letterpaper,11pt]{article}

\usepackage{epsfig, amsmath, epsfig, amssymb,array,MnSymbol}
\usepackage{sidecap}
\usepackage{graphicx}
\usepackage{bm}
\usepackage{color}
\def\etmiss{E\!\!\!\!\slash_{T}}

\def\pslash{\not{\hbox{\kern-4pt $p$}}}
\def\qslash{\not{\hbox{\kern-4pt $q$}}}
\def\lv{\not{\hbox{\kern-4pt $L$}}}
\def\lsim{\mathrel{\raise.3ex\hbox{$<$\kern-.75em\lower1ex\hbox{$\sim$}}}}
\def\gsim{\mathrel{\raise.3ex\hbox{$>$\kern-.75em\lower1ex\hbox{$\sim$}}}}
\def\ifmath#1{\relax\ifmmode #1\else $#1$\fi}

\newcommand{\nc}{\newcommand}
\nc{\postscript}[2]{\setlength{\epsfxsize}{#2\hsize}\centerline{\epsfbox{#1}}}

\nc{\beq}{\begin{equation}}   \nc{\eeq}{\end{equation}}
\nc{\bea}{\begin{eqnarray}}   \nc{\eea}{\end{eqnarray}}
\nc{\baa}{\begin{array}}      \nc{\eaa}{\end{array}}
\nc{\bit}{\begin{itemize}}    \nc{\eit}{\end{itemize}}
\nc{\ben}{\begin{enumerate}}  \nc{\een}{\end{enumerate}}
\nc{\bce}{\begin{center}}     \nc{\ece}{\end{center}}

\nc{\non}{\nonumber}

\begin{document}

\baselineskip=17pt


\thispagestyle{empty}
\vspace{20pt}
\font\cmss=cmss10 \font\cmsss=cmss10 at 7pt

\begin{flushright}
\today \\
UMD-PP-10-015 \\
\end{flushright}

\hfill

\begin{center}
{\Large \textbf
{ Using $M_{ T2 }$ to Distinguish Dark Matter Stabilization Symmetries}}
\end{center}

\vspace{15pt}

\begin{center}
{\large Kaustubh Agashe$\, ^{1}$, Doojin Kim$\, ^{1}$,
Devin G. E. Walker$\, ^{2}$ and Lijun Zhu$\, ^{1}$} \\
\vspace{15pt}
$^{1}$\textit{Maryland Center for Fundamental Physics,
     Department of Physics,
     University of Maryland,
     College Park, Maryland 20742, U.~S.~A.}
\\
and
\\
$^{2}$\textit{Center for the Fundamental Laws of Nature, Jefferson
  Physical Laboratory, Harvard University, Cambridge, Massachusetts 02138,
  U.~S.~A.}
\end{center}

\vspace{5pt}

\begin{center}
\textbf{Abstract}
\end{center}
\vspace{5pt} {\small \noindent
We examine the potential of using colliders to distinguish models with parity ($Z_2$) stabilized dark matter (DM) from models in which the DM is stabilized by other symmetries, taking the latter to be a $Z_3$ symmetry for illustration.  The key observation is that a heavier mother particle charged under a $Z_3$ stabilization symmetry can decay into {\em one or two} DM particles along with Standard Model (SM) particles.  This can be contrasted with the decay of a mother particle charged under a parity symmetry; typically, only one DM particle appears in the decay chain.  The arXiv:1003.0899 
studied the distributions of {\em visible} invariant mass from the 
decay of a {\em single} such mother particle in order to highlight the resulting distinctive signatures of $Z_3$ 
symmetry versus parity symmetry stabilized dark matter candidates.  We now describe a complementary study which focuses on decay chains of the
{\em two} 
%
%
mother particles which are necessarily present in these events.
%
%
We also
include in our analysis 
the {\em missing} energy/momentum in the event.  For the $Z_3$ 
symmetry 
stabilized mothers, the resulting inclusive final state can have two, three or 
four 
DM
%
%
particles.  In contrast, models with $Z_2$ 
symmetry can have only two.  We show that the shapes and edges of the distribution of $M_{T 2}$-type variables, along with ratio of the visible momentum/energy on the two sides of the event, are powerful in distinguishing these different scenarios.  Finally we conclude by outlining future work which focuses on reducing combinatoric ambiguities from reconstructing multi-jet events.  Increasing the reconstruction efficiency can allow better reconstruction of events with two or three dark matter candidates in the final state.}

\vfill\eject
\noindent


\section{Introduction}

A stable weakly interacting massive particle (WIMP) -- with a mass also
of order the weak scale -- is a well-motivated candidate for dark matter (DM) in the Universe since it approximately has the correct relic density upon thermal freeze-out~\cite{Bertone:2004pz}.  
Such a particle also often arises in extensions of the Standard Model (SM), especially those motivated by solutions to the Planck-weak hierarchy problem of the SM.  Finally, if the WIMP is 
a part of an extension of the SM, then it is likely to have (weak) interactions with 
SM particles. Hence, the WIMP paradigm
can be tested via {\em non}-gravitational methods, for example, direct/indirect detection of cosmic DM or production of the DM at colliders.  The latter is our interest here.

The collider searches of the DM paradigm typically involve producing a {\em heavier} particle charged under the
same symmetry which stabilizes the DM. Such a ``mother'' particle must decay to SM particles 
and DM, manifesting as missing energy, along with a SM final state.  
Reconstructing the decay chains leading to such events
will enable us to determine the masses of the DM and the mother particles;
it is advantageous to do this in a model-{\em independent} manner.
A tremendous amount of 
effort has been put-in into such a research
program, especially at the CERN Large Hadron Collider (LHC), see, for 
example~\cite{Lester:1999tx, Barr:2002ex,
Cho:2007qv, Gripaios:2007is, Barr:2009jv, Lester:2007fq, Cheng:2008hk, Burns:2008va, Konar:2009qr, Nojiri:2008hy, 
Tovey:2008ui, Cho:2009ve, Barr:2010ii, Serna:2008zk, Cohen:2010wv,
Cheng:2007xv, Burns:2008cp, Nojiri:2010dk, Webber:2009vm, Han:2009ss, Tovey:2010de}.
%
%
See also \cite{Barr:2010zj} for a review.  Most of 
this work still 
assumes that 
%
%
a $Z_2$/parity 
symmetry stabilizes the DM (henceforth called $Z_2$ models).  
This is partly because there are several models with such a 
symmetry~\cite{Jungman:1995df, Lee:2008pc, Cheng:2003ju, Servant:2002aq, Agashe:2007jb}.  However, while $Z_2$ might be the simplest possibility for such a symmetry, it is by no means the only one \cite{Walker:2009en,Ma:2007gq, 
Agashe:2004ci,Batell:2010bp}.  So, in previous work \cite{Walker:2009ei,
Agashe:2010gt}, 
the program of instead {\em determining} the DM stabilization symmetry from collider data was started.  We focused on how to distinguish a $Z_3$ DM stabilization symmetry (henceforth called a $Z_3$ model) from a $Z_2$ model\footnote{See reference \cite{D'Eramo:2010ep} for 
how calculation of relic density and indirect detection might be modified in a $Z_3$ model vs. a $Z_2$ model.  Also see reference \cite{Walker:2009ei} for a collider study
of non-$Z_2$ models, but with 
decays of mother particles into DM occurring {\em outside} the detector 
vs. the case of such
decays taking place inside the detector studied here and in reference \cite{Agashe:2010gt}.}.  
We have focused on $Z_3$ models for simplicity and definiteness.  However, we emphasize our techniques in this and in previous papers~\cite{Walker:2009ei,Agashe:2010gt} can be generalized to distinguish most other 
DM
%
%
stabilization symmetries from parity symmetry stabilized DM.
%
%

The basic idea behind distinguishing $Z_3$ from $Z_2$ models is that a 
\begin{itemize}

\item
{\em single} mother charged under a $Z_3$ symmetry is allowed (based simply on 
the symmetry) to decay into 
\textit{one} or \textit{two} DM candidates.  

\end{itemize}
This is to be contrasted with the fact that mother particles charged under a $Z_2$ symmetry have only \textit{one} DM candidate in the final state\footnote{In both $Z_3$ and $Z_2$ models, a mother can decay into {\em three} DM particles, but we will not consider this possibility here 
for simplicity and since it is expected to be phase-space suppressed compared to the 
other possibilities.}.  As discussed in reference \cite{Agashe:2010gt}, 
decays of a {\em single} 
$Z_3$-charged mother particle generate a ``double edge'' in the invariant mass distribution of the {\em visible} (SM) particles.  
This is with the condition that the intermediate particles in the decay 
chains are off-shell
and that the decay chains with one and two DM contain identical
SM particles.  For the case of on-shell intermediate particles, it was shown in \cite{Agashe:2010gt} that this invariant mass distribution has a ``cusp'' for 
certain decay topology (with two DM) of a $Z_3$ mother particle.
 In all, \cite{Agashe:2010gt} focused on new features in 
observables from a 
single decay chain only.  

In this work we consider the total inclusive event in order to glean even 
more information, recalling that there must be two such mother particles
present.  
%
%
%
For example, 
consider 
%
%
the case where there is only one visible (SM) particle in the 
decay chain of a mother particle. 
%
%
Constructing the invariant mass of the visible particle of this 
decay chain, as per the analysis of \cite{Agashe:2010gt}, is not very useful
for the purpose of reconstructing the mass of the mother
particle: 
%
%
one might have to resort to including information 
%
%
about the invisible
particle(s) in the same decay chain. 
%
%
Since we can only measure the {\em total} missing transverse momentum
in the event which is {\em shared} between
invisible particles from two mothers, we must use
measurements from both sides.  An option is to use ``$M_{ T2 }$''-type observables/variables
\cite{Lester:1999tx, Barr:2002ex,
Cho:2007qv, Gripaios:2007is, Barr:2009jv, 
Lester:2007fq, Cheng:2008hk, Burns:2008va, Konar:2009qr, Nojiri:2008hy, 
Tovey:2008ui, Cho:2009ve, Barr:2010ii, Serna:2008zk, Cohen:2010wv}
%
%
Another case where one of the analyses of 
reference \cite{Agashe:2010gt} (based on
single mother decay) might not work
is when the visible/SM particles in the decay chains
with one and two DM (of course for $Z_3$ model) are not identical (even if
they are more than one). Thus, one does not obtain a double edge
for the case of intermediate particles in the decay chains being {\em off}-shell.

With the above motivations in mind, in this paper, 
\begin{itemize}

\item
we develop techniques for 
distinguishing $Z_3$ models
from $Z_2$ models using information from {\em both} mother decays and the
{\em missing} (in addition to visible) energy/momentum in an event.

\end{itemize}
We especially study the above cases where the techniques of
reference \cite{Agashe:2010gt} might not work --
in this sense, our present work is {\em complementary} to that
of reference \cite{Agashe:2010gt}.
%
%
%
%
%
%
%
%
%
\begin{itemize}

\item
We show that shapes and edges of these $M_{ T2 }$ distributions,
along with the ratio of visible momentum/energy on the two sides of the event, act as
powerful
discriminants between $Z_3$ and $Z_2$ models (including the case of a neutrino, i.e.,
mass{\em less} invisible particle -- in addition to DM, in the final
state for $Z_2$ models).

\end{itemize}
An outline of this paper is as follows: In the next section,
we begin with a review of $M_{T2}$ variable in $Z_2$ models. We present some
important formulas such as the location of maximum value of the $M_{T2}$
distribution, and discuss some interesting features
%
%
such as a ``kink'' in the maximum
$M_{T2}$ as a function of the ``trial'' DM mass in the case where more than one visible particle is involved in each decay chain. In Sec.~\ref{sec:mt2z3} we move on to the $M_{T2}$ variable in $Z_3$ models. We define three different
event types arising in $Z_3$ models based on the number of dark matter particles in the final state and provide their corresponding theoretical predictions 
of the maximum $M_{T2}$.  In particular, we discuss the conditions to have a kink in maximum $M_{ T2}$ as a function of trial DM mass for the cases where there exist more than two DM particles in the final state: such situations do not arise in $Z_2$ models. We further show some simulation
results for $M_{ T2}$ distributions for the new types of events in  $Z_3$ models and discuss some notable features to be used for distinguishing $Z_3$ models from $Z_2$ models.

In the following two sections, we provide detailed
applications 
of our results of the
previous two sections for distinguishing
$Z_3$ from $Z_3$ models. First we consider the easier case where the decay chain with one DM contains visible/SM particle(s) which are
{\em not} identical to the ones in the two DM decay chain
mentioned above (in $Z_3$ models). Based on the theoretical considerations
given in Secs.~\ref{sec:review} and~\ref{sec:mt2z3}, we provide ways of distinguishing $Z_3$ models from $Z_2$
models, as well as measuring
the mother and DM masses. We do it 
for both the  case of one visible/SM particle
in the two decay chains and more than one
visible/SM particle case. In the next section we deal with the case where one DM and two DM decay chains contain identical visible/SM particle(s), and
discuss {\em additional} techniques required in this case
to distinguish $Z_2$ and $Z_3$ models.  In all of the examples above, we make the simplifying assumption that the intermediate particles in the decay chain are \textit{off}-shell, i.e., they are heavier than the mass of their mother particle.  In section~\ref{sec:onshell}, we 
briefly mention some aspects 
%
%
of 
the case of intermediate particles being on-shell.
We next conclude and show details of some of the calculations in the Appendix.


\section{A Review of $M_{T2}$ for $Z_2$ Models}
\label{sec:review}

For simplicity, in this paper
\begin{itemize}

\item
we consider pair-production of a \textit{single} type of
mother particle which is charged under the DM stabilization symmetry.

\end{itemize}
We also assume that the {\em total} transverse momentum
of the two mother particles produced in an event is zero, for
example, we neglect any initial/final state radiation. 
In $Z_2$ models, each such mother decays into SM/visible particle(s) and one DM/invisible
particle\footnote{We assume that there is only one type of DM particle in this
(and similarly the $Z_3$) model so that the
invisible/DM particles in each decay chain are identical.}.
Furthermore,  it is assumed 
\begin{itemize}

\item

we know which {\em visible} particle(s) originate from which decay
chain.\footnote{Of course, for $Z_3$ models, we do not know which decay chain emits one or two
DM particles.} For example, if the pair-produced mother particles are boosted
sufficiently, their decay products are likely to be collimated so that
the visible particles coming from the same decay chain are detected in the
same hemisphere in the collider. For alternate methods
of determining the correct assignment of visible particles to the
two decay chains, see reference 
\cite{Rajaraman:2010hy} and section~\ref{sec:algorithm}.  
%
%

\end{itemize}
The $M_{T2}$ variable~\cite{Lester:1999tx, Barr:2002ex} is a generalization of
the transverse mass\footnote{Of course, the usual transverse mass assumes only a single mother particle.} to this case.
%
%
Specifically,
for each event, it is defined to be a minimization of the maximum of the two
transverse masses in each decay chain under the constraint that the sum of all
the transverse momenta of the visible and invisible particles
vanishes~\cite{Lester:1999tx, Barr:2002ex}:
\bea
M_{T2}\equiv \min_{\textbf{p}_T^{v(1)}+\textbf{p}_T^{v(2)}+
\tilde{\textbf{p}}_T^{(1)} + \tilde{\textbf{p}}_T^{(2)}
%
%
=0}\left[\max\left\{M_T^{(1)},\;M_T^{(2)} \right\} \right]
\eea
where $\textbf{p}_T^{v(i)}$ denote the vector {\em sum} of visible
transverse momenta and $\tilde{\textbf{p}}_T^{(i)}$ denote the
transverse momentum of the invisible particle
in the \textit{i}th decay chain ($\textit{i}=1,\;2$): the minimization is performed over the
latter momenta.
$M_T^{(i)}$ is the usual transverse mass:
\bea
\left(M_T^{(i)} \right)^2=\left(m_T^{v(i)} \right)^2+\tilde{m}^2+2\left(E_T^{v(i)}\tilde{E}_T^{(i)}-\textbf{p}_T^{v(i)}\cdot
\tilde{\textbf{p}}_T^{(i)} \right)
\eea
where $E_T^{v(i)}$ and
$m_T^{v(i)}$ are (respectively) the transverse energy and transverse
mass formed by all visible particles belonging to the same decay chain.
The variables with a tilde represent the corresponding quantities formed by the invisible particle in the same decay chain.
Note that the mass of the invisible particle $\tilde{m}$ should
be regarded as a unknown/free parameter because we are not aware of it in advance, and henceforth we call it  ``trial'' DM mass. In this sense $M_{T2}$ should be considered to be a function of the trial DM mass
$\tilde{m}$, and its maximum value among many events (which will be used
extensively in the following) is defined as
\bea
M_{T2}^{\max}(\tilde{m})=\max_{\hbox{many events}}\left[ M_{T2}(\tilde{m}) \right].
\eea
Obviously $M_{T2}^{\max}(\tilde{m})$ is also a function of the trial DM mass (see App.~\ref{app:locofmaxmt2} for details). An important result to be noted is that if there are a sufficient number of events and the actual DM mass is substituted into $\tilde{m}$, then the above-given $M_{T2}^{\max}$ becomes the actual mass of the pair-produced mother particles~\cite{Lester:1999tx, Barr:2002ex}:
\bea
M_{T2}^{\max}(\tilde{m}=m_{DM})=M
\eea
where $M$ and $m_{DM}$ indicate the true masses of mother and DM, respectively.

In order to see how this $M_{T2}$ analysis is applied to realistic situations, we first take the case where there exists a single visible/SM particle in each decay chain, and then move on to the case where there exists more than one visible/SM particle in each decay chain.
A similar analysis can be done for the mixed case, i.e., one visible particle on one side and more than one on the
other.

\subsection{One Visible/SM Particle in Each Decay Chain}
In this case the upper edge in $M_{T2}$ distribution is obtained by
``balanced''~\cite{Lester:1999tx, Barr:2002ex, Cho:2007qv} 
solution (see App.~\ref{app:locofmaxmt2}
for details).
\bea
M_{T2}^{\max}=M_{T2}^{\max,bal}= \sqrt{\frac{(M^2-m_{DM}^2)^2}{4M^2}}
+\sqrt{\frac{(M^2-m_{DM}^2)^2}{4M^2}+\tilde{m}^2}
\label{eq:mt2maxz2}
\eea
Here (and henceforth)
\begin{itemize}

\item
we assume that all visible particles are massless for simplicity.

\end{itemize}
As mentioned
earlier, the above-given upper edge is a function of the trial DM mass $\tilde{m}$ and one can see
that it reduces to the true mother mass $M$ with $\tilde{m}$ equal to the true DM mass $m_{DM}$.
\begin{figure}[t]
	\centering
	 \includegraphics[width=8.0truecm,height=8truecm,clip=true]{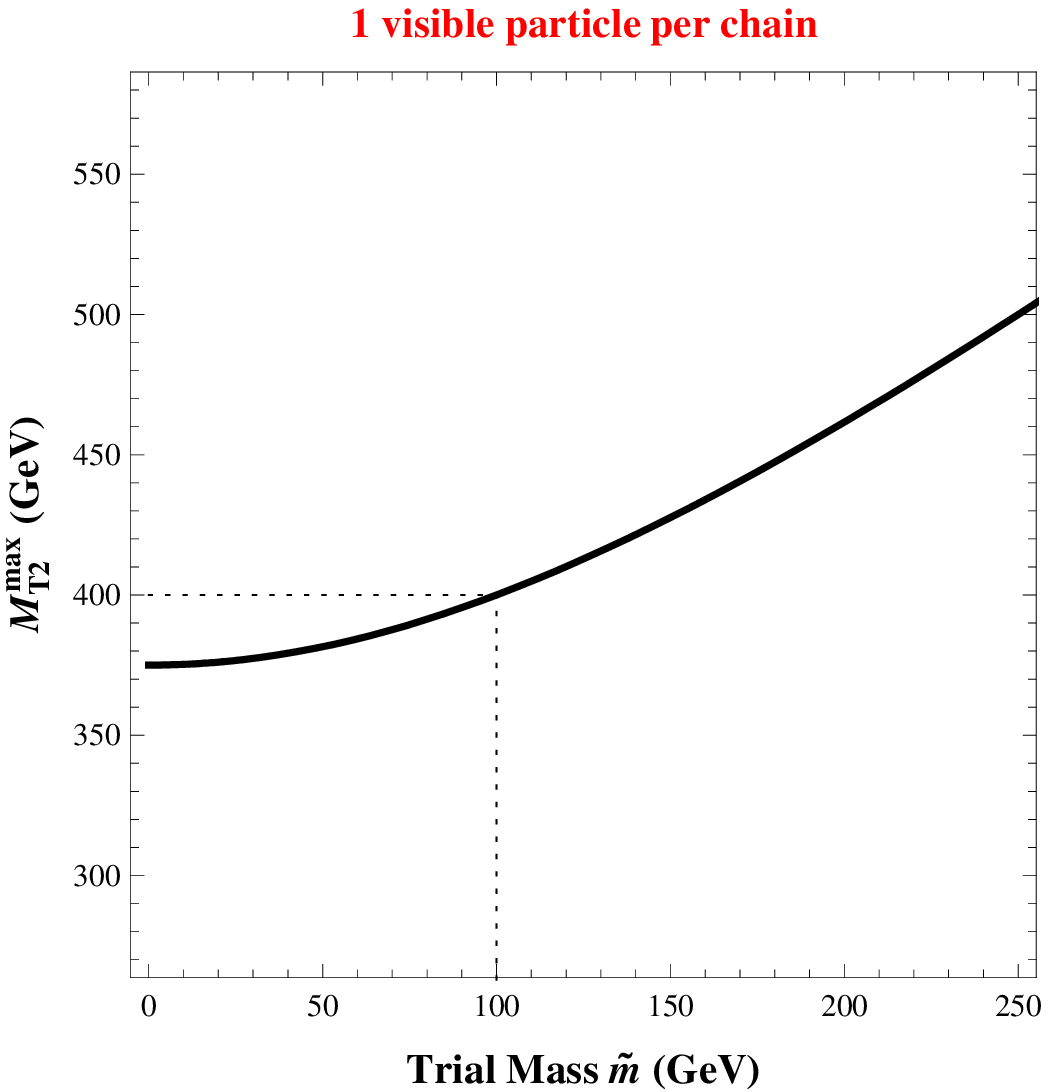}
	\hspace{0.2cm}
	 \includegraphics[width=8.0truecm,height=8truecm,clip=true]{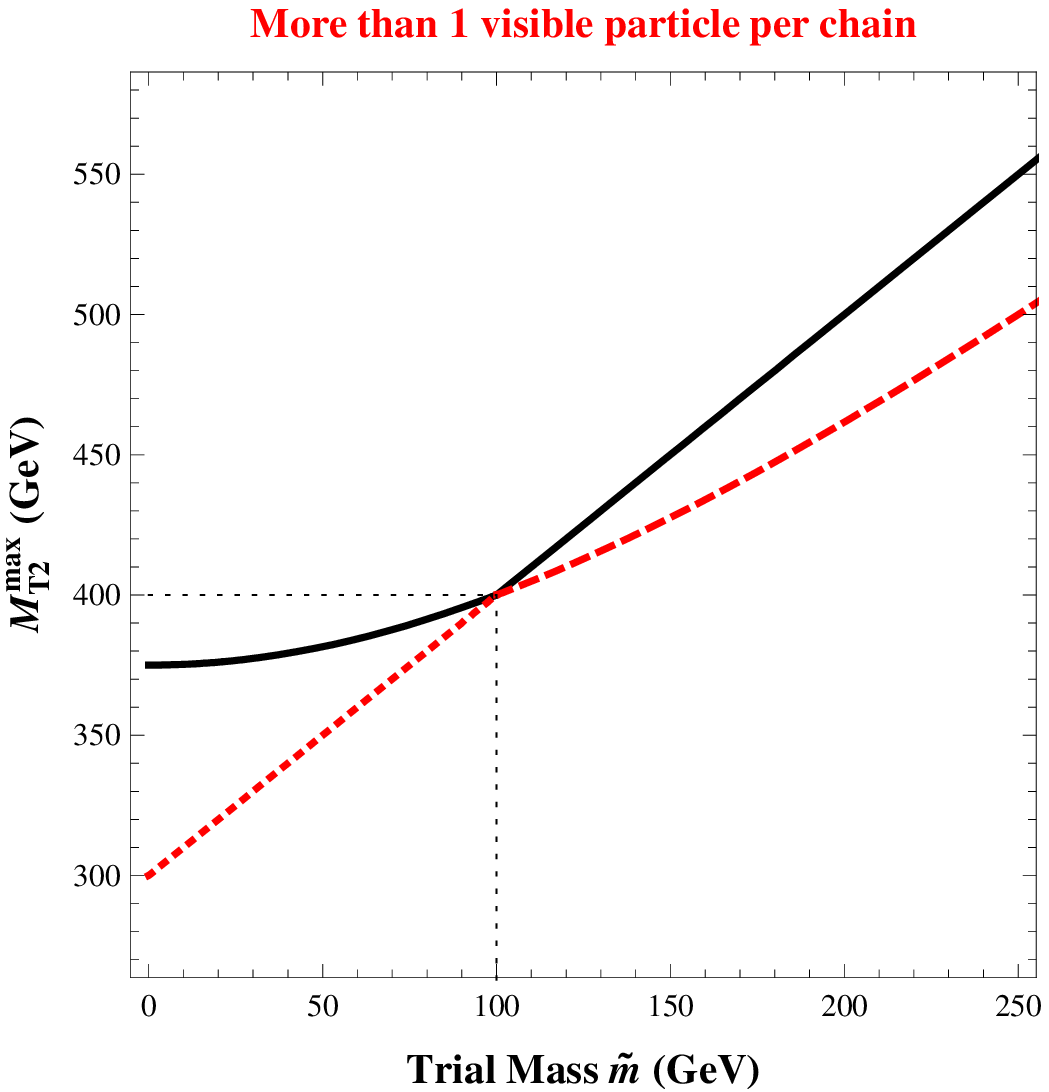}
	\caption{Theoretical expectation of $M_{T2}^{\max}$ versus the trial mass
	$\tilde{m}$ for $Z_2$ events.
The masses of mother and DM particles are 400 GeV and
	100 GeV, respectively. The left panel shows the case where there exists only a single visible particle per chain.
%
%
The right panel shows the case where there exists more than one visible particle per decay chain. In both panels,
the solid black curve represents the overall/net upper edge, $M_{T2}^{\max}$.
In the right panel,
the dotted straight line which extends into the right-hand part of the solid line
is the  $M_{T2}^{\max}$ for the unbalanced solution, whereas the dashed line which extends into the left-hand part of the solid
line is that for the balanced solution.}
%
%
%
	\label{fig:mt2maxz2}
\end{figure}
The left panel of Fig.~\ref{fig:mt2maxz2} shows the above theoretical prediction for the location
of maximum $M_{T2}$ for $Z_2$ models as a function of the trial mass $\tilde{m}$.
We used 400 GeV and 100 GeV as mother and DM particle masses. As expected from
Eq. (\ref{eq:mt2maxz2}), the curve in the figure ``smoothly'' increases with $\tilde{m}$
(cf. the following section), and that the  $M_{T2}^{\max}$ value corresponding to $\tilde{m}=m_{DM}$ (here 100 GeV) is the same as the true mother particle mass (here 400 GeV: see
the black dotted lines).

\subsection{More than One Visible/SM Particle in Each Decay Chain}
Once there exists more than one visible particle per decay chain, another type of solution to $M_{T2}$, denoted by ``unbalanced'' ~\cite{Lester:1999tx, Barr:2002ex, Cho:2007qv}, 
arises. If
\begin{itemize}

\item
we assume that the intermediate particles in the decay chains are \textit{off}-shell, i.e., heavier than
their mother particles

\end{itemize}
(as we will
for most of this paper), then the balanced solution in this case
is still given by Eq.~(\ref{eq:mt2maxz2}), and the unbalanced solution is
as follows\footnote{Of course, in general, one can find the expressions for both the balanced and the unbalanced solutions in the case of \textit{on}-shell intermediate particles~\cite{Cho:2007qv}.}:
\bea
M_{T2}^{\max,unbal}=M-m_{DM}+\tilde{m}.
\label{eq:mt2maxz2unbal}
\eea
Hence the overall upper edge in the $M_{T2}$ distribution is
determined by a ``competition'' between balanced and unbalanced solutions:
\bea
M^{ \max } _{ T2 } =\max\left[M_{T2}^{\max ,bal},\;M_{T2}^{\max, unbal} \right] =
\left\{
\begin{array}{l}
M-m_{DM}+\tilde{m} \hspace{4cm}\hbox{for $\tilde{m}\geq m_{DM}$} \cr
\cr
\sqrt{\frac{(M^2-m_{DM}^2)^2}{4M^2}}
+\sqrt{\frac{(M^2-m_{DM}^2)^2}{4M^2}+\tilde{m}^2}
\hspace{0.6cm} \hbox{for $\tilde{m}\leq m_{DM}$}.
\end{array}\right.
\label{eq:kinkZ2}
\eea
Note that the $M_{T2}^{\max}$ shows different functional behaviors depending on the relative size of the trial DM mass to the true DM mass.
As a result, $M_{T2}^{\max}$ is no longer smoothly increasing with $\tilde{m}$ in contrast to the case with one visible/SM per decay chain. Instead, there arises a ``kink'' at the location of the actual DM mass, with the
corresponding $M_{T2}^{\max}$ being the actual mother
mass~\cite{Cho:2007qv}\footnote{A similar kink
also appears for the case of {\em one} visible particle
in each decay chain {\em if} the total transverse
momentum of the two mother particles is {\em non}-zero, for example,
in the presence of initial/final state radiation (see \cite{Gripaios:2007is}), but
(as mentioned earlier) we neglect this possibility for simplicity.}.
This is illustrated in the
right panel of Fig.~\ref{fig:mt2maxz2} where the upper edges for the two possible types of solutions $M_{T2}^{\max ,unbal}$ and $M_{T2}^{\max ,bal}$ are shown
by a straight line (i.e., the dotted line, which extends into the right-hand part of
the solid line) and a dashed curve (which extends into the
left-hand part of the solid curve), respectively. The upper edge in the $M_{T2}$ distribution
is given by the larger of these two values, i.e., the black solid curve,
and shows a kink (indicated by the black dotted lines), at $\tilde{m} = 100$ GeV
and $M_{ T2 }^{\max} = 400$ GeV (as expected).


\section{$M_{T2}$ for $Z_3$ Models}
\label{sec:mt2z3}
To begin with, we would like to reiterate some of the relevant features of $Z_3$ symmetry
in order to avoid any possible confusion later. Under the $Z_3$ symmetry, a particle/field $\phi$ transforms as
\begin{eqnarray}
\phi &\longrightarrow& \phi \exp \left(\frac{2\pi i q}{3}\right)
\end{eqnarray}
where $q=0$ (i.e., neutral) or +1, +2. Suppose the lightest of the $Z_3$ charged particles (labeled $\phi_0$) has charge $q=+1$ (similar argument can be made for charge $q=+2$). Clearly, its anti-particle ($\bar{\phi}_0$) has (a different) charge $q=-1$ (which is equivalent to $q=+2$)
%
%
and has the same mass as $\phi_0$. Then, solely based on $Z_3$ symmetry considerations, all other (heavier) $Z_3$-charged particles can decay into this lightest $Z_3$-charged particle or its anti-particle (in addition to $Z_3$-neutral particles, including SM particles). To be explicit, a heavier $Z_3$-charged particle with charge $q=+1$ can decay into either (single) $\phi_0$ or \textit{two} $\bar{\phi}_0$'s (and $Z_3$-neutral particles). Taking the CP conjugate of the preceding statement, we see that a heavier $Z_3$-charged particle with the other type of charge, namely $q=2$, is allowed to decay into \textit{two} $\phi_0$'s or single $\bar{\phi}_0$. Of course, $\phi_0$ or $\bar{\phi}_0$ cannot decay and thus is the (single) DM candidate in this theory. We will denote this DM particle and its anti-particle by DM and $\overline{\textnormal{DM}}$, respectively, although we do not make this distinction in the text since DM and anti-DM particles are still degenerate.\footnote{Of course, which of the two particles is denoted anti-DM is a matter of convention. Also, as a corollary, the DM
particle should be Dirac fermion or complex scalar in a $Z_3$ model.}

According to the above-given argument, for $Z_3$ models, each mother particle can emit either one or two DM particles so that there exist two, three, or four DM particles in the final state
(for pair-production of mothers)
while there are only two DM particles for $Z_2$ models.
We therefore expect richer structures in the $M_{T2}$ distribution 
for $Z_3$ models.
Here
\begin{itemize}

\item
%
%
we take as an ansatz 
%
%
only a
{\em single} DM particle in each decay chain for the sole purpose of defining
$M_{T}$,
even if there could be two DM particles in either or both of the two decay chains.

\end{itemize}

%
%
We do so for the following two reasons.
%
%
First, in the real collider experiment, there is (a {\em priori})
no clear information on the number of invisible particles involved in the decay process of interest so that each individual decay chain 
with only one DM is a natural (starting) assumption.
%
%
Moreover, one can naturally expect (and we will show)
that decay events from $Z_3$ models will show different features in the $M_{T2}$ analysis compared with those for $Z_2$ models. Therefore, starting with a $Z_2$ assumption and deriving a
``contradiction'' in the $M_{T2}$ analysis, we can distinguish $Z_3$ models 
from $Z_2$ ones (which is our primary goal here). 
%
%
In this context, we call such an analysis imposing one-DM-per-chain assumption the ``naive'' $M_{T2}$ analysis. 

\begin{figure}

\centering
	 \includegraphics[width=5.0truecm,height=4.0truecm,clip=true]{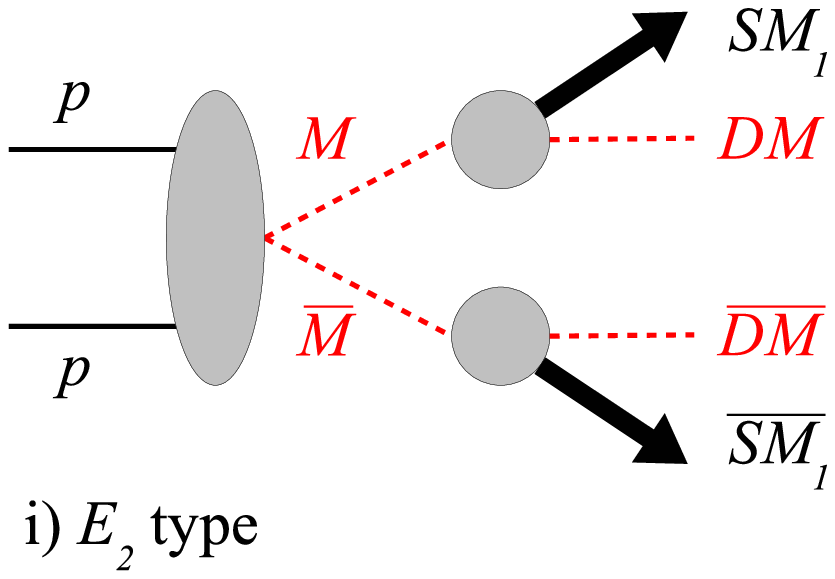}
	\hspace{0.2cm}
	 \includegraphics[width=5.0truecm,height=4.0truecm,clip=true]{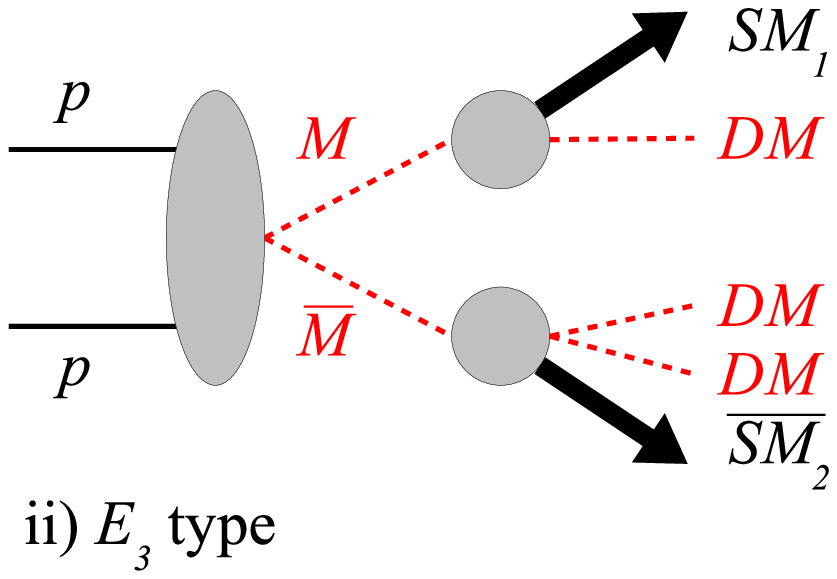}
    \hspace{0.2cm}
	 \includegraphics[width=5.0truecm,height=4.0truecm,clip=true]{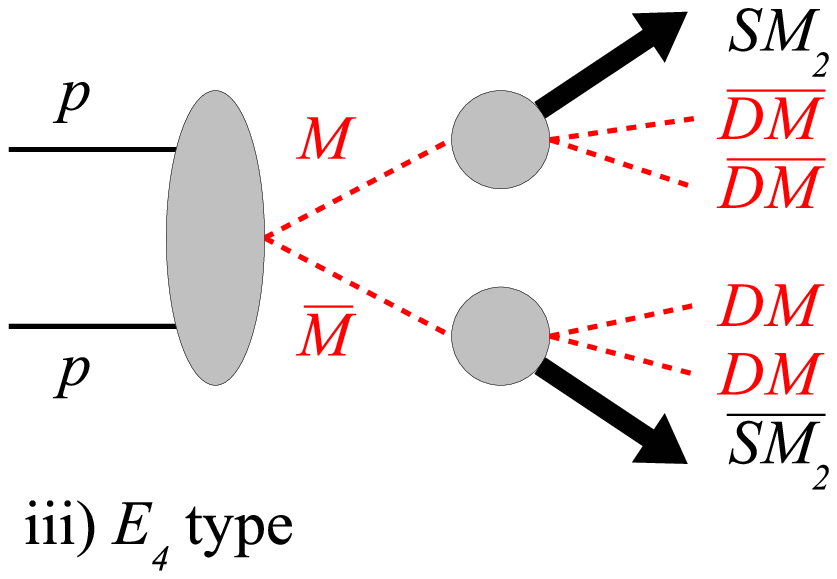}

\caption{The three types of events in $Z_3$ models, based on the total
number
of DM in the event. ``M" denotes the mother particle. Each SM final state can have more than one particle, and the subscripts 1 and 2 on SM denote the visible/SM final states in the decay chains with one and two DM, respectively.
Note that, based simply on the $Z_3$ symmetry, if a mother decays into DM,
then the {\em same} mother decays into two {\em anti}-DM in the other decay chain.
Since DM and anti-DM  have same mass
and are {\em not} detected, we neglect this distinction between the two henceforth (we
already did so thus far). For simplicity, we will also henceforth not differentiate between
SM and $\bar{\hbox{SM}}$ or between $M$ and $\bar{M}$.}

\label{E234}

\end{figure}

For a more systematic consideration let us define the three different events
having different numbers of (a single type of) DM particles as $E_2$, $E_3$, and $E_4$-type, 
respectively,
i.e., each subscript on $E$ simply implies the total number of DM particles in
the final state: see Fig.~\ref{E234}, where
SM$_{ 1, \; 2 }$ denote the visible/SM final states (irrespective
of the actual number of particles in the state) in the decay
chains with one and two DM (respectively)\footnote{These two SM final states
might not or might be identical: we will return to these two possibilities
in the next two sections (respectively).}. Here the red dashed lines denote any particles charged under dark matter stabilization symmetry (in this section $Z_3$ symmetry) while the black solid lines/arrows denote any visible/SM particles.
One should note that $E_4$ type events represent the case with 2 DM particles in each
decay chain. Also, both decay chains (with one and two DM) might
not exist for a {\em specific} mother so that all three types of events
might not occur.
%
%
Like in $Z_2$ models, we start with the case with one visible/SM particle in each decay chain, and we consider the case with more than one visible/SM particle in each decay chain in the following subsection.
While doing so, we see how the $M_{T2}$ analysis applied to $Z_3$ models contrasts
with $Z_2$ models.

\subsection{One Visible/SM Particle in Each Decay Chain}
In this case the upper edge in the (naive) $M_{T2}$ distribution
is determined only by the balanced solution like $Z_2$ models,
and the analytic expressions for the three type events are given as follows (see App.
\ref{app:locofmaxmt2} for details):
\bea
M^{ \max}_{ T2,E_2 } =
M^{ \max,bal }_{ T2,E_2 }  =\sqrt{\frac{(M^2-m_{DM}^2)^2}{4M^2}}
+\sqrt{\frac{(M^2-m_{DM}^2)^2}{4M^2}+\tilde{m}^2} \hspace{5.1cm} \hbox{for $E_2$}
\label{eq:maxcase1} \\
M^{ \max}_{ T2,E_3 } = M^{ \max,bal }_{ T2,E_3 } = \sqrt{\frac{(M^2-m_{DM}^2)(M^2-4m_{DM}^2)}{4M^2}}
+\sqrt{\frac{(M^2-m_{DM}^2)(M^2-4m_{DM}^2)}{4M^2}+\tilde{m}^2}
\hspace{0.5cm}\hbox{for $E_3$}
\label{eq:maxcase2} \\
M^{ \max}_{ T2,E_4 } = 
M^{ \max,bal }_{ T2,E_4 } = \sqrt{\frac{(M^2-4m_{DM}^2)^2}{4M^2}}
+\sqrt{\frac{(M^2-4m_{DM}^2)^2}{4M^2}+\tilde{m}^2}
\hspace{4.7cm}\hbox{for $E_4$}
\label{eq:maxcase3}
\eea
As a reminder, the events with $E_2$, $E_3$ and $E_4$ represent events with two, three and four dark matter candidates.  Note that Eq.~(\ref{eq:maxcase1}) has the same form as
Eq.~(\ref{eq:mt2maxz2}) in $Z_2$
models
because $E_2$ type events also contain two DM particles in the final state (just
like $Z_2$ models) whereas the other two types of events do \textit{not} appear in $Z_2$ models so that the corresponding
Eqs.~(\ref{eq:maxcase2}) and~(\ref{eq:maxcase3})
(and similar ones later)
are new/do not appear in previous literature.
Substituting $\tilde{m}=m_{DM}$ in Eq.~(\ref{eq:maxcase1})
gives the true mother mass $M$ for the value of $M_{T2}^{\max}$ (as expected), but the other two equations
give
a combination of the true mother and DM masses rather than the true mother mass.
Actually, this is not surprising because we have used the naive $M_{T2}$
variable for 
the $E_{ 3 \; 4 }$-type events in 
$Z_3$ models, whereas the actual physics is different from the physics under which our
$M_{T2}$ variable is defined.
For example, for $E_3$ type events there is an asymmetry between the final states of the two decay chains, which is caused by adding one more DM to either of the two decay chains.
%
%
For $E_4$ type events, even though the two decay chains have symmetric final states, the
``effective'' DM mass is twice the true DM mass so that the true mother mass
(for the value of $M_{T2}^{\max}$) is in fact
obtained by setting $\tilde{m}=2m_{DM}$ instead as clearly seen from Eq.~(\ref{eq:maxcase3}).

\begin{figure}[t]
	\centering
	 \includegraphics[width=8.0truecm,height=8truecm,clip=true]{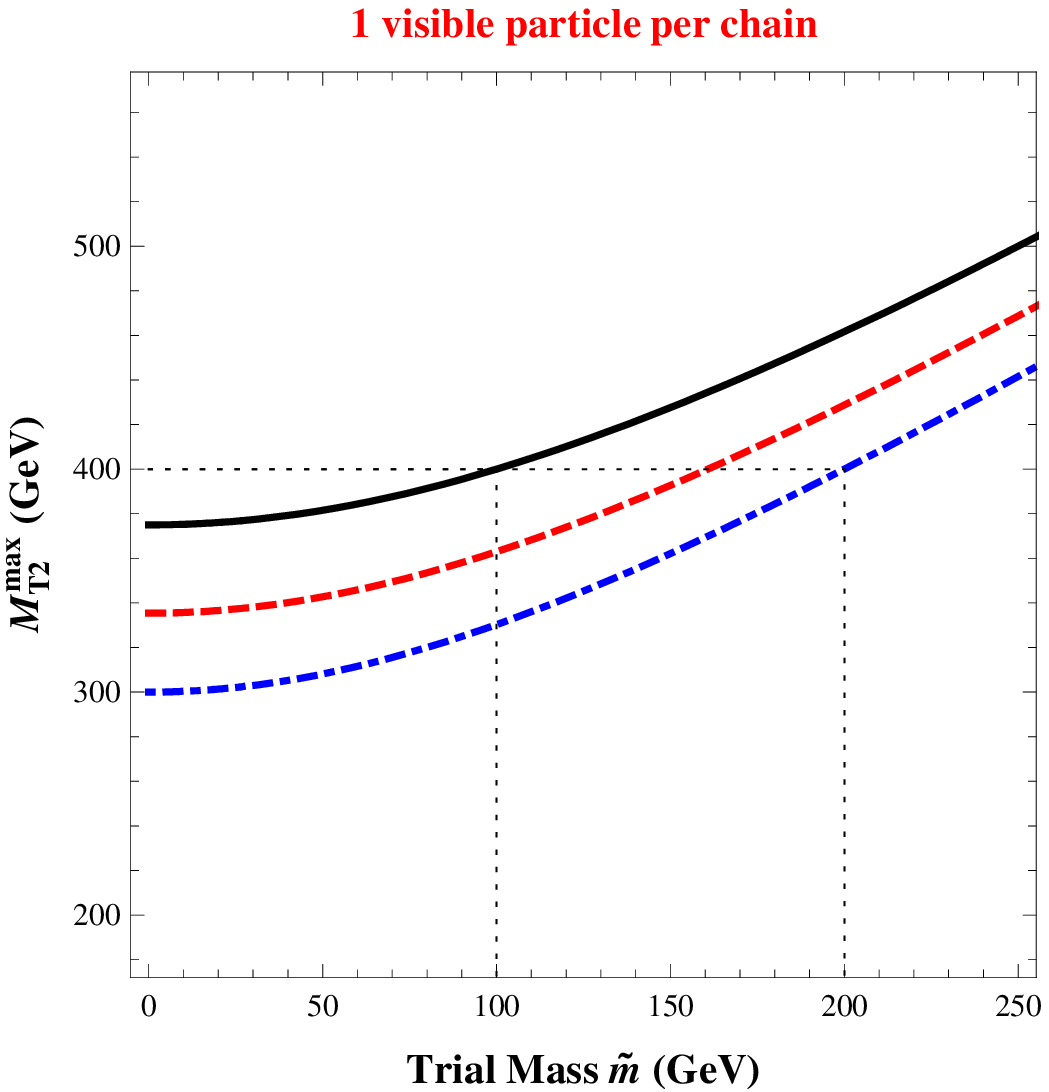}
	\hspace{0.2cm}
	 \includegraphics[width=8.0truecm,height=8truecm,clip=true]{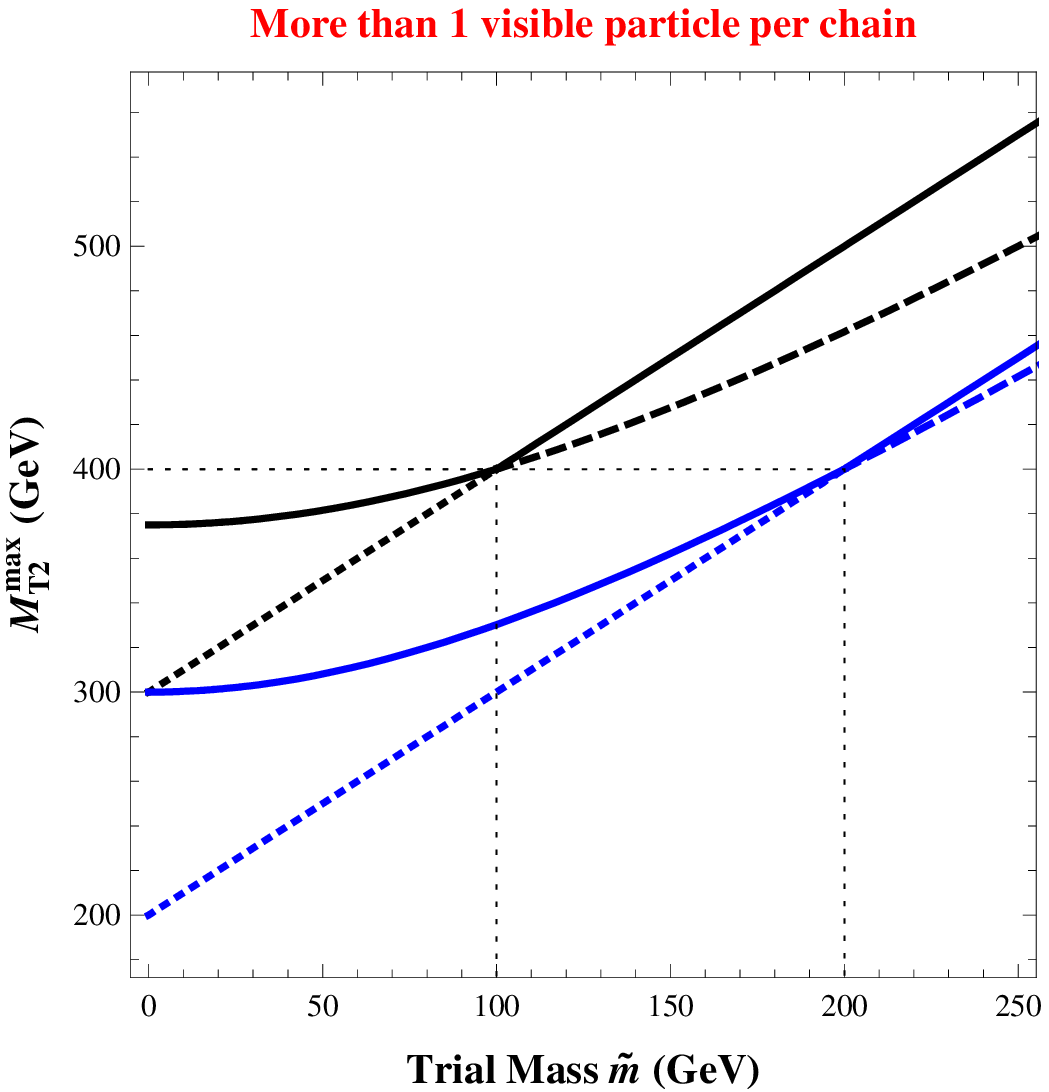}
	\caption{Theoretical expectation of $M_{T2}^{\max}$ versus the trial mass
	$\tilde{m}$ for $Z_3$ model.
The masses of mother and DM particles are 400 GeV and
	100 GeV, respectively. The left panel shows the case where there exists only a single visible particle per chain. The black, red, and blue curves are showing the corresponding $M_{T2}^{\max}$ values to $E_2$, $E_3$, and $E_4$ type events over $\tilde{m}$, respectively.
%
%
The right panel shows the case where there exists more than one visible particle per decay chain.
The overall upper edges, $M_{T2}^{\max}$
for $E_2$ and $E_4$ type events, are given by the solid black and blue curves, whereas the balanced
and unbalanced solutions are denoted by the dashed and dotted curves (respectively)
which merge into the solid curves on the right (left)-hand part.
%
%
The corresponding plot for $E_3$
type events can be found in the next figure.}
	\label{fig:mt2maxz3}
\end{figure}

All of the theoretical predictions mentioned above are demonstrated in the left panel of Fig.~\ref{fig:mt2maxz3}.
Again, we used 400 GeV and 100 GeV as mother and DM masses. The black solid, the
red dashed, and the blue dot-dashed curves represent the theoretical
expectations of $M_{T2}^{\max}$ for $E_2$, $E_3$, and $E_4$ type events, respectively
(the curve for $E_2$-type events is of course the same as in right-hand side of
Fig.~\ref{fig:mt2maxz2}).
As discussed above, the $M_{ T2 }^{ \max }$
for $E_2$ type events corresponds to
the true mass of the mother
particle (here 400 GeV) with the trial DM mass equal to the true DM mass (here 100 GeV)
whereas $E_4$ type events do it for twice the DM mass (here 200 GeV),
as shown by the dotted/black lines.

In addition, there are a couple of features
to be noted; there is no kink arising in the $M_{T2}^{\max}$ curves for $E_{ 3, 4}$ -type events
just like the case of a single visible particle per decay chain in $Z_2$ models (or an 
$E_2$-type event
in $Z_3$ models). Second
for any given $\tilde{m}$, the $M_{T2}^{\max}$ values form a hierarchy of
\bea
M^{ \max }_{ T2,E_2 } >M^{ \max }_{ T2,E_3 }> M^{ \max }_{ T2,E_4 }.
\label{eq:mt2maxhier}
\eea

\subsection{More than One Visible/SM Particle in Each Decay Chain}
\label{sec:z3morethanone}
Once more visible particle(s) are added in each decay chain, one could naturally expect that a kink appears like
in $Z_2$ models. The reason is that, just like for $Z_2$ models, the maximum unbalanced solutions take part in determining the overall upper edge in the $M_{T2}$ distribution together with the balanced solutions. It turns out, however, that this expectation
is true only for $E_2$ and $E_4$ type events which we discuss to begin with. Again, assuming the intermediate particles are \textit{off}-shell the maximum values of the balanced solutions for $E_2$ and $E_4$ type events are simply given (as for
the one visible particle case) by
Eqs.~(\ref{eq:maxcase1}) and~(\ref{eq:maxcase3}), respectively, and those of the unbalanced solutions are given as follows:
\bea
M_{T2,E_{2}}^{\max,unbal}&=&M-m_{DM}+\tilde{m}\hspace{3cm}\hbox{for }E_{2}
\label{eq:mt2unbalz3} \\
M_{T2,E_{4}}^{\max,unbal}&=&M-2m_{DM}+\tilde{m}\hspace{2.8cm}\hbox{for }E_{4}.
\label{eq:mt2unbalz34}
\eea
Here Eq.~(\ref{eq:mt2unbalz3}) is of exactly the same form as Eq.~(\ref{eq:mt2maxz2unbal}) due to the similarity between the decay structures for $Z_2$ models
and $E_2$ type events while Eq.~(\ref{eq:mt2unbalz34}) for $E_4$-type events
is relevant only for $Z_3$ models,
i.e., it is not present in $Z_2$ models.
More quantitatively, the above-given two equations differ by $m_{DM}$ for any given $\tilde{m}$ because one more DM particle is emitted in both decay chains for $E_4$ type events compared with $E_2$ type events (see Eqs.~(\ref{eq:rangeinvis}) and~(\ref{eq:mt2unbalgen}) in  App.~\ref{app:locofmaxmt2}).

As mentioned for $Z_2$ models,
the maximum $M_{T2}$ values are given by the larger of the balanced and unbalanced solutions:
for $E_2$ type events as in Eq.~(\ref{eq:kinkZ2}), and for $E_4$ type events by
\bea
M^{ \max } _{ T2,E_4 }  = \max\left[M_{T2,E_4}^{\max ,bal},\;M_{T2,E_4}^{\max, unbal} \right] =
\left\{
\begin{array}{l}
M-2m_{DM}+\tilde{m} \hspace{4.1cm}\hbox{for $\tilde{m}\geq 2m_{DM}$} \cr
\cr
\sqrt{\frac{(M^2-4m_{DM}^2)^2}{4M^2}}
+\sqrt{\frac{(M^2-4m_{DM}^2)^2}{4M^2}+\tilde{m}^2}
\hspace{0.6cm} \hbox{for $\tilde{m}\leq 2m_{DM}$}.
\end{array}\right.
\label{eq:kinke4}
\eea
Note that there is also a
kink for $E_4$-type events as seen from Eq. (\ref{eq:kinke4}), but
at $\tilde{m} = 2 m_{ DM }$ (i.e., not $m_{DM}$), because
the effective DM mass for $E_4$ is given by $2m_{DM}$.

The right panel of Fig.~\ref{fig:mt2maxz3} illustrates the above theoretical
considerations for $E_2$ and $E_4$ type events
(of course the curve for
$E_2$-type events is the same as the right-hand side of Fig. \ref{fig:mt2maxz2}).
As before, the two straight dotted lines which extend into the right-hand parts of the solid lines indicate the maximum values of
the unbalanced solutions while the two dashed curves which extend into left-hand parts of the solid curve
indicate the maximum values of the balanced solutions.
The actual upper edge in the $M_{T2}$ distribution for any $\tilde{m}$ is given by
the black (for $E_2$ type events) or blue (for $E_4$ type events) solid curves.
Identifying the location of the kink in $E_2$-type events
and its corresponding $M_{T2}^{\max}$
enables us to determine the masses of mother and DM particles separately (just like in $Z_2$ models).
The figure also shows the kink for $E_4$-type events, but located at $\tilde{m} = 200$ GeV (i.e., $2 m_{ DM }$, as expected)
and $M_{ T2 }^{ \max } = 400$ GeV.
This observation can be used as a cross-check for the determination of $M$ and $m_{DM}$
based on $E_2$-type events (again, this feature is
new in $Z_3$ models relative to $Z_2$).

On the other hand, as far as $E_3$ type events
(which are
absent in $Z_2$ models) are concerned, whether or not there exists a kink depends on the mass hierarchy between mother and DM particles
(see App.~\ref{app:kink} for details). Again, assuming \textit{off}-shell intermediate particles the
maximum balanced solution is simply given by Eq.~(\ref{eq:maxcase2}) (just like the
case of one visible particle per decay chain), whereas the maximum unbalanced solution
has the same form as that for $E_2$ type events because one of the two decay chains
still emits a single DM particle in the final state (see Eqs.~(\ref{eq:rangeinvis}) and
~(\ref{eq:mt2unbalgen}) in App.~\ref{app:locofmaxmt2}).
\bea
M_{T2,E_{3}}^{\max,unbal}&=&M-m_{DM}+\tilde{m}\hspace{3cm}\hbox{for }E_{3}
\label{eq:mt2unbalz33}
\eea
\begin{figure}[t]
	\centering
	 \includegraphics[width=8.0truecm,height=8truecm,clip=true]{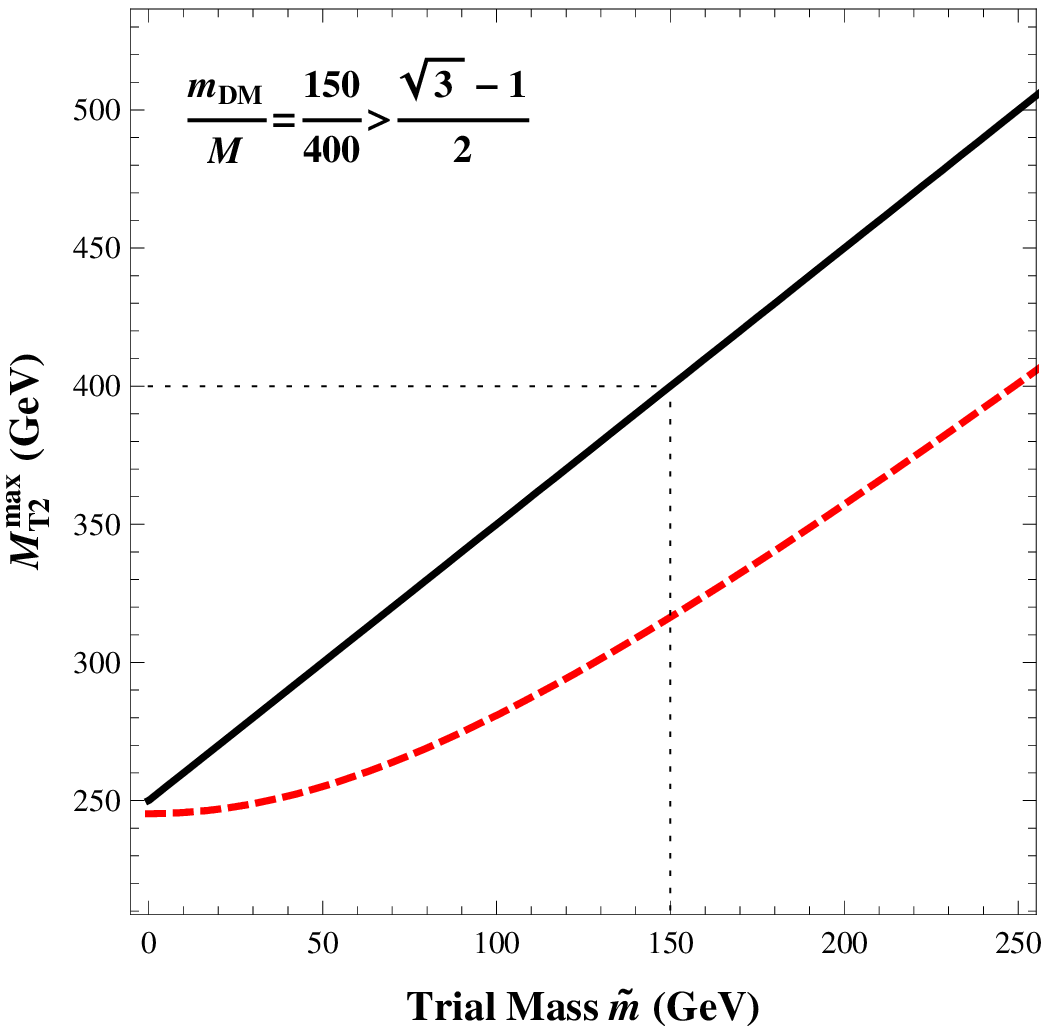}
	\hspace{0.2cm}
	 \includegraphics[width=8.0truecm,height=8truecm,clip=true]{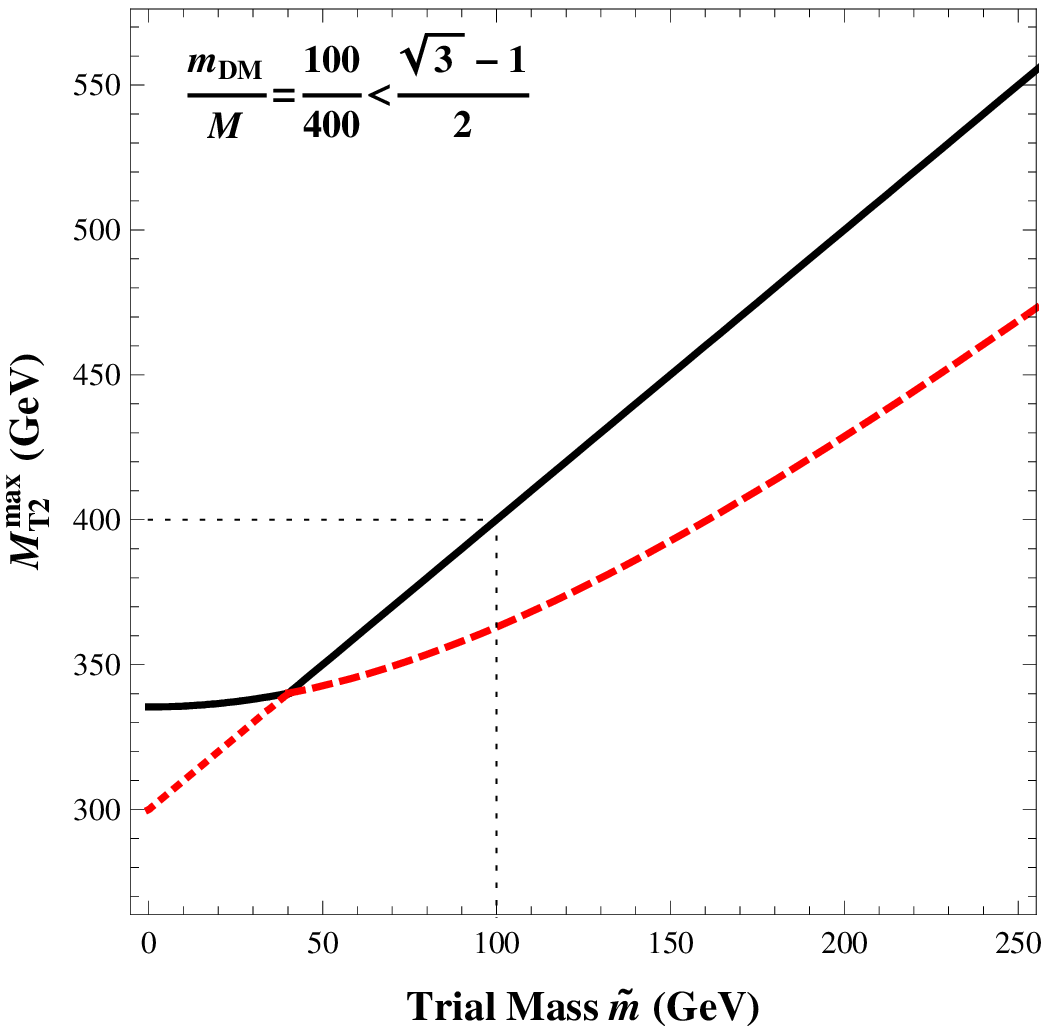}
	\caption{Theoretical expectation of $M_{T2}^{\max}$ versus the trial mass
	$\tilde{m}$ for $E_3$ type events. The mass of mother particle is 400 GeV for
	both cases, but the masses of DM particle to be used are 150 GeV and 100 GeV
	for the left panel and the right panel, respectively.
For both cases, the black solid lines give the maximum of $M_{ T2 }$, whereas the dashed
curves give the maximum for the balanced solution. The maximum for the
unbalanced solution coincides with the solid line on the left panel, whereas on the right panel, it is given by
the
dotted straight line
(which extends into the right-hand part of the solid line).
%
%
}
	\label{fig:mt2max}
\end{figure}
If the ratio of the DM mass to the mother mass is larger than $(\sqrt{3}-1)/2$,
it turns out that the maximum unbalanced solution given in Eq.~(\ref{eq:mt2unbalz33}) is always bigger than the maximum balanced solution given in Eq.~(\ref{eq:maxcase2}) so that
\bea
M^{ \max } _{ T2,E_3 }  =\max\left[M_{T2,E_3}^{\max ,bal},\;M_{T2,E_3}^{\max, unbal}\right]=
M-m_{DM}+\tilde{m} \hspace{0.6cm} \hbox{for} \;
\frac{ m_{ DM } }{M} \geq \frac{ \sqrt{3} - 1 }{2}
\; \hbox{and for all} \; \tilde{m}.
\label{eq:mt2maxe31}
\eea
The left panel of Fig.~\ref{fig:mt2max} clearly confirms our expectation
(based on above equation)
that there occurs no kink in the upper edge of $M_{T2}$ as
a function of the trial DM mass, i.e., the upper edge in the $M_{T2}$ distribution is always determined by the unbalanced solution (black solid line), not by the balanced solution (red dashed curve). Here we adopted $M=400$ GeV and $m_{DM}=150$ GeV, and thus the ratio between them is obviously larger than $(\sqrt{3}-1)/2$.

On the other hand, once the ratio of DM to mother mass
is smaller than $(\sqrt{3}-1)/2$, the competition between the balanced and the unbalanced solutions results in
\bea
M^{ \max } _{ T2,E_3 } & = &\max\left[M_{T2,E_3}^{\max ,bal},\;M_{T2,E_3}^{\max, unbal}\right] \nonumber \\
&=&
\left\{
\begin{array}{l}
M-m_{DM}+\tilde{m} \hspace{7.4cm}\hbox{for $\tilde{m}\geq m'$} \cr
\cr
\sqrt{\frac{(M^2-m_{DM}^2)(M^2-4m_{DM}^2)}{4M^2}}
+\sqrt{\frac{(M^2-m_{DM}^2)(M^2-4m_{DM}^2)}{4M^2}+\tilde{m}^2}
\hspace{0.6cm} \hbox{for $\tilde{m}\leq m'$}
\end{array}
\right.
\nonumber \\
 & &
\hbox{and for} \;
\frac{ m_{ DM } }{M} \leq \frac{ \sqrt{3} - 1 }{2}
\label{eq:kinkZ3}
\eea
where
\bea
m'=\frac{(M-m_{DM})\left(\sqrt{(M^2-m_{DM}^2)(M^2-4m_{DM}^2)}-M(M-m_{DM}) \right)}
{2M(M-m_{DM})-\sqrt{(M^2-m_{DM}^2)(M^2-4m_{DM}^2)}}.
\label{eq:kinklocz3}
\eea
We see that there is a kink at $\tilde{m}=m'$.
Here $m'$ is {\em not} simply the true DM mass but it is
given by a combination of the true mother and DM masses
(in fact, it is smaller than the true DM mass), which is clearly different from that in $Z_2$ models.
The functional behavior of $M_{T2}^{\max}$ for this case is shown in the right panel of Fig.~\ref{fig:mt2max}. Here we took $m_{DM}=100$ GeV which makes the ratio smaller than
$(\sqrt{3}-1)/2$. As before, the maximum $M_{T2}$
for the balanced and unbalanced solutions are shown by the
dashed and dotted curves (which extend into the black solid curve to the RHS and LHS). The
final maximum $M_{T2}$ is given by the larger of these two solutions (black solid curve) which clearly shows a kink at a value of $\tilde{m}$ which is
different from the actual DM mass $m_{DM}=100$ GeV (shown by the vertical black dotted line) as expected based on above discussion.
Of course, we can still evaluate the masses of mother and DM particles
(using $E_3$-type events only) by obtaining $M_{T2}^{\max}$ and $m'$ from the above $M_{T2}$ analysis, substituting them into Eqs.~(\ref{eq:kinkZ3}) and~(\ref{eq:kinklocz3}), and solving those two equations about $M$ and $m_{DM}$.

Next, let us investigate the hierarchy among the three $M_{T2}^{\max}$ values for $E_2$, $E_3$, and $E_4$ type events.
Although a bit more complicated than the one visible particle case,
it is nonetheless straightforward to derive this hierarchy based on above equations,
We have a following hierarchy among the $M_{T2}^{\max}$ values for the three types
(cf. the one visible particle case shown in Eq. (\ref{eq:mt2maxhier})):
\bea
%
\begin{array}{l}
M_{T2,E_2}^{\max}=M_{T2,E_3}^{\max}>M_{T2,E_4}^{\max} 
\hspace{2cm}\hbox{for $\tilde{m}\geq m_{DM}$} \cr
\cr
M_{T2,E_2}^{\max}>M_{T2,E_3}^{\max}>M_{T2,E_4}^{\max} \hspace{2cm} 
\hbox{for $\tilde{m}\leq m_{DM}$}.
\end{array}
%
%
\label{eq:hierz32vis}
\eea

\begin{figure}[t]
	\centering
	 \includegraphics[width=7.0truecm,height=6.0truecm,clip=true]{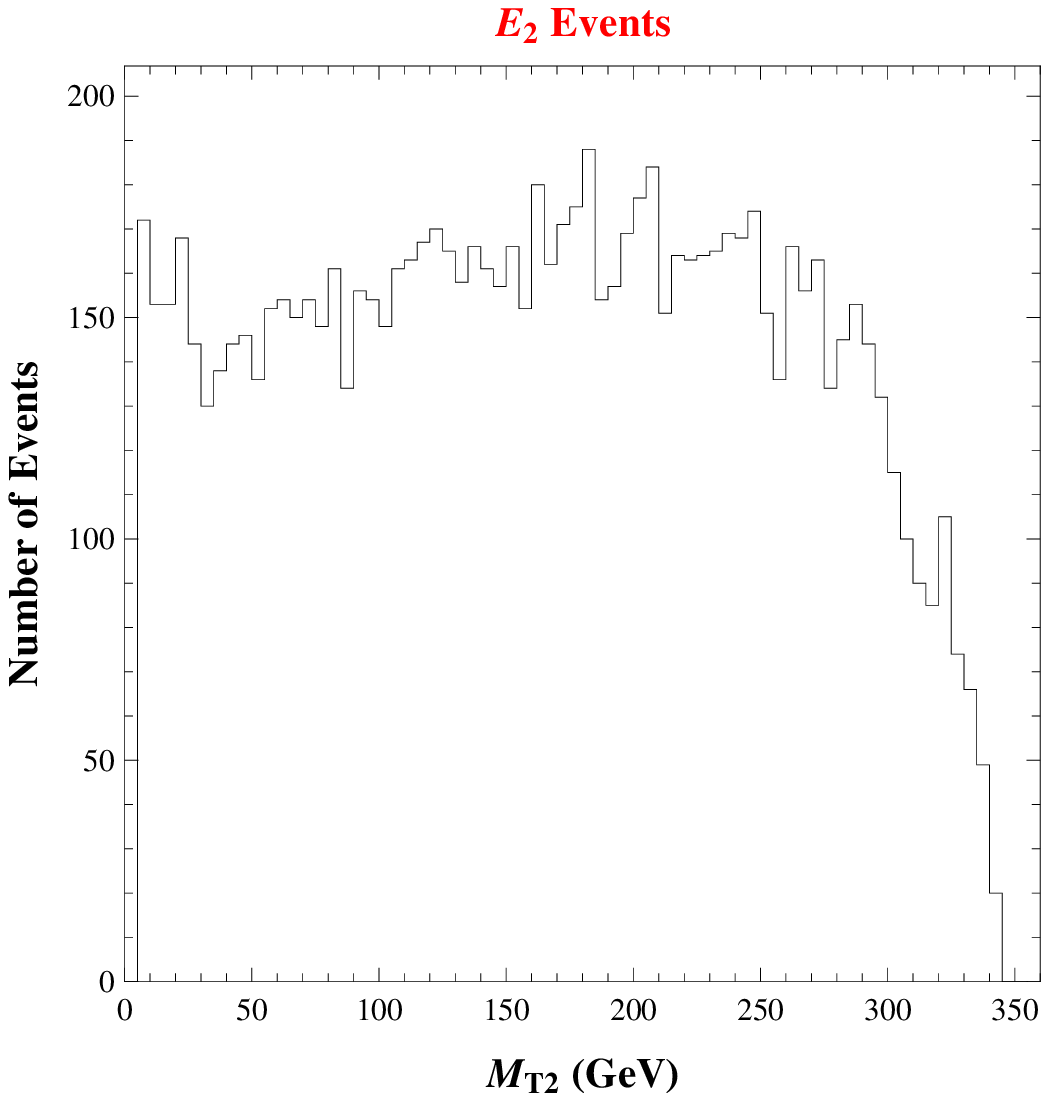}
	\hspace{0.2cm}
	 \includegraphics[width=7.0truecm,height=6.0truecm,clip=true]{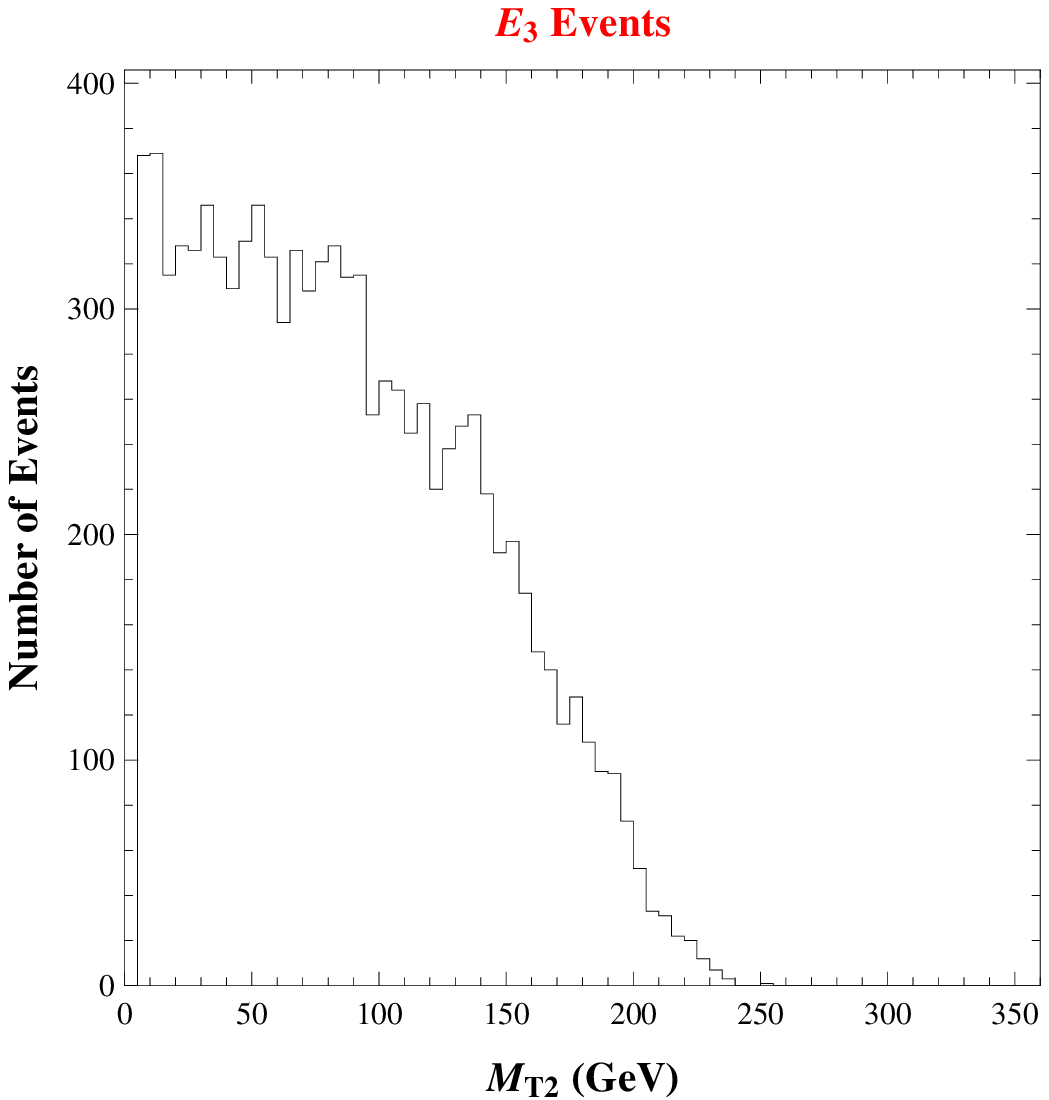}
	 \vspace{0.2cm}
	 \includegraphics[width=7.0truecm,height=6.0truecm,clip=true]{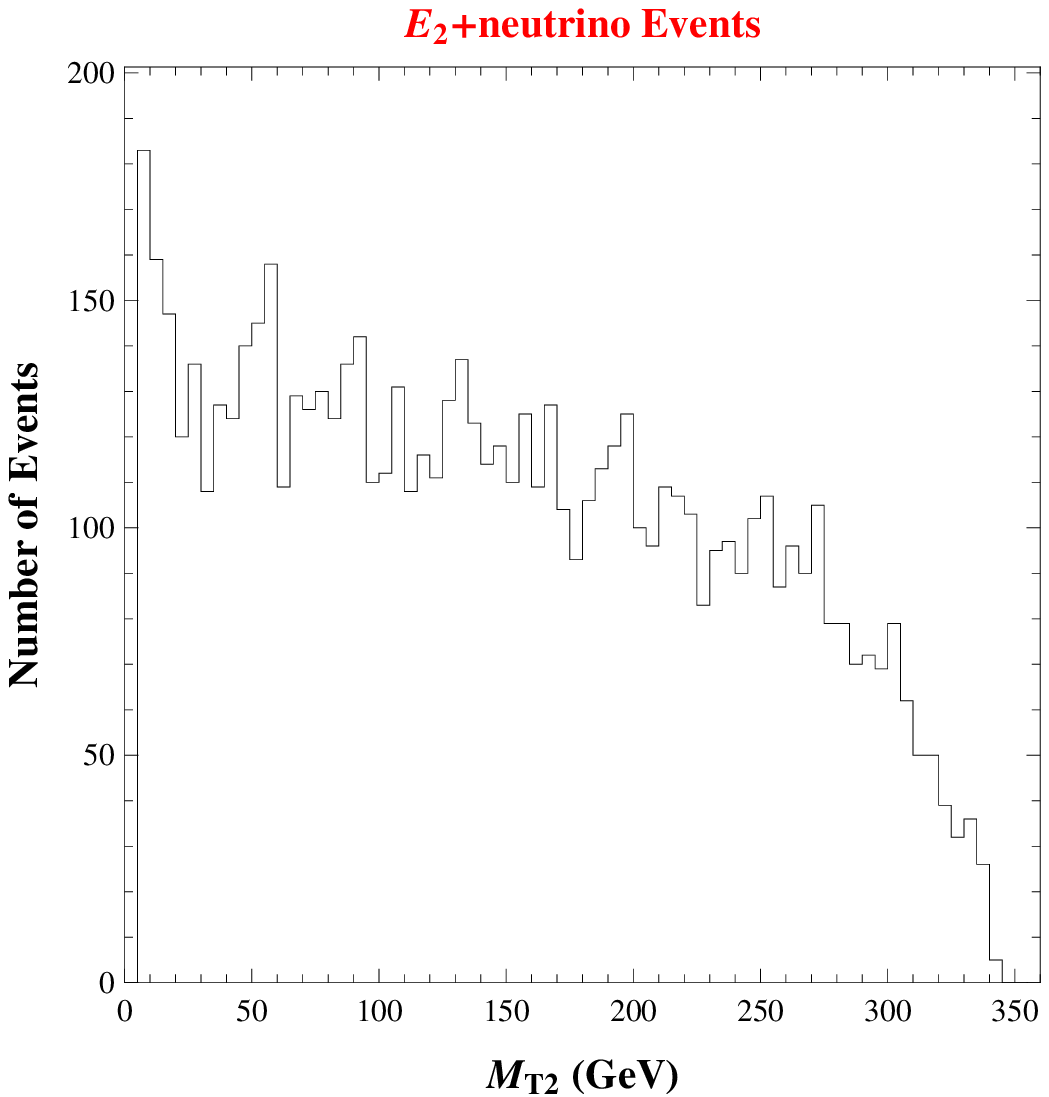}
	\caption{The $M_{T2}$ distributions for $E_2$ (top left) and $E_3$ (top right)
	type events and $E_2+\nu$ events (bottom).
	The mother and the DM particle masses are 400 GeV and 150 GeV, respectively, and
	the trial DM mass ($\tilde{m}$) used is 6 GeV.
%
%
%
}
	\label{fig:mt2diste2e3}
\end{figure}

\subsection{Shapes of $M_{ T2 }$ Distributions}
\label{sec:shape}
Before closing the present section, let us examine the shape of the $M_{T2}$ distributions for $Z_3$ cases.
For this purpose we simulated events using MadGraph/MadEvent~\cite{Alwall:2007st}. Here and 
in sections \ref{sec:onevisibleidentical} and \ref{sec:onevisiblenonidentical}, we make the following
assumptions (mostly for simplicity):
(a) effects of spin correlations are neglected in both production and decay of mother particles, i.e., we assumed scalar particles only in our simulations,; 
(b)
the beam is proton-proton with 14 TeV total
energy in the center of mass frame (motivated by
the LHC parameters); 
(c) the non-colored (scalar) mother particles are pair-produced via \textit{s}-channel exchange (of another scalar particle);
and finally 
(d)
only the {\em relative} values of the number of events (on vertical scale)
are meaningful, i.e. only the \textit{shape} of the distributions is the robust aspect of our analysis.

For all the simulations in {\em this} section, 
the masses of mother and DM particles used in the (toy) model 
are 400 GeV and 150 GeV,
respectively, we took a {\em single} visible particle per decay chain for simplicity and chose a small trial DM mass $\tilde{m}$ (here 6 GeV) for the purpose of illustration. The upper left panel of Fig.~\ref{fig:mt2diste2e3} demonstrates the $M_{T2}$ distribution
for $E_2$ type events. Obviously, this is similar to the $M_{T2}$ distribution
for events in $Z_2$ models ~\cite{Lester:1999tx, Barr:2002ex, Cho:2007qv}, which is not surprising because
$E_2$ type events also have only two dark matter particles in the final state like $Z_2$.
The (naive) $M_{T2}$ distribution for $E_3$ type events illustrated in the upper right panel
of Fig.~\ref{fig:mt2diste2e3} has two notable features. As expected
from the analytic expressions given before, first of all, the location of the upper edge is clearly lower than that for
$E_2$ type events (or for $Z_2$ models). Secondly, the shape of $M_{T2}$ distribution for $E_3$ type events shows a long tail near the upper edge compared with $E_2$ type events
(which have relatively sharp upper edge): this is 
because more physical constraints (e.g., rapidity) between decay products should be
satisfied in $E_3$ type events in order that they form a kinematic
configuration to give maximum $M_{T2}$, thereby reducing the
corresponding number of events near $M_{T2}^\text{max}$.

This feature of a (relatively) long tail is also true for a special case
in $Z_2$ models
with three invisible particles in the full event
one of which is a (massless) SM neutrino, i.e.,
with
one DM in one decay chain, but one DM and a neutrino in the other (henceforth
we call it $E_2+\nu$ events)\footnote{$M_{ T2}$ distributions for such events
have been studied in \cite{Barr:2002ex} and in
\cite{Barr:2009mx}.}. 
The bottom panel of Fig.~\ref{fig:mt2diste2e3} demonstrates the $M_{T2}$ distribution for $E_2+\nu$ events, assuming intermediate 
particles are off-shell~\footnote{As
for the events with three (or more) DM, we assume here that there
is a single (massive) invisible particle in each decay chain
for the purpose of defining $M_T$.}. At the upper edge
it shows a tail which is longer than that for $E_2$ type 
events (or neutrino-less events in $Z_2$ models), but
which is not as long as that for $E_3$ type events\footnote{This 
%
%
feature of a long tail in $E_2 + \nu$-type events is valid
even for the case of 
more than one visible particle in each decay chain.}. Such a 
shape is not surprising because the additional invisible particle (neutrino) in
$E_2 + \nu$ vs. $E_2$-type events
is massless so that $E_2+\nu$ events can be understood as a transitional type between $E_2$ and $E_3$ type events. However, it is crucial
to note that the location of the upper edge for the $E_2+\nu$ events is the
{\em same} as that for usual (i.e.,
neutrino-less)
events in $Z_2$
models with only a single massive invisible particle per decay chain, because the effective DM mass in former decay chains
is also $m_{DM}$. We can therefore distinguish the $E_2+\nu$ events
from the $E_3$-type events in $Z_3$ models by observing the location of the upper edge: the latter events will have a smaller edge.


\section{Applications: Non-Identical Visible Particles
in the Two Decay Chains}
\label{sec:nonidentical}
Next, we apply the theoretical observations on the $M_{T2}$ technique for $Z_2$ and
$Z_3$ models, which are described in the previous sections, for distinguishing $Z_3$ models from $Z_2$ ones in some \textit{specific} cases.
Like in
the previous sections, we assume (for simplicity) pair-produced (same) mother particles
and that all visible particles are massless and
use the naive $M_{ T 2 }$ analysis for all events.
As mentioned earlier,
in $Z_3$ models each mother particle can decay into either one or two dark matter particles along with visible/SM particle(s). Here, 
\begin{itemize}

\item
we consider a mother particle
in a $Z_3$ model
for which {\em both} these decay chains (with one and two DM, respectively) 
exist.

\end{itemize}
Similarly,
in $Z_2$ models the mother particles can only decay into a single dark matter particle along with visible/SM particle(s). Here, we assume two such decay chains for a $Z_2$-mother
which have the same visible final states as the above two $Z_3$-mother decay chains
(respectively).
The idea behind this choice is that such a
$Z_2$ model could easily fake a $Z_3$ one (at least based on the identity of the visible states).
This motivates us to distinguish these two types of DM stabilization symmetry using the $M_{T2}$ variable.
%
%

For later convenience, we divide the discussion into two cases based on
whether or not the visible state in the decay chain with one DM
is identical to the one in the decay chain with
two DM in the $Z_3$ model.
%
%
We begin with the case where the SM final states in the two decay chains
%
%
are not identical (this includes the case of partial overlap between these final states). Following
the notation of previous section, 
let us denote SM$_{ 1, \; 2 }$ to be these SM final state particles
-- whether they consist of one or more SM particles -- in the two decay 
chains. And, for the $Z_3$ model, assume that SM$_1$ comes with one DM particle and SM$_2$ is associated with two DM particles.\footnote{Since
a $Z_2$ model does not allow two DM particles in each decay chain,
SM$_1$ and SM$_2$ are both emitted with only one DM in the final state
in this model.}
We thus have three {\em distinct} (based simply on identity of visible states)
types of events in both the $Z_3$ and $Z_2$ models, denoted by SM$_{ 1 1 }$, i.e., SM$_1$ on
each side/from each mother and similarly SM$_{ 1 2, \; 22 }$.
Clearly, for $Z_3$ models,
these three types of events correspond (respectively) to $E_{ 2,3, 4}$-type 
events mentioned in the previous section and shown in Fig.~\ref{E234}
(with SM$_1$ $\neq$ SM$_2$).
%
%
Hence, we can apply the formulas for the
theoretical predictions of $M_{T2}^{\max}$ for $E_2$, $E_3$, and $E_4$ type
events
derived in the previous section to the SM$_{ 11, \; 12, \; 22 }$ events.


\subsection{One Visible/SM Particle in Each Decay Chain}
\label{sec:onevis}

As a further subcase, we assume that SM$_{ 1, 2 }$ consist of only one
particle.
%
%
%
Clearly, the upper edges of the $M_{T2}$ variable for SM$_{11}$, SM$_{12}$,
and SM$_{22}$ are given by Eqs.~(\ref{eq:maxcase1}),~(\ref{eq:maxcase2}),
and~(\ref{eq:maxcase3}), respectively.
As is obviously seen from the left panel of Fig.~\ref{fig:mt2maxz3} or equivalently
Eq.~(\ref{eq:mt2maxhier}), the location of
the upper edge for SM$_{12}$ or SM$_{22}$ is lower (for {\em all}
trial DM mass $\tilde{m}$: cf. the case of more than one visible particles
below)
than that for SM$_{11}$
(for the same mother and DM masses).
In contrast, in the $Z_2$ model we have the same expression for $M_{ T2 }^{ \max }$
given by Eq.~(\ref{eq:mt2maxz2}) for {\em all} of SM$_{11}$,
SM$_{12}$, and
SM$_{22}$ because they all involve two DM particles in the final state. Thus,
\begin{itemize}

\item
different edges
for the SM$_{ 11, \; 12, \; 22 }$ events
(in particular, larger for SM$_{11}$)
can be evidence for $Z_3$ models, i.e.,
they provide discrimination between $Z_2$ and $Z_3$ models\footnote{unless multiple mother particles in the $Z_2$ models decay
into the identical final state.}.
\end{itemize}

We can be further quantitative:
\begin{itemize}

\item
for $Z_3$ models we can measure the masses of mother and DM separately as follows (in spite of
absence of kink in left panel of Fig.~\ref{fig:mt2maxz3}).
\end{itemize}
Note that the theoretical formulas for
$M_{ T2 }^{\max}$ in Eq.~(\ref{eq:maxcase1}) through Eq.~(\ref{eq:maxcase3}) --
considered as a function of trial mass $\tilde{m}$ -- 
have a structure of $\sqrt{C}+\sqrt{C+\tilde{m}^2}$
where $C$ is a constant.
So, the idea is to choose an arbitrary trial mass, then
calculate the corresponding $M_{T2}^{ \max}$ from the {\em experimental} data
and thus determine the above-defined $C$.
Also, in our specific case where visible particles are
assumed massless, the  {\em theoretical} formula for
each $C$ is written only in terms of
the mother mass and the DM mass:
\bea
\frac{(M^2-m_{DM}^2)^2}{4M^2}&=&
\frac{\left((M_{T2,E_2}^{\max})^2-\tilde{m}^2\right)^2}{4(M_{T2,E_2}^{\max})^2}
\equiv C_{ E2 } \label{eq:c1} \\
\frac{(M^2-m_{DM}^2)(M^2-4m_{DM}^2)}{4M^2}&=&
\frac{\left((M_{T2,E_3}^{\max})^2-\tilde{m}^2\right)^2}{4(M_{T2,E_3}^{\max})^2}
\equiv C_{ E3 } \label{eq:c2}
\eea
Solving the above-given two equations, we obtain {\em both} mother and DM
masses \footnote{The upper edge from SM$_{ 2 2 }$ provides redundant information, but
of course can be a cross-check.}:
\bea
M&=&\frac{2}{3\sqrt{ C_{ E2 } }}(4C_{ E2 }-C_{ E3 }) \label{eq:M} \\
m_{DM}&=&\frac{2}{3\sqrt{C_{E2}}}\sqrt{(4C_{ E2 }-C_{ E3 })(C_{ E2 }-C_{ E3 })} \label{eq:mDM}
\eea
This situation is  somewhat like the double-edge signal for single
mother decay studied in reference \cite{Agashe:2010gt}, where it was again
possible to obtain mother and DM masses from the two edges in invariant mass
distribution of visible/SM final state.

On the other hand, for $Z_2$ models, we obtain
only a {\em combination} of mother and DM masses
from the (single) measurement of $M_{ T2 }^{ \max }$,
given by
\bea
\frac{(M^2-m_{DM}^2)^2}{4M^2}&=&
\frac{\left((M_{T2}^{\max})^2-\tilde{m}^2\right)^2}{4(M_{T2}^{\max})^2}
\equiv C,
\label{eq:c}
\eea
and thus it is {\em not} possible to determine mother and DM masses separately.


\subsection{More than One Visible/SM Particle in Each Decay Chain}
\label{subsec:morethan1}

In this case, there is more interesting behavior of $M_{ T2 }^{\max}$
than the case of one visible particle per chain \footnote{In the SM$_{ 12 }$
events, the endpoints
of the {\em visible}
invariant mass distributions for the two sides of the event/decay chains
will be different in $Z_3$ models, i.e., in $E_3$-type events
(vs. being the same in $Z_2$ models), already providing a discrimination
between the two types of models \cite{Agashe:2010gt}. 
However, developing another technique for distinguishing 
$Z_3$ from $Z_2$ models based on $M_{ T 2 }$
can still be useful.}.
The upper edges of $M_{ T2 }$ for SM$_{ 11, \; 12, \; 22 }$ events are now obviously given by
Eqs.~(\ref{eq:kinkZ2}), (\ref{eq:mt2maxe31}), (\ref{eq:kinkZ3})
and (\ref{eq:kinke4}), respectively.
As discussed in Eq.~(\ref{eq:hierz32vis}), the upper edge for SM$_{12}$ is the
{\em same} as that for SM$_{11}$ (cf. one visible particle case above) for $\tilde{m}>m_{DM}$, but is lower for $\tilde{m}<m_{DM}$ than that for SM$_{11}$. And,
the upper edge for
SM$_{22}$ events is always lower
than SM$_{11}$.
This fact enables us to {\em distinguish}
$Z_3$ models from $Z_2$ ones because in $Z_2$ models
the upper edges for SM$_{11, \; 12, \;
22}$ coincide for {\em all} $\tilde{m}$ (just like the case with one visible particle in each decay chain).

Moreover, there occurs a kink in the
upper edge of $M_{T2}$ as a function of the trial DM mass
as discussed in the previous sections. Because of the existence of this
kink structure, SM$_{11}$ itself is sufficient for the purpose of determining
mother and DM masses (again, unlike one visible case): the trial mass which gives rise to a kink is the true
DM mass and its corresponding $M_{T2}^{\max}$ is the true mother mass.
(Of course this is how one can measure the masses of mother and DM particles separately
even in $Z_2$ models.)
Such a direct measurement of mother and DM masses leads us to
\begin{itemize}

\item
a {\em prediction} (cf. one visible case) on the location of the upper edges for
the {\em other} two types of events, namely SM$_{12}$ and SM$_{22}$, a confirmation of which
can provide evidence for $Z_3$ symmetry as underlying
physics\footnote{We can also predict (and then verify) location of kink in
SM$_{ 22}$ events.
Alternatively, we can use kink in SM$_{ 22 }$, i.e., $E_4$-type
events to determine the mother and DM masses and then make predictions.}.
\end{itemize}

For SM$_{12}$, i.e., $E_3$-type events, actually, there are more interesting aspects
of the kink structure in $M_{ T2 }^{\max}$ due to the dependence on the ratio of DM and mother masses:
%
%
as discussed in Sec.~\ref{sec:mt2z3}, the critical ratio is given by
\bea
\frac{m_{DM}}{M}=\frac{\sqrt{3}-1}{2}.
\eea
The kink is present only when the ratio of DM and mother masses
is less than the above-given critical ratio. In this case,
the kink location can be predicted by substituting mother and DM masses measured from the kink in SM$_{11}$ into Eq.~(\ref{eq:kinklocz3}) so that it can provide a
{\em
further} verification for $Z_3$ symmetry\footnote{Alternatively, this kink can be used to
determine mother and DM masses, which are then used to predict edges/kinks in other events.}.

Finally,
note that another way to distinguish $Z_3$ from $Z_2$ models in this case was discussed in \cite{Agashe:2010gt}.
The idea is to use SM$_{12}$, i.e., $E_3$-type, events in $Z_3$ models, where edges in invariant mass distributions of
{\em visible} particles
on each side are different, i.e., $\left( M- m_{ DM} \right)$ and
$\left( M - 2 \; m_{ DM } \right)$
(vs. the two edges being the same for $Z_2$ models).

\subsection{Signal Fakes by an (Effective) 2nd DM Particle
%
%
}
\label{sec:sigfake}
In the two previous sections we have focused on decay processes with a single 
type of DM particle in the final state (for both $Z_3$ and $Z_2$ models). 
The crucial observation for the sake of discriminating $Z_3$ 
from $Z_2$ models is that $Z_3$ models have more event-topologies (i.e., $E_2$, $E_3$, and $E_4$ type events with 
different upper edges in $M_{ T 2 }$
distributions) than the case of $Z_2$  (which has a single upper edge), 
regardless of the number of 
visible/SM particles in each decay chain. 
In turn, this contrasting feature is  
due to the different possibilities in {\em each} decay chain in $Z_3$ models, i.e., presence of 
one or two DM (unlike only one DM in 
$Z_2$ case).

However, $Z_2$ models can also acquire such different possibilities for decay chains
(and thus fake $Z_3$ signals in the $M_{T2}$ analysis) if we assume that
there is a
{\em second} DM (obviously $Z_2$-odd)
particle (with larger mass) 
denoted by DM$^{ \prime }$ into which the $Z_2$-mother can decay, i.e.,
there are actually 
two (absolutely) stable DM particles in a $Z_2$ model \cite{multiple}.
Clearly, even with only one DM in a $Z_2$ model,  
a similar effect can arise from a mother decaying into an ($Z_2$-odd) 
{\em on}-shell 
color/electrically neutral particle which {\em decays} 
(into DM particle and SM, possibly visible), but outside the detector (i.e., 
there exists a $Z_2$-odd particle -- other than the DM -- which is 
stable and invisible as far as the detector is concerned).
Another 
related possibility is that there is 
a $Z_2$-odd (on-shell) neutral particle which decays {\em in}side the detector, but invisibly,
i.e., into DM and invisible SM, for example, neutrino.
A classic example of the last type is found in supersymmetry 
where sneutrino decaying into neutrino and lightest neutralino (which is
assumed to be the lightest supersymmetric particle, i.e., DM). 
Even in the latter two cases, there is ``effectively'' (i.e., as
far as the collider analysis is concerned) 
a second ``DM'' and so we will denote it also by DM$^{ \prime }$.
In particular, in the last case mentioned above, i.e., even if there is
an (on-shell) neutral particle decaying invisibly inside detector, 
the theoretical prediction of the $M_{T2}$ variable is the same as with a 
DM$^{ \prime }$ of the same mass
as this neutral mother particle.

Here, we note that the  
reference~\cite{Konar:2009qr} has studied such (asymmetric) events using 
an $M_{T2}$ 
type 
analysis, 
in particular, variants of the usual $M_{ T 2 }$ variable have been developed. 
As before, we will instead apply the naive/usual $M_{ T 2 }$ variable, i.e., assume (again, just for the
purpose of constructing $M_{ T2 }$) that there is a single and same DM 
in both decay chains.

In more detail, the above case in $Z_2$ models gives rise to ``$Z_3$-faking'' signals 
is as follows.
%
%
As before, consider pair production of a single mother such that decay chains with DM and
DM$^{ \prime }$ are both allowed.
Consequently, we obtain three distinct decay topologies for the full event:
two DM, one DM and one DM$^{ \prime }$ and two DM$^{ \prime }$.
We will denote these three types of events by
$E_2'$, $E_3'$, and $E_4'$ since they obviously resemble (and thus can fake) $E_2$, $E_3$, and $E_4$ type 
events being found in $Z_3$ models, respectively. 
In particular, 
we can expect 
three different upper edges for $M_{ T 2 }$ in 
$E_2'$, $E_3'$, and $E_4'$-type 
events\footnote{An extreme case is when DM$^{ \prime }$ is massless,
for example, SM neutrino. However, in this case, as mentioned in section
\ref{sec:shape}, the upper edges for $E_3'$ (denoted by
$E_2 + \nu$ event in section \ref{sec:shape}) and $E_4'$-type
events will be {\em same} as for $E_2'$-type events.
Thus this case can be easily distinguished from $Z_3$ models.}.
Explicitly, 
the maximum balanced $M_{T2}$ solutions for them (for both
the cases with one visible particle per decay 
chain and more than one visible particle per decay chain) are simply given as follows:
\bea
M^{ \max,bal }_{ T2,E_2' }  =\sqrt{\frac{(M^2-m_{DM}^2)^2}{4M^2}}
+\sqrt{\frac{(M^2-m_{DM}^2)^2}{4M^2}+\tilde{m}^2} \hspace{4.7cm}\hbox{for $E_2'$}
\label{eq:maxcasefake1} \\
M^{ \max,bal }_{ T2,E_3' } = \sqrt{\frac{(M^2-m_{DM}^2)(M^2-m_{DM}'^2)}{4M^2}}
+\sqrt{\frac{(M^2-m_{DM}^2)(M^2-m_{DM}'^2)}{4M^2}+\tilde{m}^2}
\hspace{0.5cm}\hbox{for $E_3'$}
\label{eq:maxcasefake2} \\
M^{ \max,bal }_{ T2,E_4' } = \sqrt{\frac{(M^2-m_{DM}'^2)^2}{4M^2}}
+\sqrt{\frac{(M^2-m_{DM}'^2)^2}{4M^2}+\tilde{m}^2}
\hspace{4.8cm}\hbox{for $E_4'$},
\label{eq:maxcasefake3}
\eea
and the maximum unbalanced $M_{T2}$ solutions (only for the case with more than one visible particle 
per decay chain) are given as follows:
\bea
M_{T2,E_{2}'}^{\max,unbal}&=&M-m_{DM}+\tilde{m}\hspace{3cm}\hbox{for }E_{2}'
\label{eq:mt2unbalz32fake} \\
M_{T2,E_{3}'}^{\max,unbal}&=&M-m_{DM}+\tilde{m}\hspace{3cm}\hbox{for }E_{3}' 
\label{eq:mt2unbalz33fake} \\
M_{T2,E_{4}'}^{\max,unbal}&=&M-m'_{DM}+\tilde{m}\hspace{3cm}\hbox{for }E_{4'}.
\label{eq:mt2unbalz34fake}
\eea
Here $m'_{DM}(>m_{DM})$ denotes the mass of the second DM-like particle.
Again, all three types actually contain \textit{two} DM/DM-like 
particles, i.e., the subscripts on $E'$ 
do not imply the number of DM particles in a full decay chain but rather indicate the respective  
topologies in $Z_3$ models which they fake.
Note that if we set $m'_{ DM } = 2 m_{ DM }$, then the above edges
are exactly the ones in a $Z_3$ model (see Eqs.~(\ref{eq:maxcase1}) through~(\ref{eq:maxcase3}) and Eqs.~(\ref{eq:mt2unbalz3}),~(\ref{eq:mt2unbalz34}), and~(\ref{eq:mt2unbalz33}) from previous sections.
This feature is as expected
since for $E_{ 3, \; 4 }$-type events (in $Z_3$ models) which are at the edge
of the respective $M_{ T 2 }$ distributions, the 
two DM from the {\em same} mother are collinear so that their invariant mass is 
$2 m_{ DM }$, i.e., 
the decay chain with two DM effectively has {\em single} DM of this mass as far as $M_{ T2 }$-edge is concerned.

Despite the fact that such $Z_2$ events with a second DM-like particle can introduce 
three decay topologies, we can still differentiate $Z_3$ and $Z_2$ models.
However, the 
strategies to be applied depend on the number/identity of visible particles in each decay chain. 
In this section, we consider the case where the visible particles in the decay chain with DM 
(denoted by SM$_1$, following the notation used earlier) are different than the visible particles (denoted by SM$_2$)
in decay chain with DM$^{ \prime }$ in the $Z_2$ model or two DM (in the $Z_3$ model).
Thus, the three types of events
SM$_{ 11 }$, SM$_{ 1 2 }$ and SM$_{ 2 2 }$, i.e., distinguishable from the
identity of SM visible particles, 
have different edges since they correspond to the $E_2'$, $E_3'$, and $E_4'$-type
events in a $Z_2$ model or (as mentioned in previous section) $E_2$, $E_3$, and $E_4$-type events
in a $Z_3$ model.
This case can be further subdivided into one and more than one visible particles
in each decay chain.

In the case 
with one visible particle per decay chain, one may distinguish $Z_2$ and $Z_3$ models by examining the {\em shape} 
of the above three $M_{T2}$ distributions. 
The idea is that,
as explicitly mentioned above, 
$E_2'$, $E_3'$, and $E_4'$ all have only \textit{two} DM/DM-like particles, i.e., two DM, one DM and 
one DM$^{ \prime }$, and two DM$^{ \prime }$, and therefore, 
%
%
they have a similar shape of the
$M_{ T 2 }$ distribution as the $E_2$-type event.
The implication of this observation is $E_2'$, $E_3'$, 
and $E_4'$ all give a sharp upper edge in the $M_{T2}$ distribution.
On the other hand, in the decay chain with two DM in 
$E_{ 3, \; 4}$-type events of $Z_3$ models, in general (i.e., 
away from edge of $M_{ T 2 }$), the two DM are not collinear so that their invariant mass of two DM is not 
$2 m_{ DM }$, 
in fact, this invariant mass is not even fixed.
Thus, 
even if the above three $M_{ T 2 }$-{\em edges} for $Z_3$ models are identical to those for $Z_2$ models
with two different DM (with the
second one being twice as heavy as first one), the shapes are not
expected to be similar.
In fact,
$E_3$ and $E_4$ type 
events in $Z_3$ models give a (relatively) longer tail as already discussed in Sec.~\ref{sec:shape}. 
Hence, 
if one of the $M_{T2}$ distributions for SM$_{ 11, \; 12, \; 22 }$ events -- again,
corresponding to the three different topologies -- has a sharp upper edge and two of 
which have a longer tail, then it is likely that such events 
originate from $Z_3$ 
models.

On the other hand, once there exists more than one visible particle in each decay chain, the shape is 
no longer a useful discriminator. 
The reason is that, in general, clear sharp edges in the $M_{T2}$ distributions are not 
expected here (unlike the cases with one visible particle per decay chain), i.e.,
the number of events/statistics at the $M_{ T 2 }$-upper edge is small in this case: 
in turn, this feature is 
due to more constraints which need be satisfied 
(see Sec.~\ref{sec:mt2z3}). 
Instead, we can take the advantage of ``kink'' in the plot of 
$M_{T2}^{\max}$ versus the trial DM mass, which allows us to determine the masses of 
mother and DM particles separately. 
Using the
SM$_{ 11 }$
events, one can 
evaluate $M$ and $m_{DM}$
as mentioned in the previous section,
{\em assuming} that it is a $Z_3$ model.
Then we predict the locations of the upper edge and 
the locations of kink for the SM$_{ 12, \; 22 }$ events.
If the underlying physics is 
a $Z_2$ model (with two different DM particles) instead, then
these predictions do not match with the experimental results from the associated $M_{T2}$ 
analysis. This is because, in general, the mass of the second DM-like particle, $m_{DM}'$ 
is not equal to twice of the DM mass, $2m_{DM}$. In other words, the cross-checking of
mother and DM masses 
between SM$_{ 11 }$ and SM$_{ 12, \; 22 }$ events
enables us to separate $Z_3$ models from $Z_2$.


\section{Identical Visible Particle(s) in the Two Decay Chains}

\label{identical}

Next, we consider the case of the visible particle(s) in the two decay chains
with one and two DM (for $Z_3$ models) being the same. In this case, in the $Z_3$ models,
we can not separate $E_{2,3,4}$ type events using simply the identities of the visible particles, i.e.,
SM$_1$ $=$ SM$_2$ in Fig.~\ref{E234} 
(unlike
in the previous section).
%
%
Obviously, we add the three (i.e.,
$E_2$, $E_3$, and $E_4$-type)
 distributions of $M_{T2}$, whose
behaviors were discussed above (for non-identical
case), to obtain the observable $M_{ T2 }$ distribution in $Z_3$ models. Of course, for $Z_2$ models (which could potentially fake the $Z_3$ models),
there are then only $E_2$-type and possibly $E_2$ + $\nu$ events that we discussed
earlier.

If we have only one visible particle in each decay chain,
the $M_{ T2 }$ distribution for the $E_3$ type events  \textit{always}
(i.e., for all $\tilde{m}$, cf. more than one visible case
discussed below) has a lower $M_{T2}^\text{max}$
than for the $E_2$ type events (see the left panel of Fig.~\ref{fig:mt2maxz3}),
so that in principle their addition/combination
would give rise to a ``kink'' in the $M_{ T2 }$ 
distribution\footnote{Such a kink in the $M_{ T2 }$ {\em distribution}
is not to be confused with that in the plot of $M_{ T2 }^{ \max }$ 
as a function
of $\tilde{m}$.}
(again for $Z_3$ model, but not for $Z_2$ model).\footnote{Adding events of $E_4$
type, i.e., two DMs in each decay chain, will introduce another, but even less visible, kink.}
It turns out, however, that the visibility of this kink is 
not clear because the $M_{T2}$
distribution for $E_3$ type events has a longer tail 
(that for $E_2$-type events) as discussed in Sec.~\ref{sec:mt2z3}
(see the right panel of Fig.~\ref{fig:mt2diste2e3}). It also turns out that
the kink will get further smeared out once uncertainties in measurements
are taken into account. In other words, this kink is not
evidence for $Z_3$ models
since it could be
faked by statistical fluctuations in the distribution or experimental errors.

On the other hand, if there exists more than one visible particle in each decay chain,
the $M_{T2}$ distribution for the $Z_3$ model shows somewhat different behavior.
As discussed in detail in Sec.~\ref{subsec:morethan1},
for trial DM mass above the true DM mass, the upper edge of $M_{T2}$ distribution for $E_2$ and $E_3$ type events is the same.
However,
the upper edge of $M_{T2}$ distribution for  $E_3$ type events is {\em increasingly} lower than that for  $E_2$ type for trial DM mass
below the true DM mass\footnote{To be more precise,
the gap between the two edges {\em relative} to mother/DM masses
increases.}.
We therefore expect a ``moving'' kink (as we vary $\tilde{m}$)
in the $M_{T2}$
distribution --
such a kink starts to appear for trial mass below the true DM mass
(i.e., no kink would appear in the total $M_{T2}$ distribution for larger trial masses)
and the gap between the kink
position (i.e., corresponding to the $E_3$ edge) and the overall upper (i.e., $E_2$) edge is increasing as the trial mass becomes smaller\footnote{In the one visible particle case,
the gap between the edges in $E_{ 2, \; 3 }$-type events,
again relative to mother/DM masses,
is roughly constant with trial mass so that kink
in $M_{ T2 }$ distribution does not move.}.
This ``moving'' feature of the kink in the total $M_{T2}$ distribution can
be further (i.e., beyond simply {\em existence} of kink)
evidence for the existence of $E_3$ type events and thus a proof of $Z_3$ models. However, even though the kink is ``moving'', it is still hard to identify it in the $M_{T2}$ distribution, and thus we do not rely on these kinks as a way to distinguish between $Z_2$ and $Z_3$ models.

These observations motivate us to introduce  new methods to
separate $E_3$ type events from $E_2$ type events.\footnote{Once we separate these two types, we can repeat the program described in detail in
Sec.~\ref{sec:nonidentical}, i.e.,
either determine the masses of mother and DM particles from the the upper edges
of $M_{ T 2 }$ in these two types of events for the
case of one visible particle in each decay chain
or {\em predict} the upper edge for $E_3$-type events (using
measurements in $E_2$-type events) for the more than one visible particle
case.}
In the following (two) subsections,
we develop such a method, and then
we apply them to the two specific cases (i.e., one visible particle and
more than one visible particles in each decay chain), and see how to use
them to distinguish between $Z_2$ and $Z_3$ models.

\subsection{Separating $E_2$ and $E_3$ Type Events using $P_t/H_t$ Ratio}
\label{sec:pthtcut}

To separate $E_2$
and $E_3$ types of events, we can utilize the fact that $E_2$ type events have one DM per decay chain, and $E_3$ type events have one DM in one decay chain and two DMs in the other decay chain. In other words, for $E_3$ type events, 
the visible particle(s) in the decay chain having two DMs in the final state 
carry less momentum/energy (than in the other decay chain).
Thus the
ratio between the momentum/energy of visible particle(s) on the two decay chains is expected to
be (relatively) sizeable on average (compared to $E_2$ type events).
In order to find out how this intuition plays out in real situations,
we begin with the case where there exists only a single visible particle per decay chain, and then move on to the case where
there exists more than one visible particle per decay chain.

\subsubsection{One Visible/SM Particle in each Decay Chain}
\label{sec:onevisibleidentical}

\begin{figure}[ht]
\centering
\includegraphics[scale = 0.615]{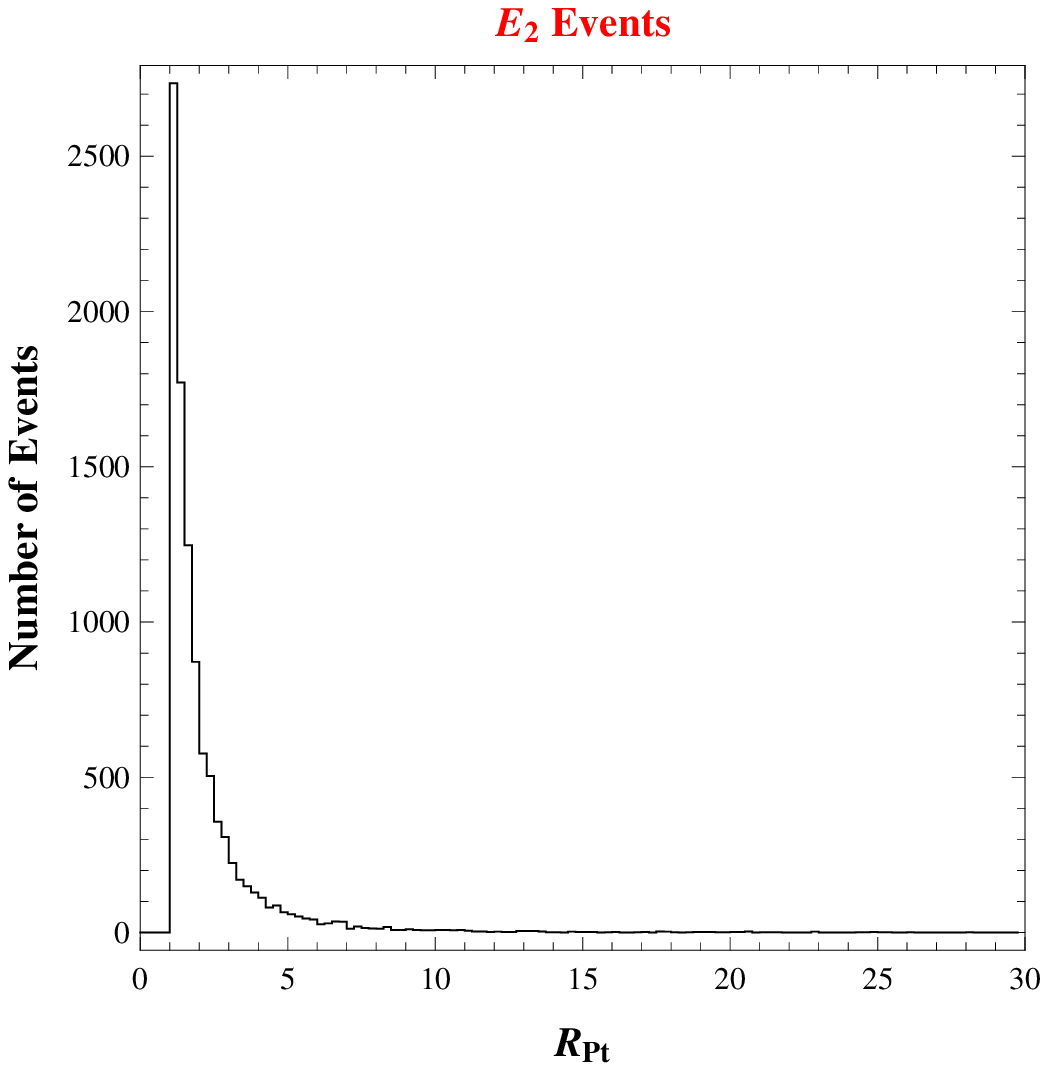}
\includegraphics[scale = 0.615]{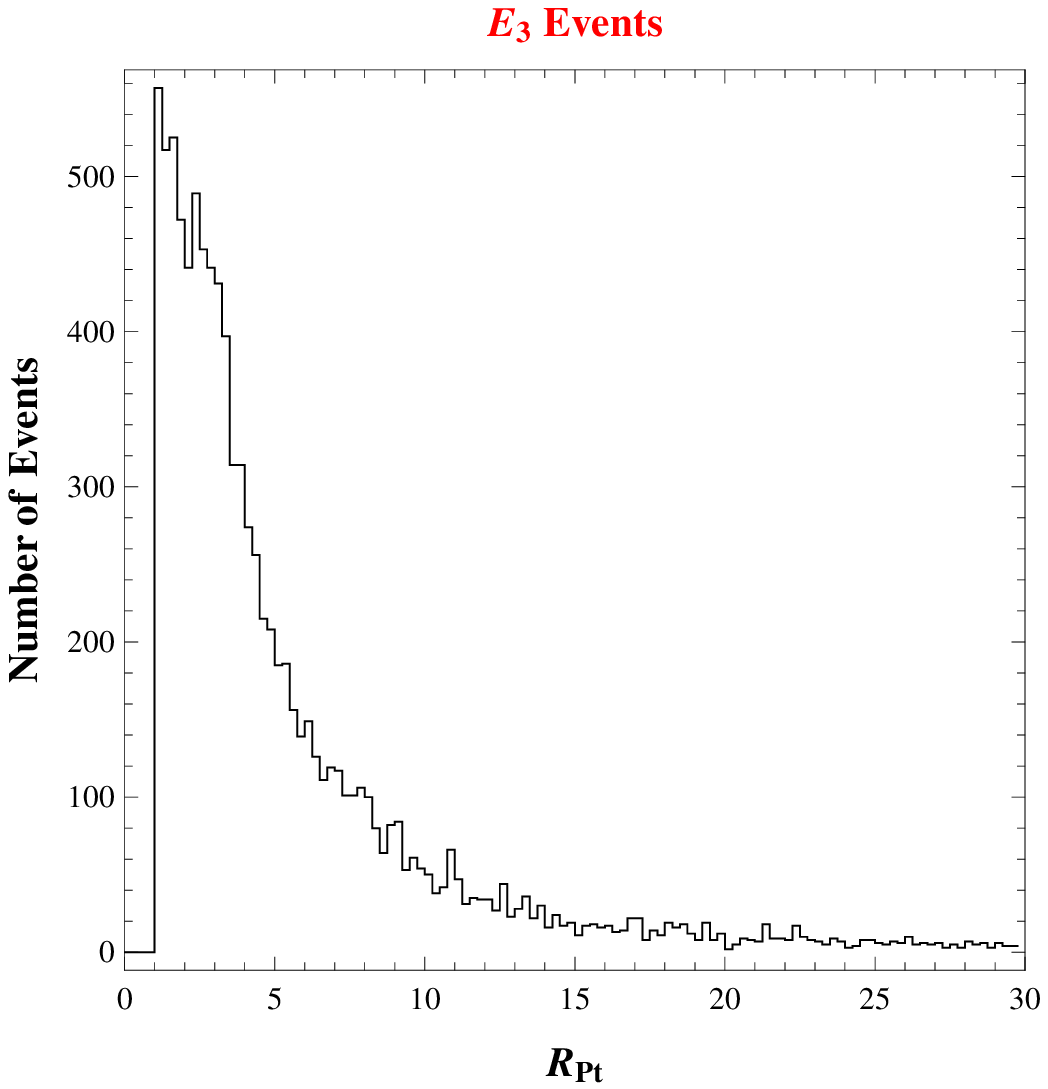}
\includegraphics[scale = 0.615]{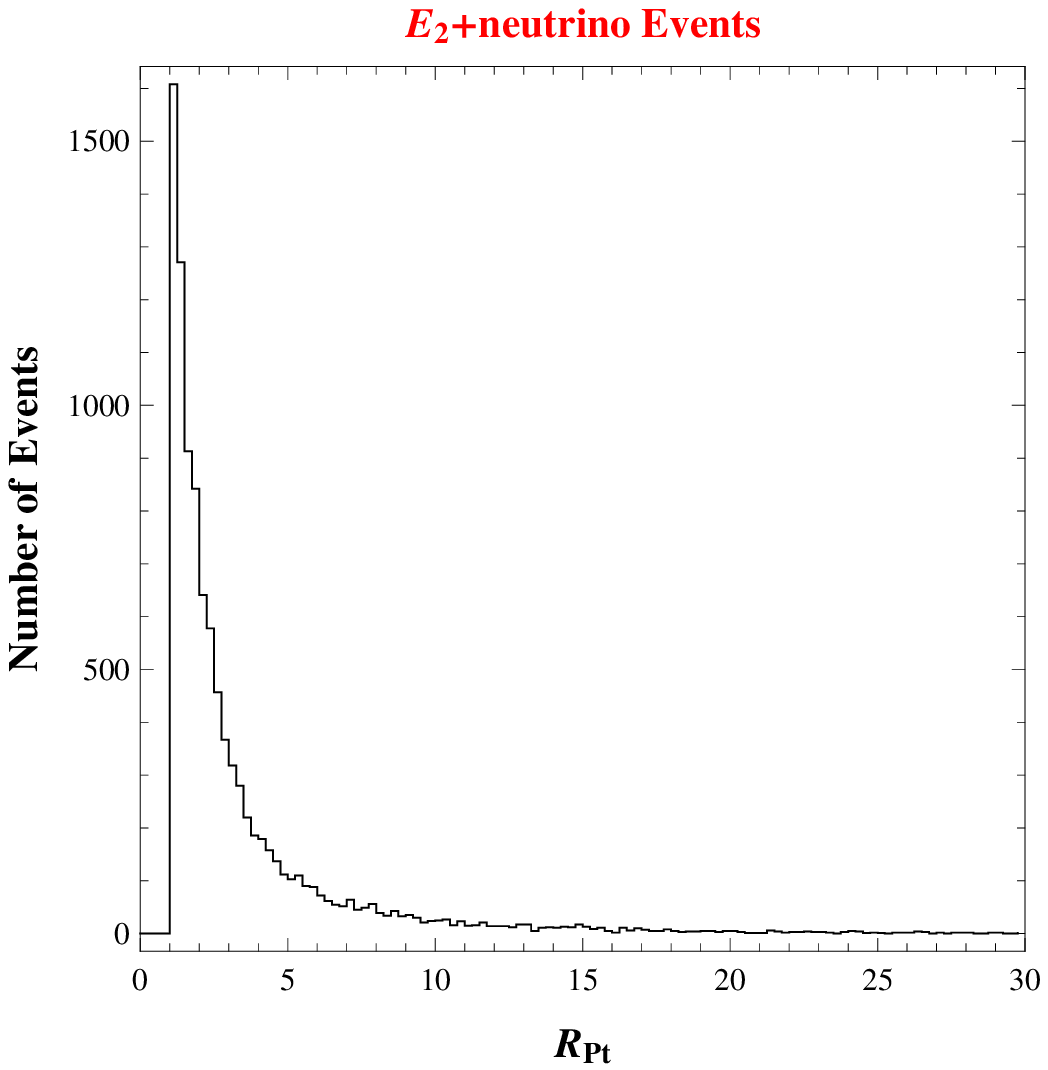}
	\caption{$R_{P_t}$ distributions for $E_2$ type events (upper-left panel),
	 $E_3$ type events (upper-right panel) and $E_2+\nu$ events (lower panel) for the case
	 with one visible particle on each decay chain.
	 The mother mass is $M = 400$ GeV; the DM mass is $m_{DM} = 150$ GeV. }
	\label{fig:ptRatio1v}
\end{figure}

For the case with one visible particle per decay chain, we consider the $P_t$ ratio of the two visible particles as follows:
\begin{eqnarray}
R_{P_t} = \frac{P_t^{\text{max}}}{P_t^{\text{min}}},
\end{eqnarray}
where $P_t^{\text{max}}$ is the larger $P_t$ of the two visibles coming from two separate
decay chains, and $P_t^{\text{min}}$ is the smaller one.
From our physical intuition mentioned above, we expect the
$R_{P_t}$ for $E_3$ type events to be larger (on average)
than that for $E_2$ type events\footnote{We neglect
$E_4$-type events and comment on this issue
later.
}. To verify this expectation,
we did a simulation using MadGraph/MadEvent for a (toy) model with mother mass $M = 400$ GeV and DM mass
$m_{DM} = 150$ GeV. 
%
%
The $R_{P_t}$ distributions for $E_2$ and
$E_3$ type events are shown in Fig.~\ref{fig:ptRatio1v}. We can see clearly that $R_{P_t}$
for $E_3$ type events is generally larger than that for $E_2$ type events. For comparison, we also
included the $R_{P_t}$ distribution for $E_2+ \nu$ events (i.e., two DM particles
and an extra invisible, but massless, particle in the full event) in Fig.~\ref{fig:ptRatio1v}.
One can easily see that the $E_2+\nu$ events have on \textit{average} larger $R_{P_t}$ than that for pure $E_2$ type events, but smaller $R_{P_t}$ than that for $E_3$ type events. This observation agrees with our
expectation: in the $E_2+\nu$ events, we have only one extra {\it{massless}}
invisible
particle (relative to $E_2$-type events), while in $E_3$ type events, we have one extra {\it{massive}} invisible particle so that the disparity
between the visible particle momenta on the two sides in the former case should be relatively
smaller.

\begin{table}
\begin{center}
\begin{tabular}{|c|c|c|c|c|c|c|c|c|c|}
\hline $R_{P_t}^\text{min}$  & 2 & 3 & 4 & 5 & 6 & 7 & 8 & 9 & 10\\
\hline $E_2$ & 0.3375 & 0.1629& 0.0957& 0.0612 & 0.0413 & 0.0285 & 0.0225 & 0.0178 & 0.0144 \\
\hline $E_3$ & 0.7929 & 0.6105 & 0.4649 & 0.3696 & 0.303 & 0.2525 & 0.21 & 0.1774 & 0.1522\\
\hline $E_2$ + neutrino & 0.5366 & 0.3323 & 0.2319 & 0.1733 & 0.1342 & 0.1101&  0.0887& 0.0738 & 0.0628 \\
\hline
\end{tabular}
\end{center}
\caption{The percentage of surviving events in $E_2$, $E_3$ and $E_2 + \nu$ events for different
choice of $R_{P_t}^{\text{min}}$ for the case with one visible
particle per decay chain. The mother mass is $M = 400$ GeV and the DM mass is $m_{DM} = 150$ GeV.\label{tab:cut1v}}
\end{table}

Because of different $R_{P_t}$ distributions of $E_2$ and $E_3$ type events, we can try to
distinguish them by doing a cut $R_{P_t} > R_{P_t}^{\text{min}}$. The percentage of
``surviving''
events in $E_2$ and $E_3$ type events according to different choice of $R_{P_t}^{\text{min}}$ is shown in
Table~\ref{tab:cut1v}.
For comparison, we also include $E_2+\nu$ events in Table~\ref{tab:cut1v}. We can see that the survival rates for $E_2$ type events are fairly independent of mother and DM masses, since in this type of event
the energies of the visible particles in two decay chains are
always comparable. Therefore,
\begin{itemize}

\item
if the survival rates
for the events (after the $R_{ P_t }$ cut)
are much larger than that of $E_2$ values
shown in Table ~\ref{tab:cut1v}, then we can conclude that the events are {\em not}
purely $E_2$ type, i.e., it is an evidence for existence of another/third invisible particle (whether massless or massive).

\end{itemize}
In general, the
survival rates for $E_3$ type events are much larger than those of $E_2$ type events,
but the survival rates of $E_3$ type events depend on the mother and DM masses. 
In addition, the survival rates for $E_2+ \nu$ events are larger than that of $E_2$ type events as well (even though they are generally smaller than that of $E_3$ type events). In this sense, an observation of large survival rates might
not (by itself)
provide a strong support that there exist $E_3$ type events in the
sample.


To get further confirmation of
$E_3$-type events
(and thus to distinguish $Z_2$ and $Z_3$ models), we can employ the $R_{P_t}$ cut
as above and then study the $M_{T2}$ distribution of the {\em surviving} events. The key idea is to compare the upper edges of the $M_{T2}$ distributions before and after the $R_{P_t}$ cut. If the underlying physics model is $Z_2$ type, then clearly
we can only obtain $E_2$ type events (or $E_2+\nu$) events before and after the cut, and the upper edge of its
$M_{T2}$ distribution is not altered. However, if the underlying physics model is $Z_3$, then
(before the cut) the total events are an {\em admixture} of $E_2$ and $E_3$ type.
Since the upper edge of $E_3$ (and $E_4$)-type events is smaller
than those of $E_2$-type events, the upper edge
of $M_{T2}$ distribution (again before the $R_{P_t}$ cut)
%
%
should be that of $E_2$ type events.
On the other hand, {\em after} the cut the upper
edge of $M_{T2}$ distribution should be lower than before since the surviving
events are mostly of $E_3$ type.

\begin{figure}[ht]
\centering
\includegraphics[scale = 0.538]{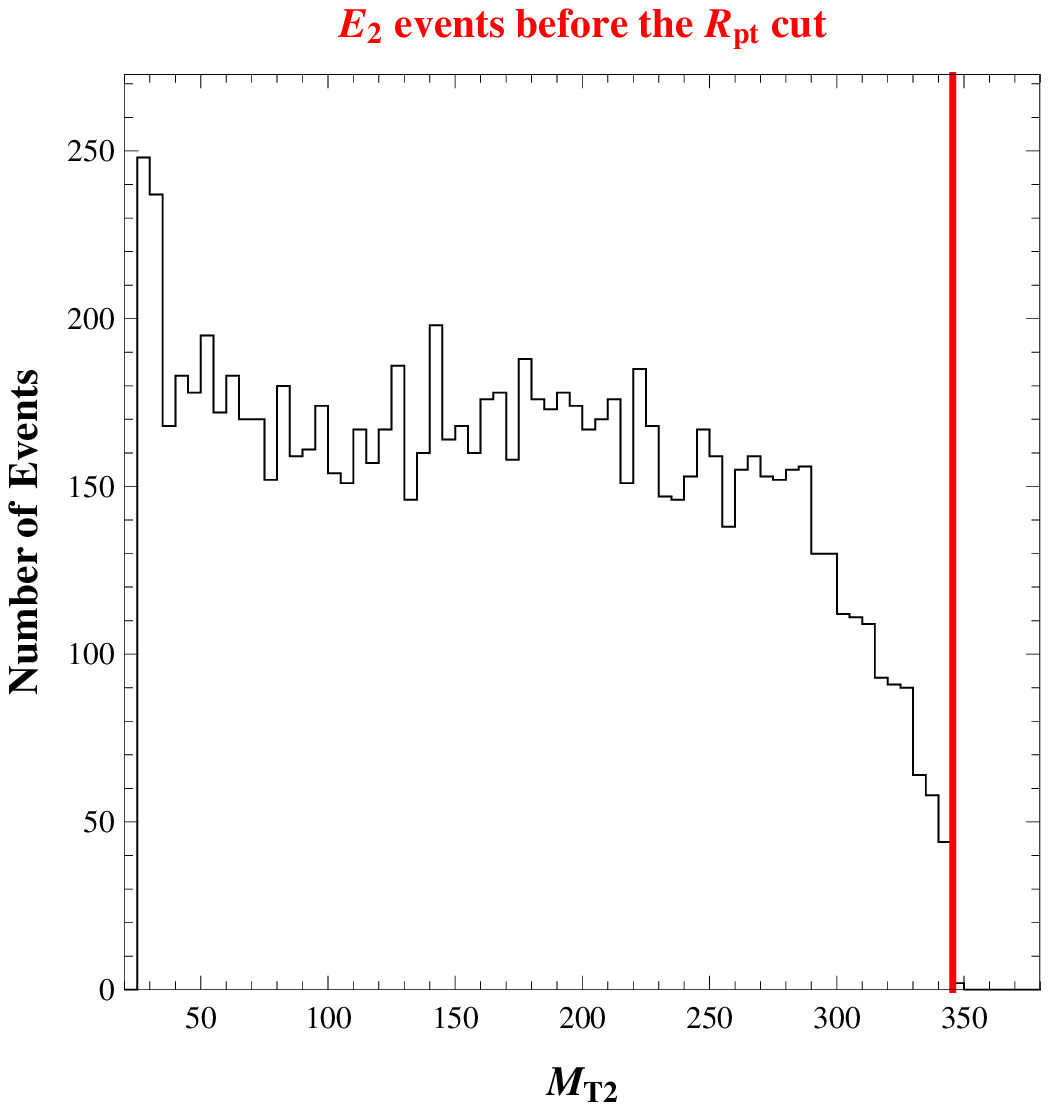} \hspace{0.5cm}
\includegraphics[scale = 0.538]{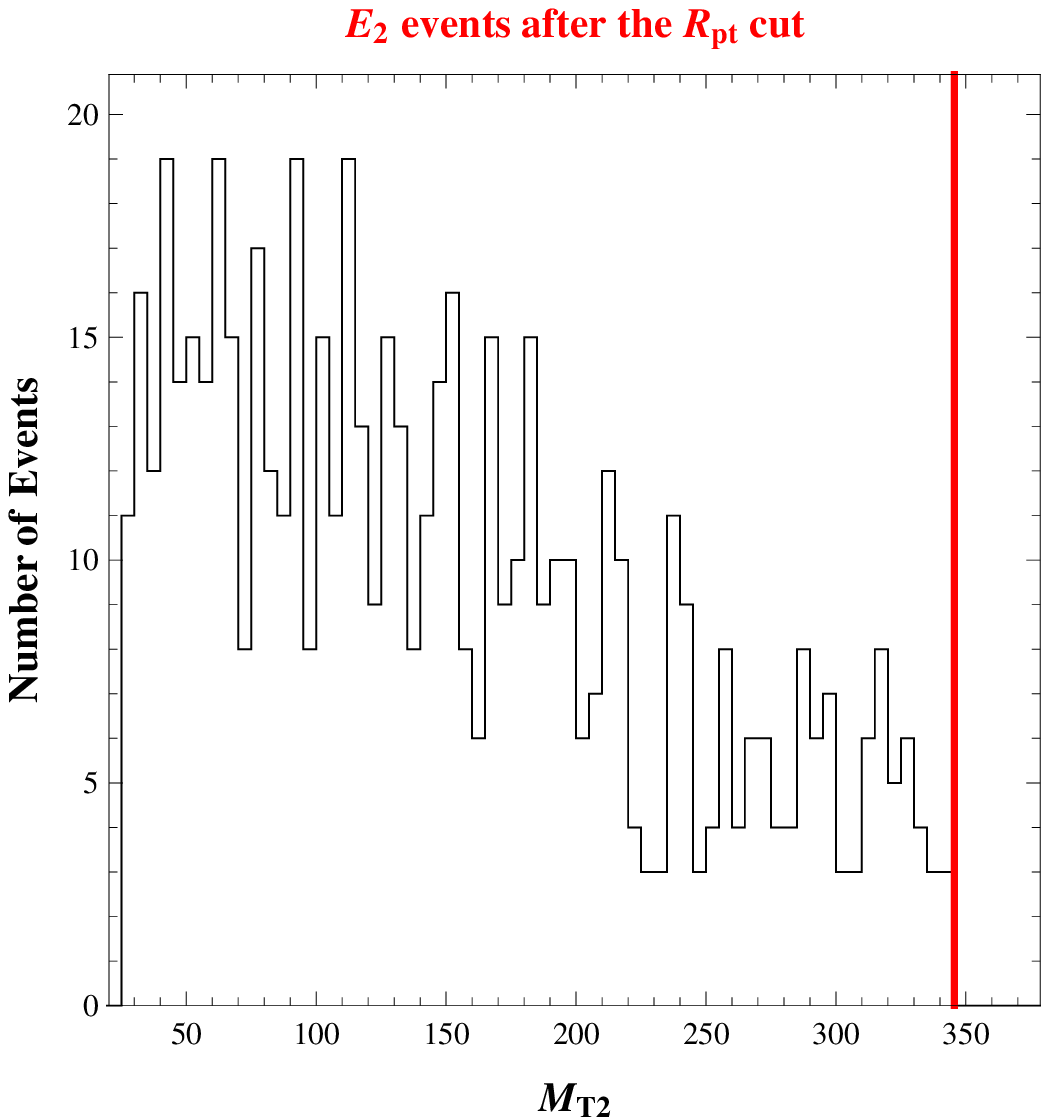} \\
\includegraphics[scale = 0.538]{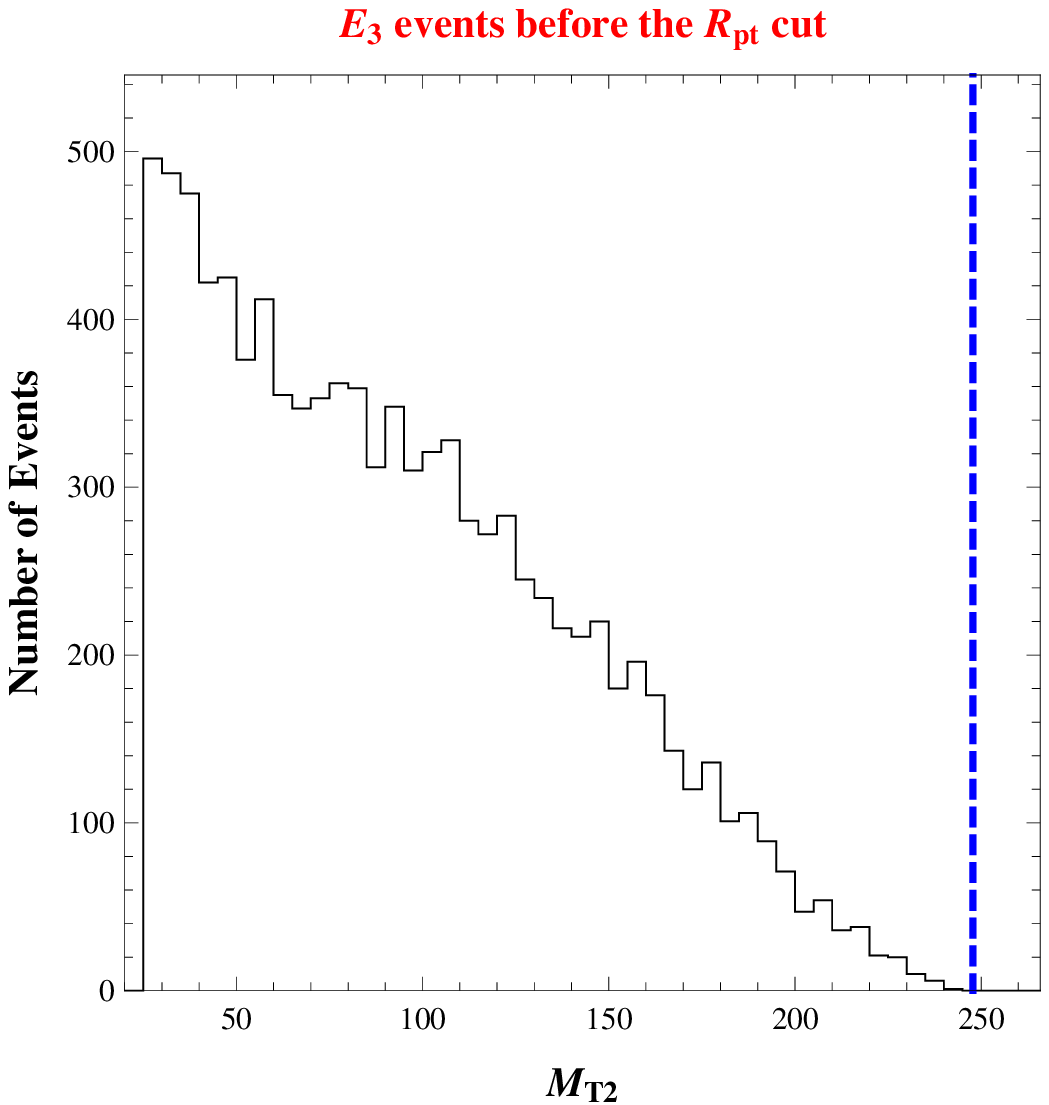} \hspace{0.5cm}
\includegraphics[scale = 0.538]{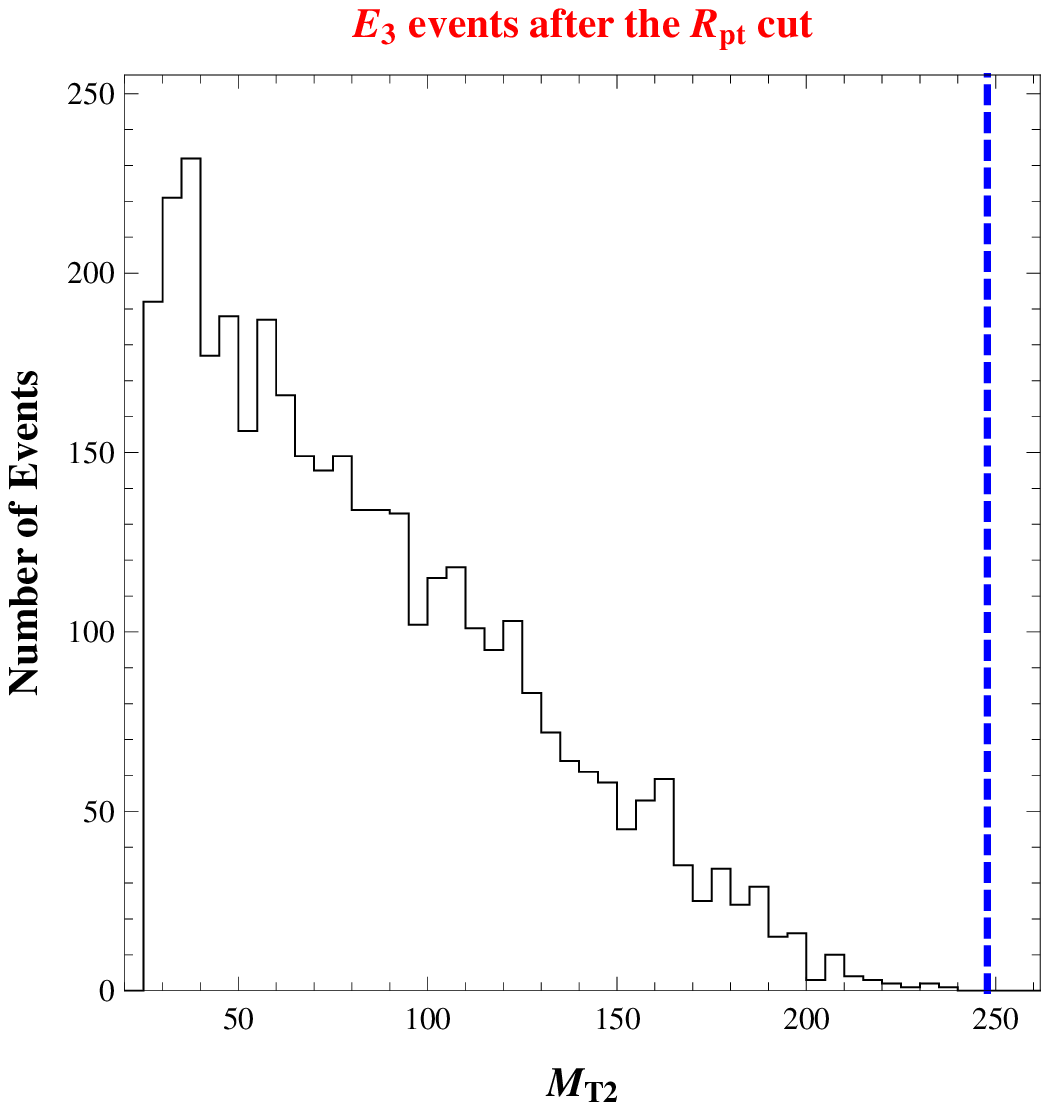}
	\caption{ $M_{T2}$ distribution for $E_2$ (top two panels) and $E_3$ type events
	(lower two panels) for simulated events using a model with $M = 400$ GeV and $m_{DM} = 150$ GeV and
	one visible particle per decay chain. The trial mass is chosen
	to be $\tilde{m} = 25$ GeV.
	 The left panels are the $M_{T2}$ distributions before the $R_{P_t}$
	cut, and the right panels are the $M_{T2}$ distributions after the $R_{P_t}$ cut.
%
%
The solid red (dashed blue) lines are the theoretical
	prediction for the upper edges of $M_{T2}$ distribution for $E_2$ and $E_3$
	type events, respectively. \label{fig:MT2_E2E3}}
\end{figure}

To illustrate this technique,
we apply the analysis outlined above to
the previously simulated events using a model with $M = 400$ GeV, $m_{DM} = 150$ GeV, and we pick the trial
mass to be $\tilde{m} = 25$ GeV for the purpose of illustration. Based on the survival rates shown in Table~\ref{tab:cut1v}, we choose $R_{P_t}^{\min} = 5$.  Before we do the analysis,
we need to investigate whether the $R_{P_t} > 5$ cut is ``biased,'' i.e., does the cut tend to
remove more events with a high $M_{T2}$ value?\footnote{If the answer is affirmative,
then $M_{ T2 }^{ \max }$ after the cut will be reduced even for {\em purely} $E_2$-type
events.} For this we consult Fig.~\ref{fig:MT2_E2E3}, which
shows {\em separately} the $M_{T2}$ distributions for both
pure $E_2$ and $E_3$ type events before and after
the $R_{P_t}$ cut. By comparing the left panels and right panels in Fig.~\ref{fig:MT2_E2E3} we can
easily see that the upper edges for both $E_2$ and $E_3$ type events do not get modified after the
$R_{P_t}$ cut, which suggests that the $R_{P_t}$ cut is not ``biased''. \footnote{But a choice of larger $R_{P_t}^\text{min}$ would introduce bias in the cut.}.
In addition, we see that these upper edges in simulated events approximately agree with
the theoretical predictions (shown by vertical lines)\footnote{But note that in experiments these predictions are {\it{a priori}} unknown since we do not know the masses of the mother and DM separately.}.

\begin{figure}
\centering
\includegraphics[scale = 0.7]{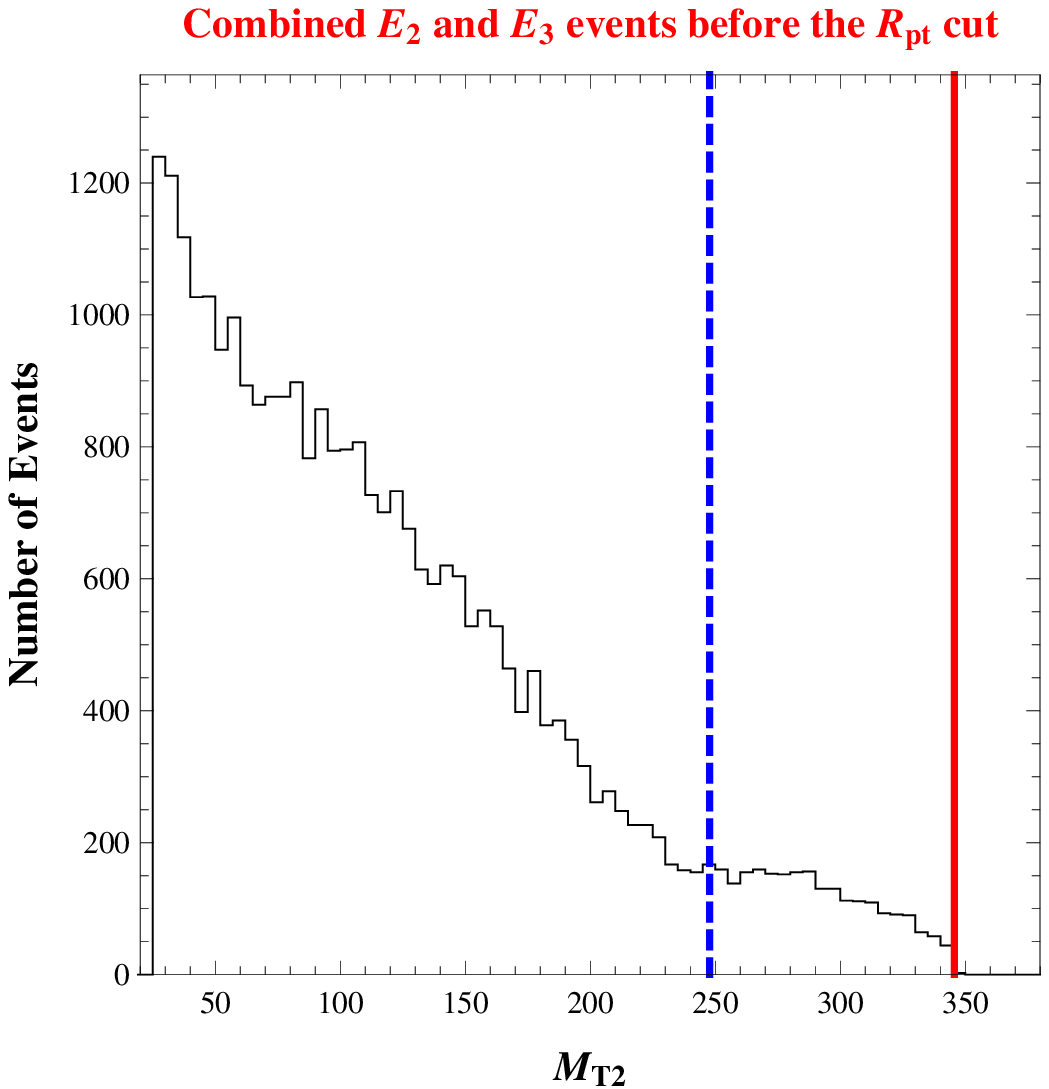} \hspace{0.5cm}
\includegraphics[scale = 0.7]{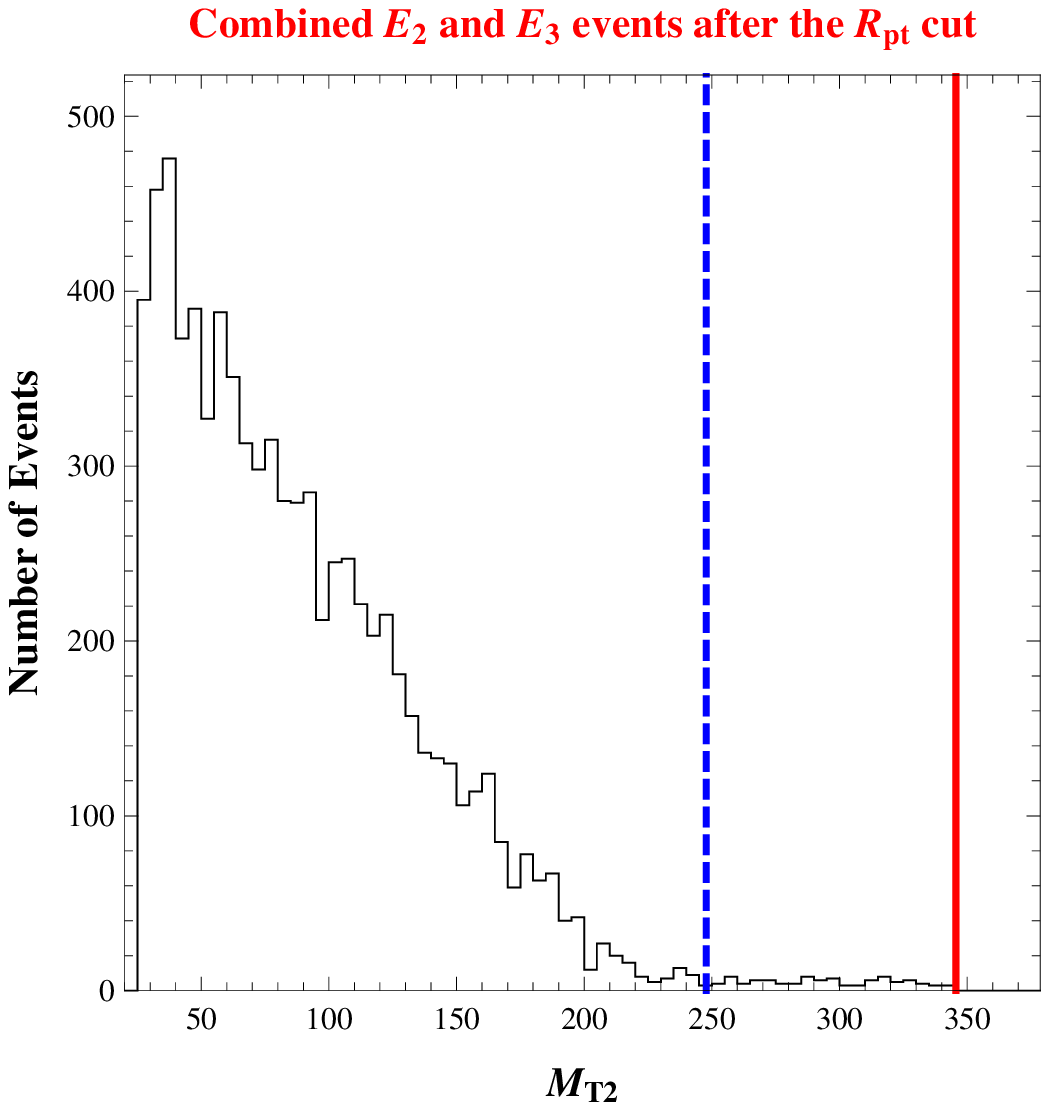}
	\caption{ $M_{T2}$ distribution for combined $E_2$ and $E_3$ type events ($1:2$ ratio) before (left panel)
	and after (right panel) the $R_{P_t} > 5$ cut for the case with one visible particle
	per decay chain. The mother mass is $M = 400$ GeV and the DM mass
	$m_{DM} = 150$ GeV. The trial mass is chosen
	to be $\tilde{m} = 25$ GeV. The solid red (dashed blue) lines represent the theoretical
	predictions for the upper edges of $M_{T2}$ distributions for $E_2$ and $E_3$-type events, respectively.
%
%
\label{fig:combinedE2E3}  }
\end{figure}

Now we can demonstrate how to distinguish $Z_2$ and $Z_3$ models using
a {\em combination} of $R_{P_t}$ cut
and $M_{T2}$ distributions. First we consider a $Z_3$ model, where
we assume that the branching ratios for the mother to decay into one DM (and visible particle) and into
two DMs (and visible particle) are both $50\%$.\footnote{In general, the decay into two DM should be phase-space suppressed relative to the decay into one DM. However, in some specific models, this suppression (for the decay chain with two DM) could be compensated by larger effective couplings in
that chain so that the two decay processes can have comparable branching ratio as 
assumed here.} Since we assumed that the visible particles in these
two decays are identical, we have to combine the $M_{ T2 }$ distributions
for $E_2$ and $E_3$ type events in a $1:2$ ratio to get
the {\em total} $M_{T2}$ distribution. The result is shown in the left panel of Fig.~\ref{fig:combinedE2E3}.
As expected, we can see from this figure that the combined $E_2$ and $E_3$ type events have an upper edge in $M_{T2}$
distribution that agrees with the theoretical expectation for $E_2$ type events
(the red solid line).
As discussed earlier,
we can also see that the total $M_{T2}$
distribution has a kink near the theoretical $M_{T2}^\text{max}$ for $E_3$ type events, but it is hard
to identify such a kink because of statistical fluctuations. The right panel
of Fig.~\ref{fig:combinedE2E3} shows the $M_{T2}$ distribution for the combined events {\em after}
the $R_{P_t}$ cut. It can be seen clearly that the upper edge of $M_{T2}$ distribution gets
reduced. In fact, the new edge
agrees with the theoretical expectation of the $M_{T2}^\text{max}$ of $E_3$ type events (the blue solid line). This confirms our expectation
that the $R_{P_t}$ cut discards most of $E_2$ type events while retaining
a sizeable fraction of $E_3$ type events, i.e., the events which pass the $R_{P_t}$ cut are mostly $E_3$ type.

Of course, we do not know {\textit{a priori}} where
the $M_{T2}^\text{max}$ for $E_3$ type events (solid blue line)
lies due to the lack of knowledge of the mother and DM masses. Rather the idea is that we can simply {\em measure} the upper edge in $M_{T2}$ distributions after the $R_{P_t}$ cut
(again, this approximately corresponds to that of $E_3$-type events). Combining this edge with that before the cut
(i.e., corresponding to $E_2$-type event)
then allows us to evaluate the masses of mother and DM particles
as described in detail in section~\ref{sec:onevis}\footnote{Note that
this cannot be done in $Z_2$ models as discussed in section ~\ref{sec:onevis}.}.
A complication arises (as follows)
in obtaining the $M_{T2}^\text{max}$ for events after the $R_{P_t}$ cut. As can be seen in the right panel of Fig.~\ref{fig:combinedE2E3}, there are still some events beyond the theoretical value of $M_{T2}^{\max}$ for $E_3$ type events (the blue solid line in the plot), for example from
(a small number of) $E_2$-type events which passed the cut. So we need an algorithm to
get rid of those ``background'' events and
do a fit to the $M_{ T2 }$ distribution in order to find $M_{T2}^{\max}$. The details of the method we employed are discussed in App.~\ref{app:algorithm}. We apply the above techniques to the simulated events. The values of mother and DM masses we obtained from
this analysis are $394 \pm 8$ GeV and $142 \pm 13$ GeV, which agree quite well with the theoretical values ($400$ GeV and
 $150$ GeV). However, we expect that uncertainty in energy measurements
would introduce
 additional errors so that a more thorough analysis taking into account these effects 
 (which is beyond the scope of this paper) is needed in order to be more realistic.

\begin{figure}
\centering
\includegraphics[scale = 0.7]{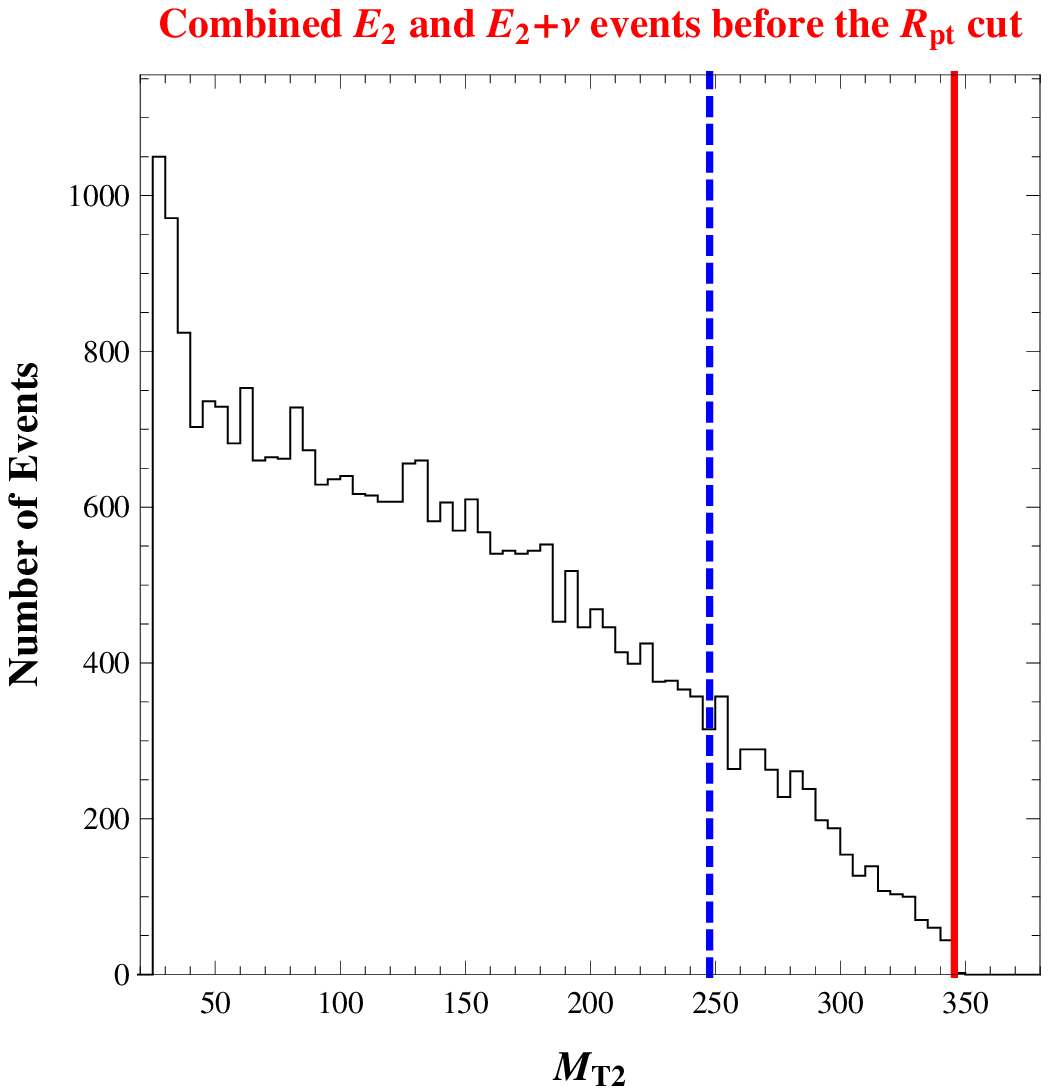} \hspace{0.5cm}
\includegraphics[scale = 0.7]{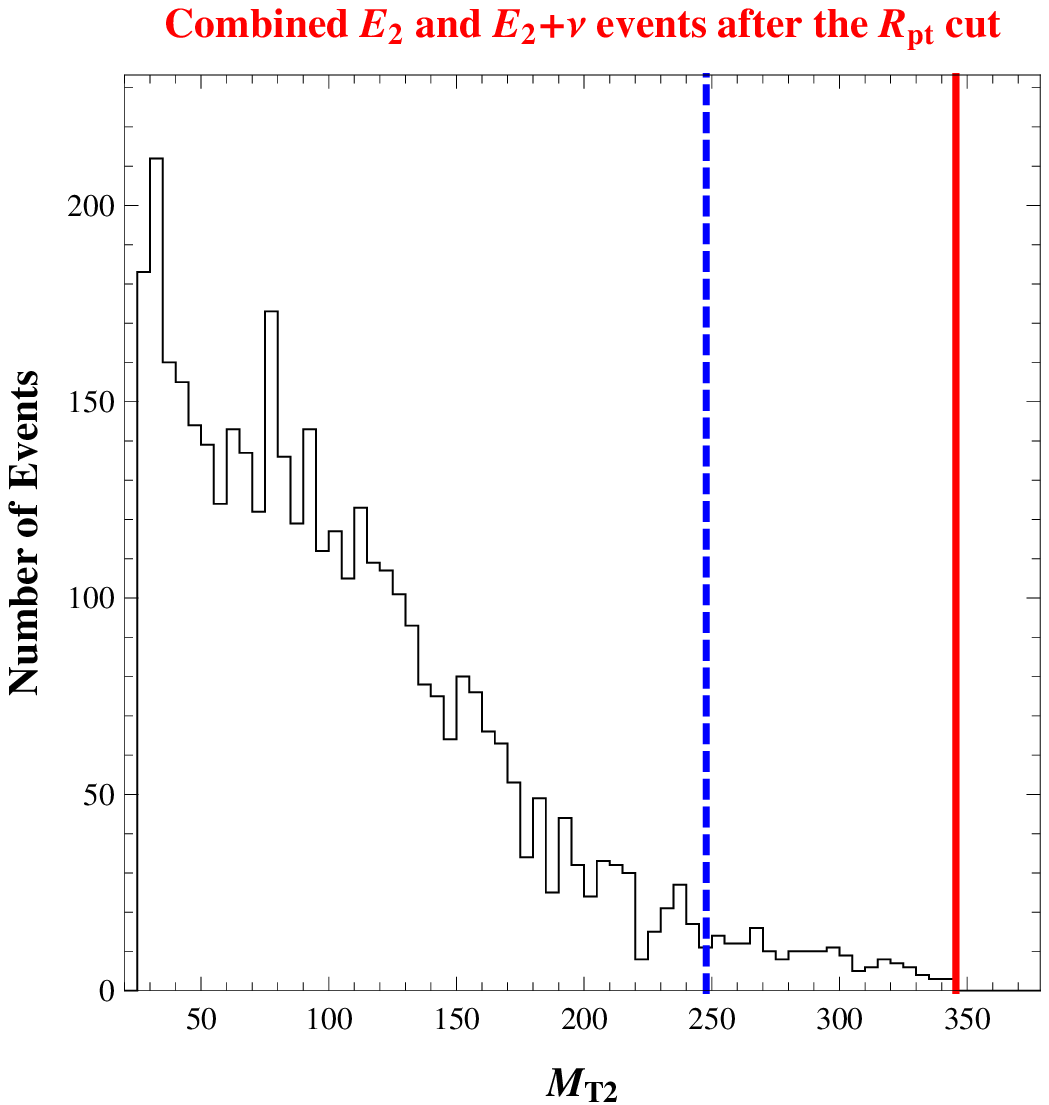}
	\caption{ $M_{T2}$ distribution for combined $E_2$ and $E_2+\nu$ events ($1:2$ ratio) before (left panel)
	and after (right panel) the $R_{P_t} > 5$ cut for the case with one visible
	particle per decay chain. The mother mass is $M = 400$ GeV and the DM mass
	$m_{DM} = 150$ GeV. The trial mass is chosen
	to be $\tilde{m} = 25$ GeV. The solid red (dashed blue) lines represent the theoretical
	predictions for the upper edges of $M_{T2}$ distributions
	for $E_2$ and $E_3$-type events, respectively.
%
%
\label{fig:combinedE2E2n}  }
\end{figure}

For comparison, we consider now $Z_2$ models. In these models, we
have either $E_2$ or $E_2+\nu$ events.
For pure $E_2$ type events, we have already shown in the upper panels of
Fig.~\ref{fig:MT2_E2E3} that the upper edge of $M_{T2}$ distribution is not reduced after the $R_{P_t}$ cut (note that this would not be true if the $R_{P_t}$ cut is ``biased''). For completeness,
we also consider
$Z_2$ models where the mother can decay into one DM or one DM plus neutrino,
with the visible particle in the two decay chains being identical.
We again assume that both branching ratios are $50\%$.
Thus, we will obtain a {\em combination} of $E2$ and $E_2 + \nu$-type events.
The $M_{T2}$ distributions in this case are shown in Fig.~\ref{fig:combinedE2E2n}.
The left panel shows that
before the $R_{P_t}$ cut, the upper edge of $M_{T2}$ distribution agrees with
the theoretical
prediction of $E_2$ type events and does not have a kink.
And,
by comparing with the right panels of this figure (i.e., after the $R_{P_t}$
cut, when mostly $E_2 + \nu$-type events survive) we can see that the location of
the upper edge for $M_{T2}$ distribution also does not change.
These two observations
are easily explained by the fact that, as
discussed earlier, the $M_{ T2 }^{ \max }$ for
$E_2 + \nu$-type events is the {\em same }as for (purely) $E_2$-type events.
Based on the above discussions
for $Z_2$ and $Z_3$ models, we conclude that
\begin{itemize}

\item
by observing whether the upper edge of $M_{T2}$ distribution changes (in particular,
reduces) {\em after
the $R_{P_t}$ cut}, we can distinguish between $Z_2$
(including those with neutrino appearing in decay of a mother) and $Z_3$ models.

\end{itemize}

\subsubsection{More than One Visible/SM Particles in Each decay Chain}
\label{sec:onevisiblenonidentical}

\begin{figure}
\centering
\includegraphics[scale = 0.6]{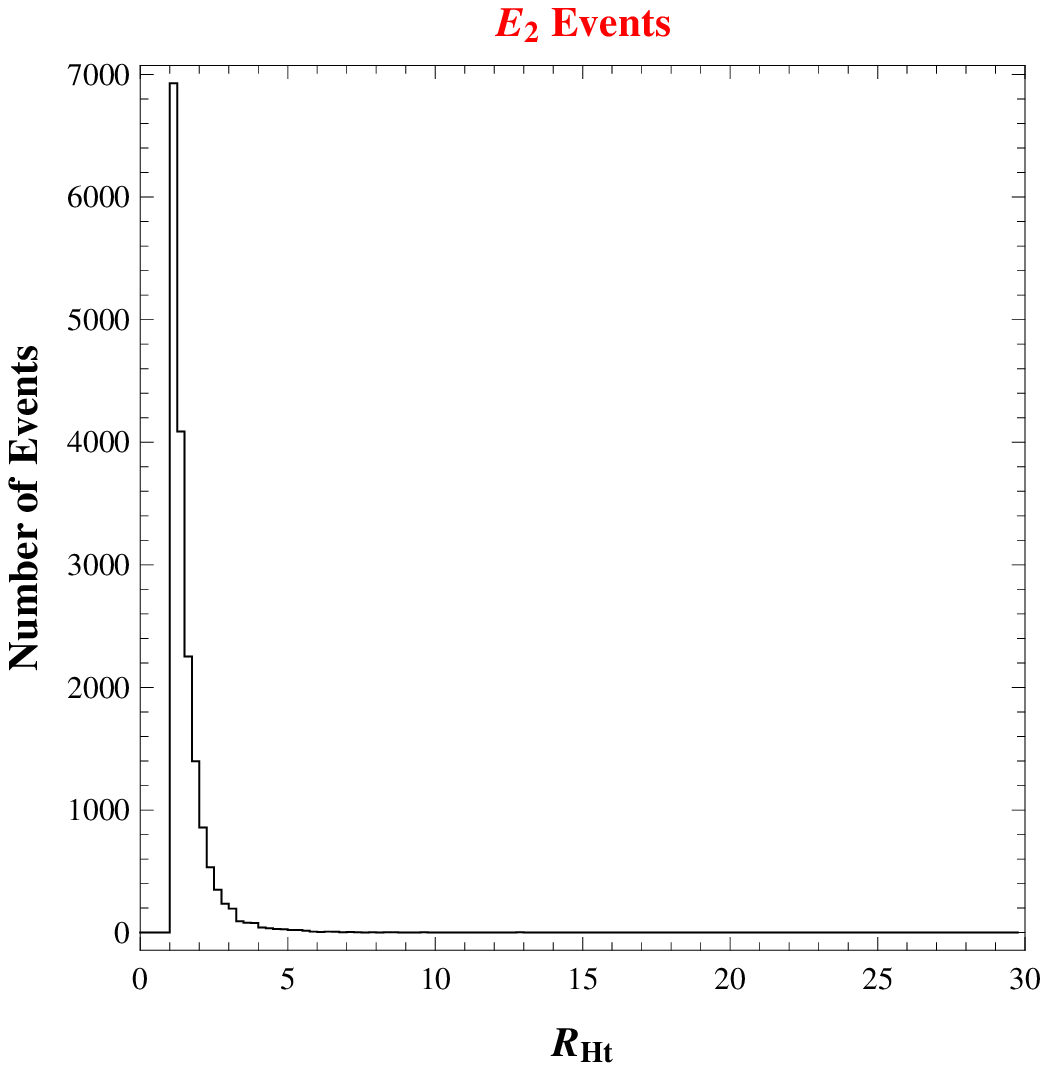}
\includegraphics[scale = 0.6]{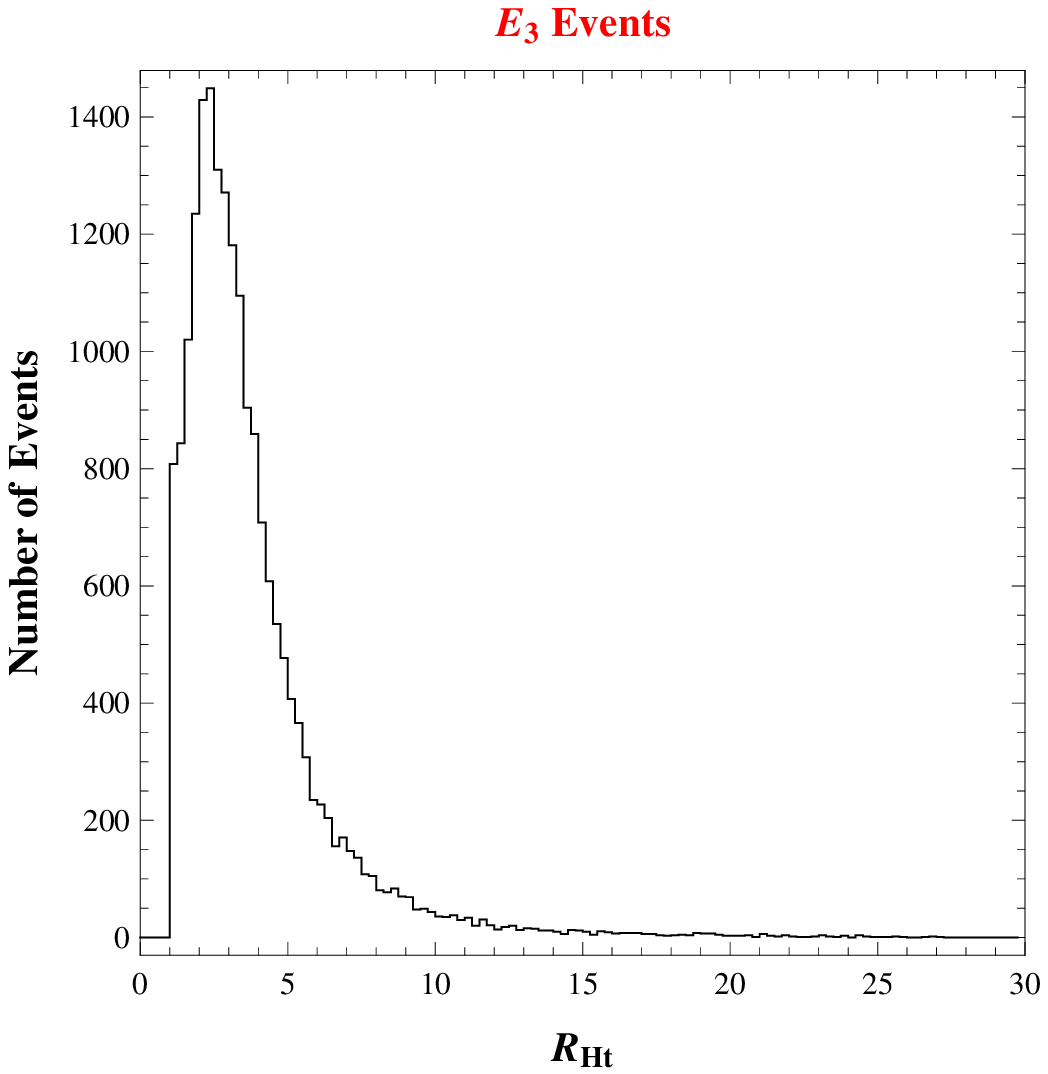}
\includegraphics[scale = 0.6]{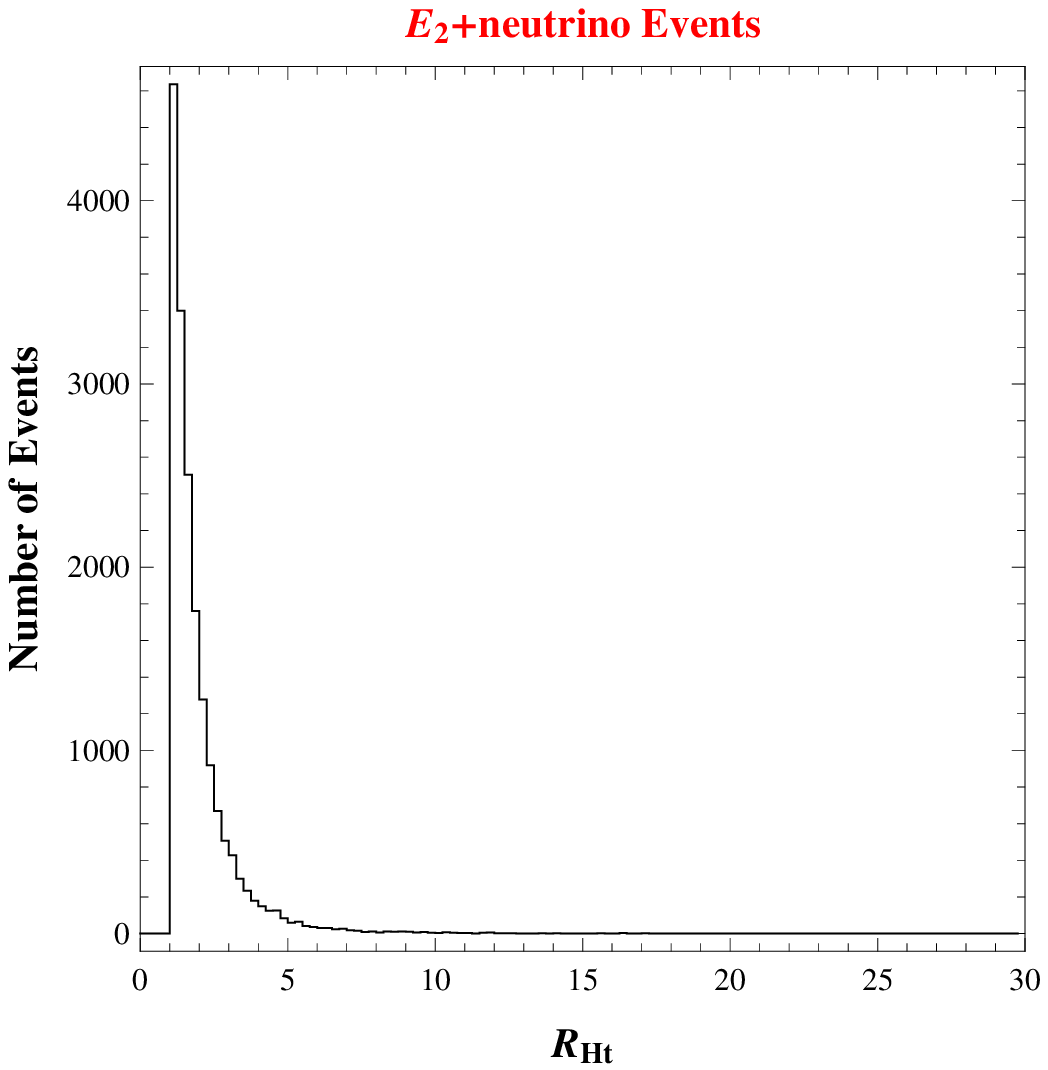}
	\caption{$R_{H_t}$ distributions for $E_2$ type events (upper-left panel),
	 $E_3$ type events (upper-right panel) and $E_2+\nu$ events (lower panel) for the case
	 with two visible particles on each decay chain.
	 The mother mass is 400 GeV, the DM mass is 150 GeV. }
	\label{fig:HtRatio2v}
\end{figure}

Next let us consider the case with more than one visible/SM particles per decay 
chain\footnote{Note that in this case, we will get a double-edge in the
visible invariant mass distribution 
from a {\em single} mother decay, which can already be used to distinguish
$Z_3$ from $Z_2$ models \cite{Agashe:2010gt}. However, it is always
useful to have more techniques -- such as the
one, using decays of 
{\em both} mothers in the event, that we are developing here -- for such discrimination.}
To be specific, we consider the case with two visible particles per decay chain. A similar analysis can be done if there are more than two visible particles. To separate $E_2$ and $E_3$ type events, we consider the ratio of $H_t$, where $H^i_t = \sum_a |P_t^{v^i_a}|$ is the scalar sum of $P_t$'s of visible particles in the same decay chain (assuming we know which particles come from which decay chain), and $i = 1,\;2$ is the index for the decay chains. $H_t$ gives a measure of how energetic the visible particles are in each decay chain. We define the $H_t$ ratio as follows
\begin{eqnarray}
R_{H_t} = \frac{H_t^\text{max}}{H_t^\text{min}},
\end{eqnarray}
where $H_t^\text{max} = \text{max}(H_t^1, H_t^2)$ and $H_t^{min} = \text{min}(H_t^1, H_t^2)$.
From similar reasons to the one visible particle case discussed above, we
expect $R_{H_t}$ for $E_3$ type events to be larger than that for $E_2$ type events on average. To illustrate this feature,
we simulate $E_2$, $E_3$ and $E_2+\nu$ events for a model with $M = 400$ GeV and $m_{DM} = 150$ GeV using MadGraph/MadEvent.
%
%
The results for the $R_{H_t}$ distribution for different types of events are shown in Fig.~\ref{fig:HtRatio2v}. It can be seen that these distributions are very similar to the $R_{P_t}$ distributions for the one visible particle case shown in Fig.~\ref{fig:ptRatio1v}. The $R_{H_t}$ for $E_3$ type events is on average larger than that of $E_2+\nu$ events, which in turn is on average larger than that for $E_2$ type events. The survival rates for $E_2$, $E_3$ and $E_2+\nu$ events for the cut $R_{H_t} > R_{H_t}^{\text{min}}$ with different choices of $R_{H_t}^\text{min}$ are shown in Table~\ref{tab:cut2v}. As in the one visible case discussed before, if the survival rates for the observed events are much larger than the $E_2$ value shown in Table~\ref{tab:cut2v}, then we can conclude that the events cannot be purely $E_2$ type. But large survival rates alone cannot be used as a convincing evidence
for $Z_3$ models.

\begin{table}
\begin{center}
\begin{tabular}{|c|c|c|c|c|c|c|c|c|c|}
\hline $R_{H_t}^\text{min}$  & 2 & 3 & 4 & 5 & 6 & 7 & 8 & 9 & 10\\
\hline $E_2$ & 0.1542 & 0.0401 & 0.0142& 0.0065& 0.0028&
0.0015& 0.0009& 0.0006& 0.0005 \\
\hline $E_3$ & 0.7990& 0.5181& 0.3103& 0.1905& 0.1228& 0.0838& 0.0583& 0.0422& 0.0314\\
\hline $E_2$ + neutrino & 0.3086& 0.1192& 0.0551& 0.0279& 0.0166& 0.0104& 0.0074& 0.0052& 0.0035\\
\hline
\end{tabular}
\end{center}
\caption{The percentage of survival events in $E_2$, $E_3$ type events and $E_2+\nu$ events for different
choice of $R_{H_t}^{\text{min}}$ for the case with two visible particles
per decay chain. The mother mass is $M = 400$ GeV and the DM mass is $m_{DM} = 150$ GeV.\label{tab:cut2v}}
\end{table}

Just like in the case with one visible particle per decay chain, we again use a combined analysis of $R_{H_t}$ cut and upper edges of $M_{T2}$ distributions to distinguish between $Z_2$ and $Z_3$ models. However, there is one major difference between the case with two visible particles per decay chain and the case with one visible particle per decay chain. In the latter case, we cannot find the mother and DM mass just based on the $M_{T2}$ upper edges for events
before $R_{P_t}$ cut. On the other hand, in the case at hand,
there is a kink structure
in the $M_{T2}^{\max}$ vs $\tilde{m}$ plot for $E_2$-type events
which tells us both the mother and the DM masses
(see section~\ref{sec:review}).
And 
$M_{T2}^{\max}$ for $E_3$ type events is smaller than that of $E_2$
type events for $\tilde{m}<m_{\textnormal{DM}}$ and the same for $\tilde{m} \geq m_{\textnormal{DM}}$ (see section~\ref{sec:z3morethanone}). Thus, before the $R_{H_t}$ cut, one expects 
that $M_{T2}^{\max}$ for the {\em combined} events is always
(i.e., irrespective of the trial DM mass) given by that of $E_2$-type events
(just like for the case of one visible particle per decay chain
discussed earlier).
Therefore, in the present scenario, we can find out the mother and DM masses
before we do any $R_{H_t}$ cut.
We can then {\em predict} the edge in $M_{ T2 }$
for $E_3$-type events, i.e.,
after the $R_{H_t}$ cut (again, the events surviving the cut will
be mostly $E_3$-type).
%
%

\begin{figure}
\centering
\includegraphics[scale = 0.6]{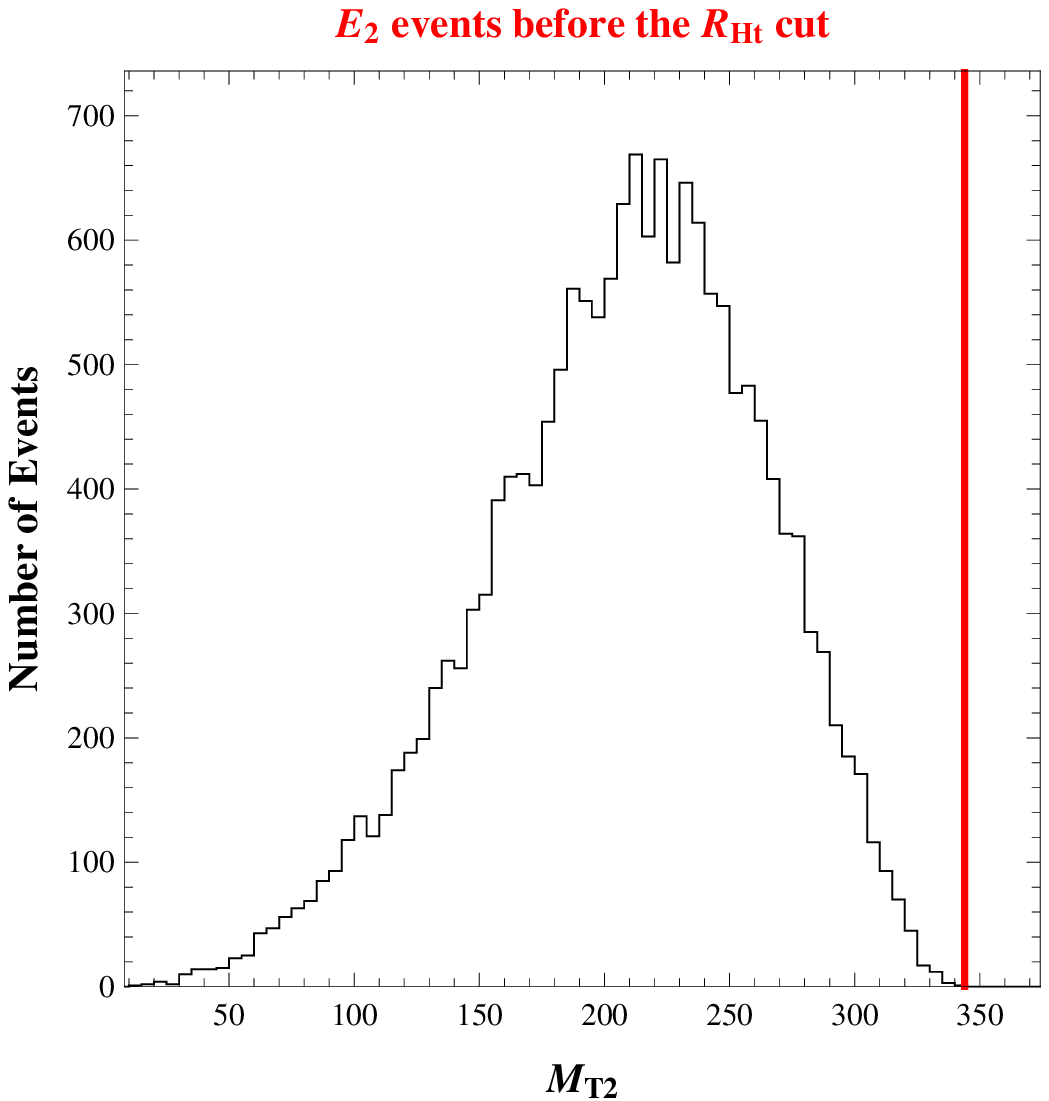} \hspace{1.0cm}
\includegraphics[scale = 0.6]{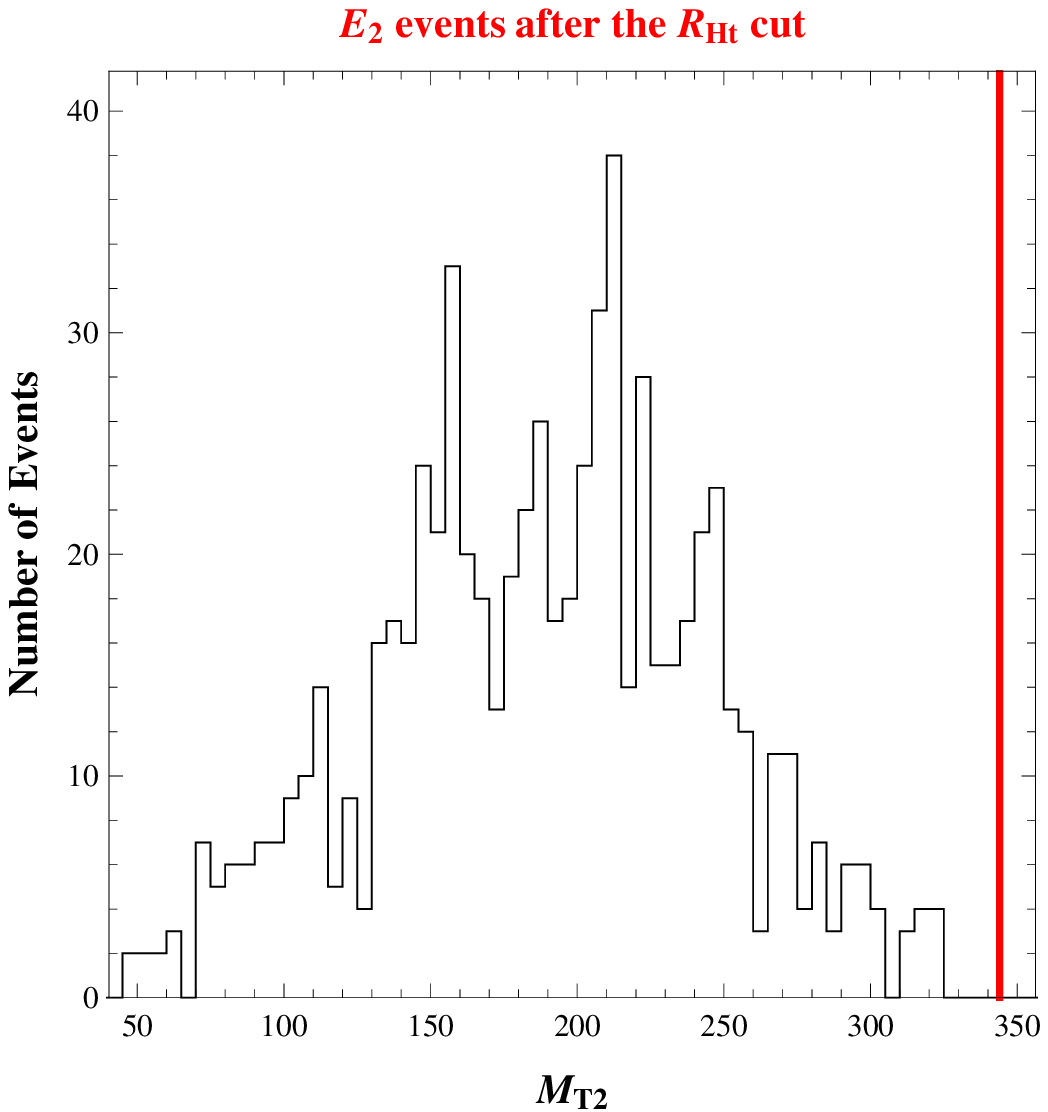} \\
\includegraphics[scale = 0.6]{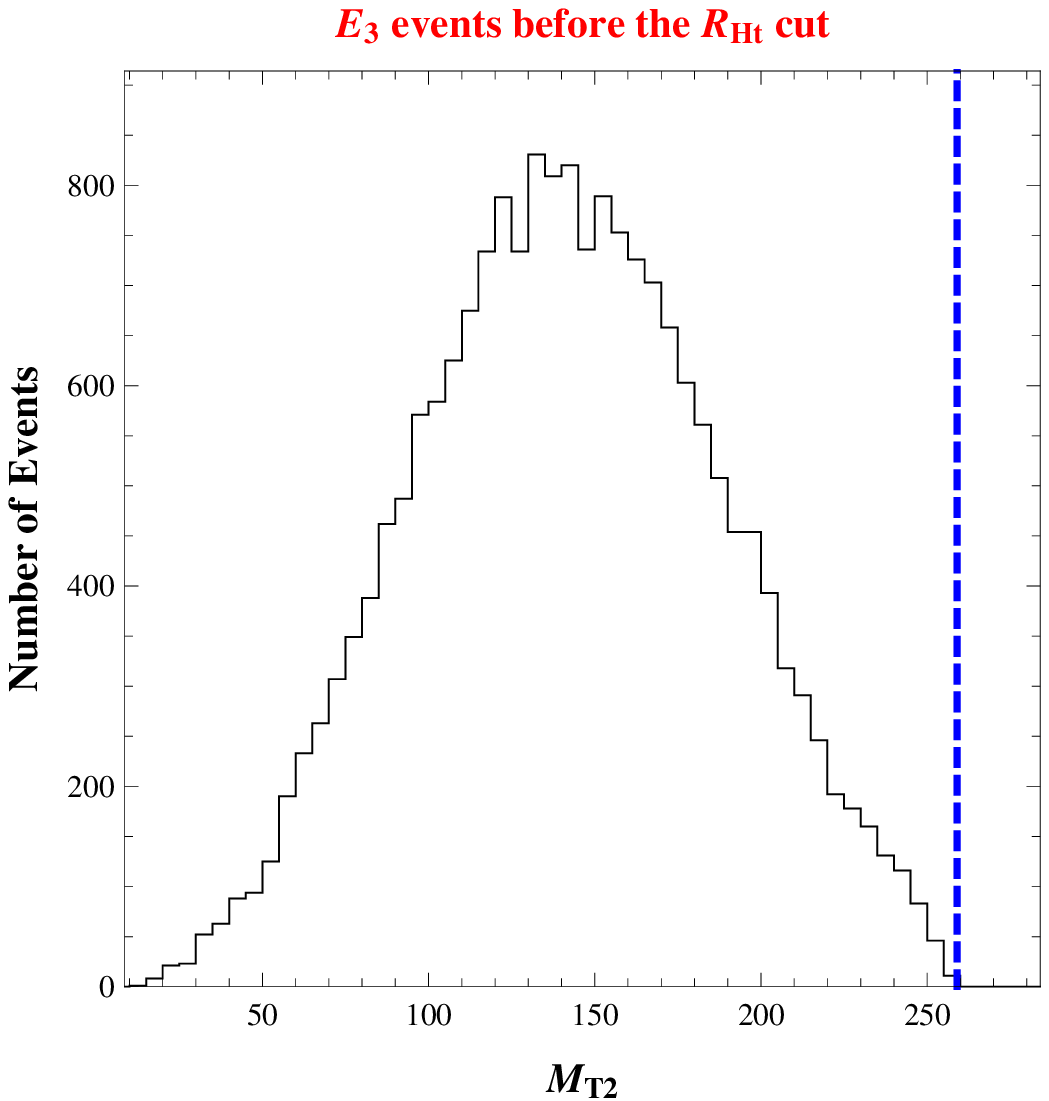}\hspace{1.0cm}
\includegraphics[scale = 0.6]{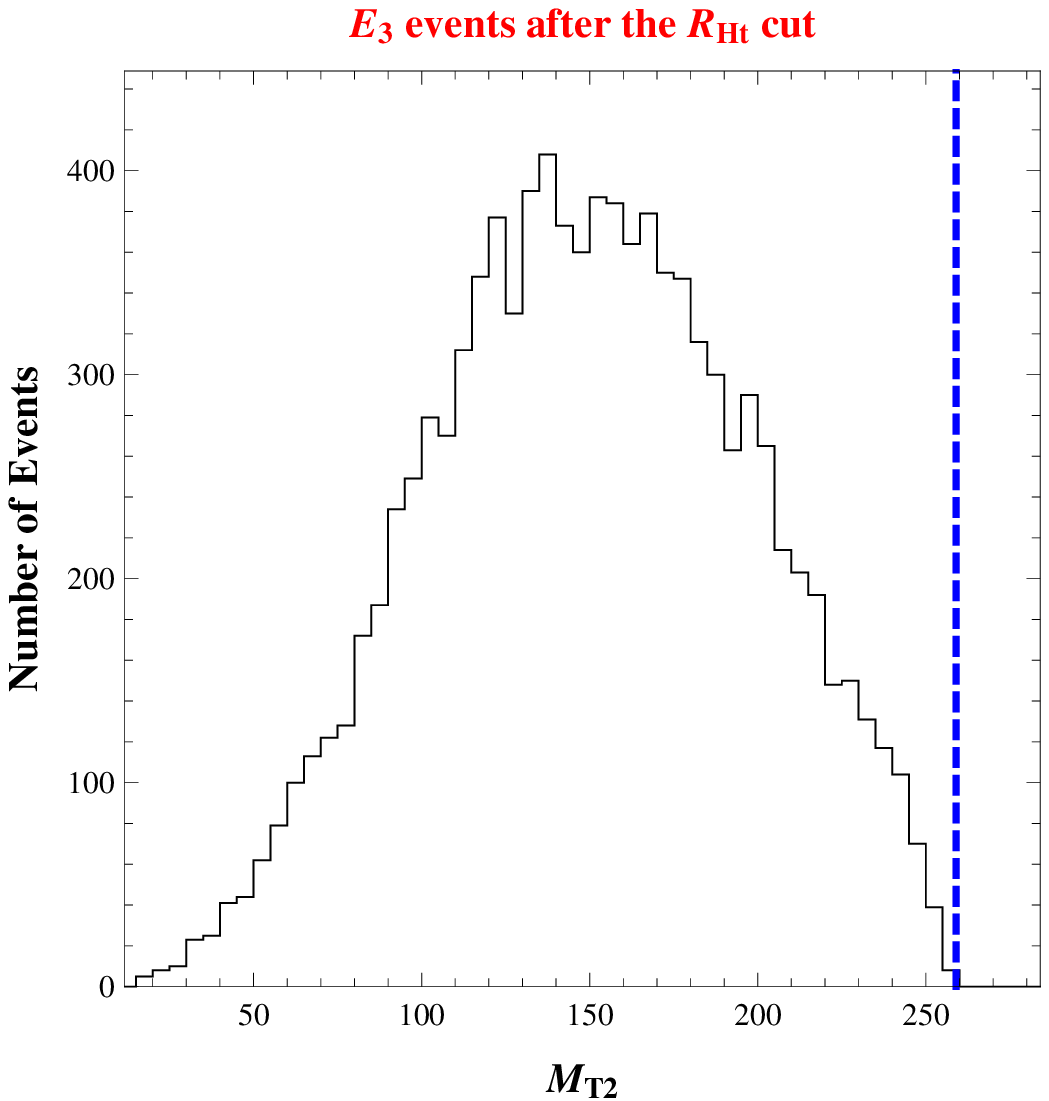}
	\caption{ $M_{T2}$ distribution for $E_2$ (top two panels) and $E_3$ type events
	(lower two panels) for simulated events using a model with $M = 400$ GeV and $m_{DM} = 150$ GeV and two
	visible particles per decay chain. The trial mass is chosen
	to be $\tilde{m} = 9$ GeV.
	 The left panels are the $M_{T2}$ distributions before the $R_{H_t} > 3$
	cut, and the right panels are the $M_{T2}$ distributions after the $R_{H_t} > 3$ cut.
%
%
The solid red (dashed blue) lines are the theoretical
	prediction for the upper edges of $M_{T2}$ distribution for $E_2$ and $E_3$
	type events. \label{fig:MT2_E2E3_2v}}
\end{figure}

\begin{figure}
\centering
\includegraphics[scale = 0.6]{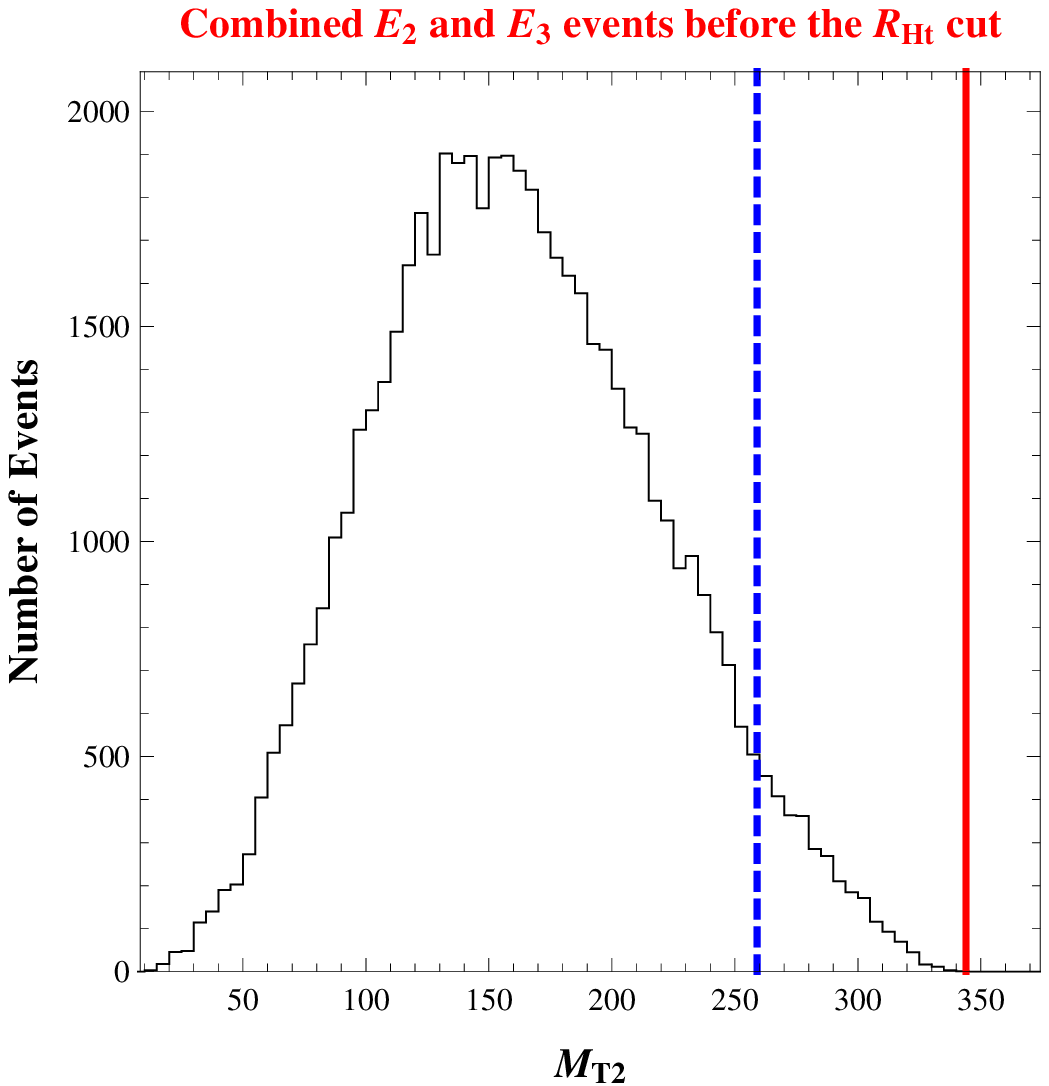} \hspace{1.0cm}
\includegraphics[scale = 0.6]{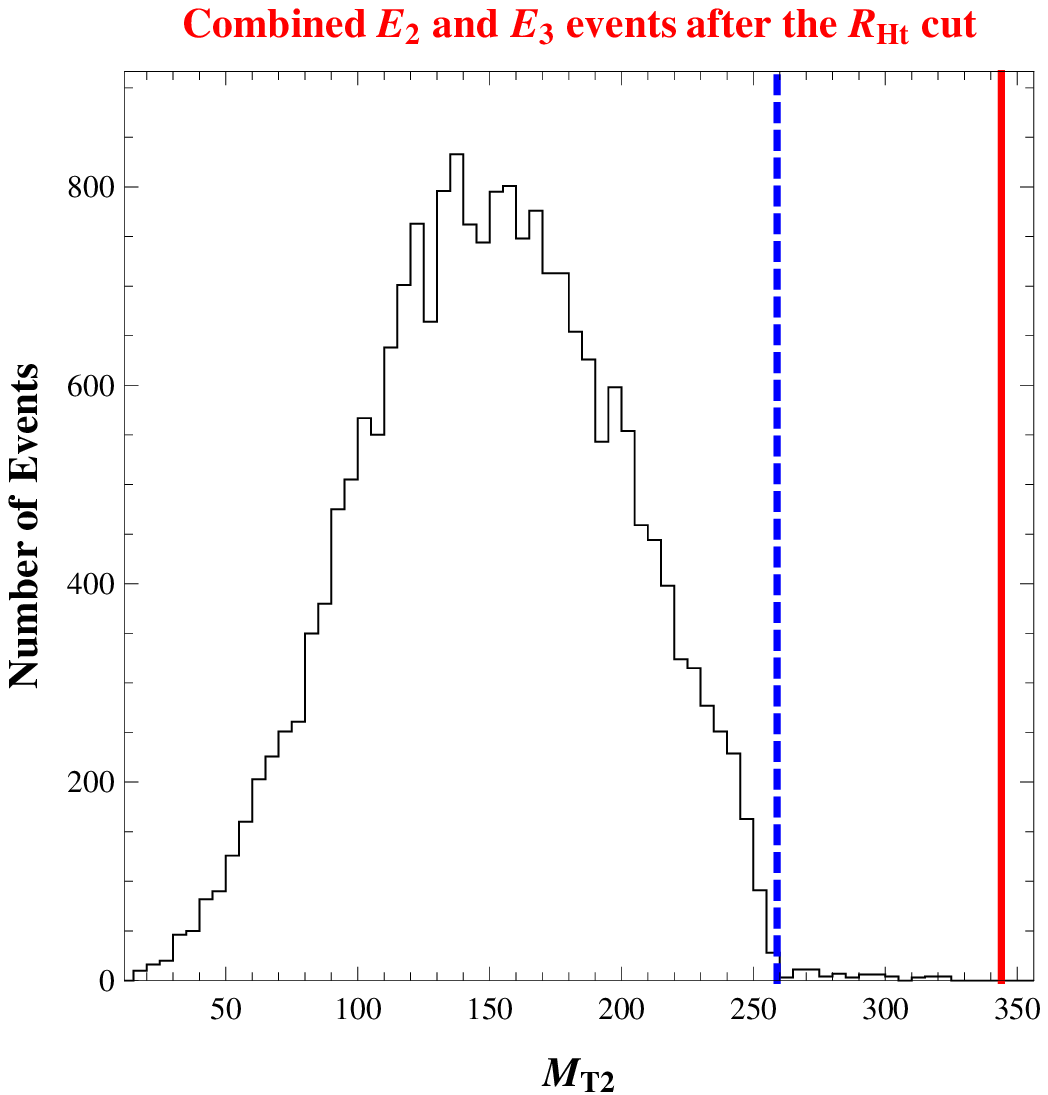}
	\caption{ $M_{T2}$ distribution for combined $E_2$ and $E_3$ type events ($1:2$ ratio) before (left panel)
	and after (right panel) the $R_{H_t} > 3$ cut for the case with two visible particles
	per decay chain. The mother mass is $M = 400$ GeV and the DM mass
	$m_{DM} = 150$ GeV. The trial mass is chosen
	to be $\tilde{m} = 9$ GeV. The solid red (dashed blue) lines represent the theoretical
	predictions for the upper edges of $M_{T2}$ distributions
	for $E_2$ and $E_3$-type events, respectively.
%
%
\label{fig:combinedE2E3_2v}  }
\end{figure}

We now demonstrate an application of the general strategy outlined above.
Based on the survival rates shown in Table~\ref{tab:cut2v}, we
choose $R_{H_t}^\text{min} = 3$ in this case. Fig.~\ref{fig:MT2_E2E3_2v} shows the $M_{T2}$ distributions
for the simulated pure $E_2$ and $E_3$ type events before and after the $R_{H_t}$ cut. 
%
%
By comparing the left and the right panels in Fig.~\ref{fig:MT2_E2E3_2v}, we can see that the $R_{H_t}$ cut does not alter the upper edge of $M_{T2}$ distribution for both $E_2$ and $E_3$ type events. Therefore the $R_{H_t}$ cut is not ``biased'' \footnote{However, a choice of higher $R_{H_t}^\text{min}$ will give rise to bias.}.
We then consider a $Z_3$ model where the branching ratios of mother decaying into two DMs (plus two visible particles) and into one DM (plus two visible particles) are both $50\%$.  We show the $M_{T2}$ distributions for the combined events before (left panel) and after (right panel) the $R_{H_t} > 3$ cut in Fig.~\ref{fig:combinedE2E3_2v}.
As expected, we see that before the $R_{H_t}$ cut, the upper edge of $M_{T2}$ distribution agrees with the theoretical
prediction of $E_2$ type events (shown by the red line).
And, the upper edge for the $M_{T2}$ distribution gets reduced
after the $R_{H_t}$ cut\footnote{Here, we have chosen $9$ GeV as the trial DM mass, i.e.,
(much) smaller than the actual DM mass so that the difference between upper edges of $M_{T2}$ for $E_2$ and $E_3$ type events, i.e., the reduction of the edge after cut, is ``exaggerated'': see  
section~\ref{sec:z3morethanone}.},
which can serve as an evidence for $Z_3$ model (cf. discussion of
$Z_2$ model below).
In addition, as mentioned earlier,
knowing the mother and DM masses from the kink in the plot of
$M_{ T2 }^{ \max }$ before the cut as a function of trial DM mass ,
we can predict (shown by the blue line)
the upper edge for the $M_{T2}$ distribution for the events that passed
the $R_{H_t}$ cut (it is just the theoretical $M_{T2}^{\text{max}}$
for $E_3$ type events).
From the right panel of Fig.~\ref{fig:combinedE2E3_2v} we see that
\begin{itemize}

\item
the observed $M_{T2}^{\text{max}}$ for events that passed the $R_{H_t}$ cut does agree with the {\em prediction} (cf. one visible particle
case above), thus providing
additional evidence that the underlying physics model is $Z_3$.

\end{itemize}
%
%

\begin{figure}
\centering
\includegraphics[scale = 0.6]{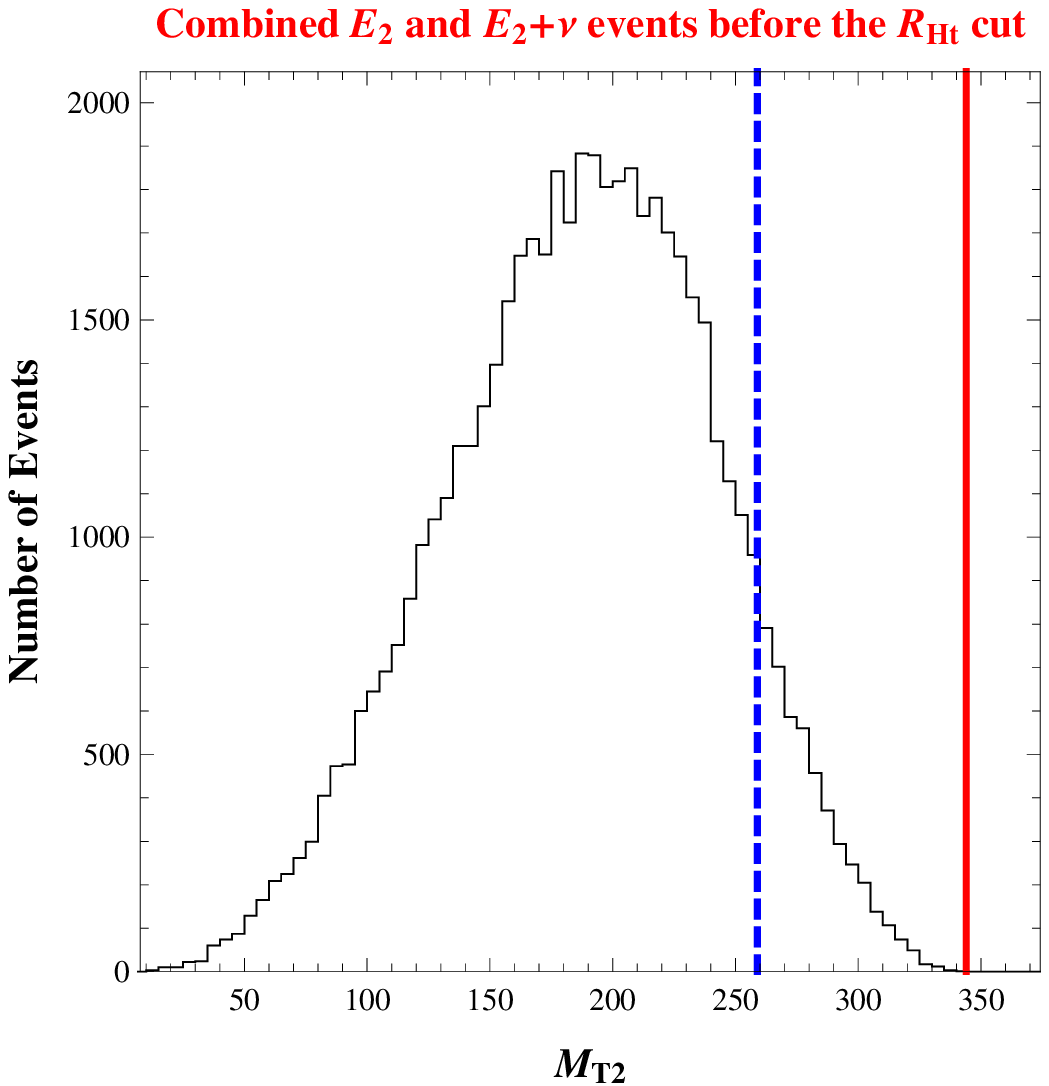}\hspace{1.0cm}
\includegraphics[scale = 0.6]{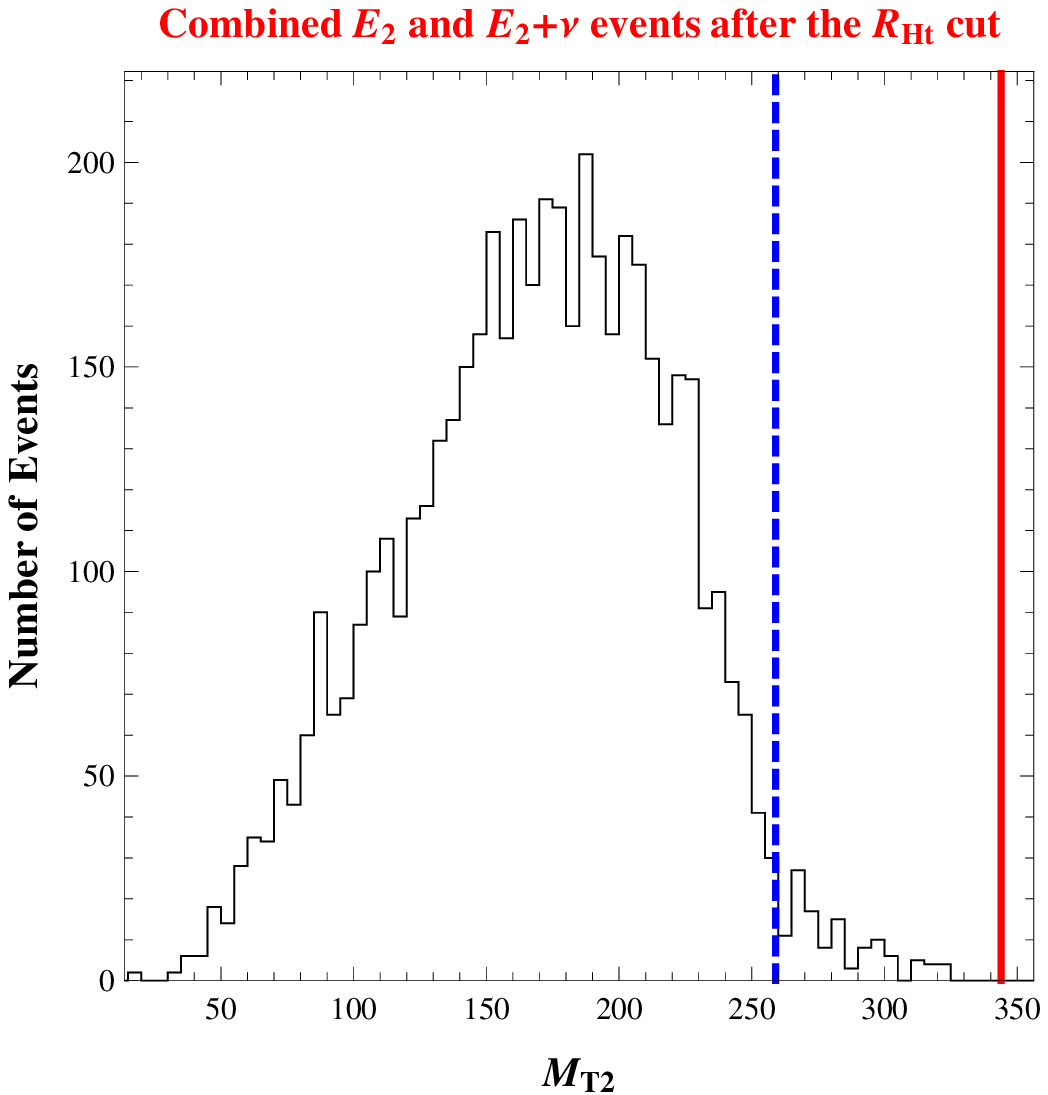}
	\caption{ $M_{T2}$ distribution for combined $E_2$ and $E_2+\nu$ events ($1:2$ ratio) before (left panel)
	and after (right panel) the $R_{H_t} > 3$ cut for the case with two visible
	particles per decay chain. The mother mass is $M = 400$ GeV and the DM mass
	$m_{DM} = 150$ GeV. The trial mass is chosen
	to be $\tilde{m} = 9$ GeV. The solid red (dashed blue) lines represent the theoretical
	predictions for the upper edges of $M_{T2}$ distributions for $E_2$ and $E_3$-type events, respectively.
%
%
\label{fig:combinedE2E2n_2v}  }
\end{figure}

We can compare the above result with $Z_2$ models.
If we just have pure $E_2$ type events, then the $R_{H_t}$ cut does not change the upper edge of $M_{T2}$ distribution, as already seen
in the upper panels of Fig.~\ref{fig:MT2_E2E3_2v}. As in the one visible
particle case, we can also consider the case where there are $E_2+\nu$
events in addition to
$E_2$ events: we assume that the branching ratios for mother to decay into one DM plus
neutrino (plus two visible particles) and into one DM (plus two visible
particles) are both $50\%$. We show the $M_{T2}$ distribution before (left
panel) and after (right panel) the $R_{H_t}$ cut for this case in
Fig.~\ref{fig:combinedE2E2n_2v}, from which we can see that
the upper
edge of $M_{T2}$ distribution
before the cut agrees with the theoretical prediction for $E_2$-type events
and that it again
does not change after the cut (as expected: see similar discussion
for the one visible particle case).

\subsection{A Summary of the Analysis and its Limitations}

Now we summarize the analysis needed to be carried out to distinguish between $Z_2$ and $Z_3$ models when the visible particles on each decay chain are identical.

For the case with one visible particle per decay chain:
\begin{itemize}
\item  We first find $M_{T2}^{\max}$ with different trial DM masses ($\tilde{m}$) for {\em all} the events. We can then substitute
this value into $\sqrt{C_{ E_2 }}+\sqrt{C_ { E_2 } +\tilde{m}^2}=M_{T2}^{\max}$ to find the parameter $C_ { E_2 } $ (see the details in section~\ref{sec:onevis}).
\item We apply the cut $R_{P_t} > 5$ and  find  $M_{T2}^{\max}$ with different trial masses ($\tilde{m}$) for the  events that passed the cut. If we observe that the $M_{T2}^{\max}$ is reduced {\em after} the $R_{P_t}$ cut
(as compared to before), then we can conclude that the underlying physics model is $Z_3$ type, otherwise it is $Z_2$ type.
On the other hand, if the $M_{T2}^{\text{max}}$ is not changed after the $R_{P_t}$ cut, then we conclude that the underlying physics model is $Z_2$ type.

\item If we confirmed (as above) that the physics model is $Z_3$,
we can then substitute the $M_{T2}^{\max}$ (for various trial DM masses) for events after the $R_{P_t}$ cut into $\sqrt{C_{ E_3 }}+\sqrt{C_{ E_3 }+\tilde{m}^2}=M_{T2}^{\max}$ to find the parameter $C_{ E_3 }$. Based on the values of $C_{ E_2 }$ and $C_{ E_3 }$, we can find the mother and DM masses simultaneously  (again see the details in section~\ref{sec:onevis}).
\end{itemize}

For the case with two visible particles per decay chain:
\begin{itemize}
\item We first find $M_{T2}^{\max}$ with different trial masses ($\tilde{m}$)
for all events. We then draw a $M_{T2}^{\max}$ versus $\tilde{m}$ plot and find the location of the kink. This can give us both the mother and DM masses.

\item We calculate the theoretical predictions of $M_{T2}^\text{max}$ for $E_3$ type events using the mother and DM masses found in the first step.

\item We apply the cut $R_{H_t} > 3$ and find $M_{T2}^{\text{max}}$ with different trial masses ($\tilde{m}$) for events that passed the $R_{H_t}$ cut.
If the edge in $M_{ T2 }$ reduces as a result of the cut,
then we conclude that the underlying physics model is $Z_3$. Otherwise, it is a $Z_2$ model.
Furthermore,
if the new $M_{T2}^{\text{max}}$ agrees with the theoretical prediction for $E_3$ type events found in the second step, then we have additional evidence
that it is a $Z_3$ model.

\end{itemize}

In the above analysis, we have ignored the $E_4$ type events. However, including these events would not affect our analysis. Specifically, the $M_{T2}^\text{max}$ for $E_4$ type events are always smaller than that of $E_2$ and $E_3$ type events (see Eqs.~(\ref{eq:mt2maxhier}) and \ref{eq:hierz32vis}) so that they would not affect the upper edges of $M_{T2}$ distribution for events {\em both} 
before and after the $R_{P_t}/R_{H_t}$ cut. However, the survival rates for events after the cuts might be modified.
In any case, we did not use the survival rates alone to distinguish between $Z_2$ and $Z_3$ models.

The above method of separating $E_2$ and $E_3$ type events using the $R_{P_t}$ or $R_{H_t}$ cut has its limitations. If the DM mass is very light compared to the mother mass, then the emitted extra DM might not carry away as much energy.
Thus,
in $E_3$-type events, the visible particles in the decay chain with two DM particles can 
%
%
be closer (relative to the heavy DM case) in 
energy 
%
%
to those in  
the other decay chain. In fact,
the DM becomes similar to a neutrino in this case so that 
the $R_{P_t}$ or $R_{H_t}$ distributions for $E_3$ type events should 
be similar to those for $E_2 + \nu$-type events, in turn,
not much different from that of $E_2$ type events (cf.
heavy DM case), and the distinguishing power of the $R_{P_t}$ or $R_{H_t}$ cut is reduced. Therefore, in order for the $R_{P_t}$ or $R_{H_t}$ cut to efficiently separate $E_2$ and $E_3$ type events
(and hence to distinguish between $Z_3$
and $Z_2$ models), we need the mass ratio between DM ($m_{DM}$)  mother ($M$) masses $\frac{m_{DM}}{M}$ to be
sizeable\footnote{Of course, we need
$m_{ DM } / M < 0.5$ in order for the decay chain with two DM
to be kinematically allowed.}.

Finally, we note that the cut on the ratio of
momentum/energy on the two sides of the full event
can also be used -- either by itself
or in conjunction with edges in $M_{ T2 }$
-- for the non-identical visible particles case
(discussed in the previous section)
in order to distinguish 
$E_2$ and $E_3$-type events.
%
%
Of course, in that case, just the {\em identity} of the visible particles
was enough to separate $E_3$ from $E_2$-type events.

\subsection{Signal Fakes by 
an (Effective) 2nd DM Particle
%
%
}

Next, we discuss the strategy to distinguish $Z_2$ models with two different DM
particles) from $Z_3$ models, similar to the
discussion in Sec.~\ref{sec:sigfake}. 
However, now we consider the two 
decay chains with one and two DM, respectively, in $Z_3$ models or with the
two different DM particles in $Z_2$ models having identical visible particles
(unlike in Sec.~\ref{sec:sigfake}).
In this case, there is a  
modification from the case discussed earlier: we obtain one $M_{T2}$ distribution 
by simply 
adding $M_{T 2 }$ distributions
for $E'_{ 2, \; 3, \; 4 }$-type events in $Z_2$ models
(and similarly, $E_{2, \; 3, \; 4 }$-type events in $Z_3$ models).
Let us consider first the case 
with a single visible particle per decay chain.
As 
%
%
mentioned before, $E_2'$, $E_3'$, and 
$E_4'$-type events, 
i.e.,
``sub''-distributions in the case of 
$Z_2$ models, 
all give a sharp upper edge in the $M_{T2}$ distribution. This observation leads to 
the expectation of two sharp ``kinks'' -- at the
location of the ``would-be'' (smaller) edges of $E_3'$, and $E_4'$-type events -- in the middle of the 
{\em combined}
$M_{T2}$
distribution\footnote{not to be confused with kink in edge of $M_{ T 2}$
as a function of trial mass!},
in addition to the overall upper edge resulting from $E_2'$ type 
events.\footnote{The 
situation is similar to the double edge signal in the case of off-shell intermediate particles 
studied in the reference~\cite{Agashe:2010gt}}.

Note that we had a similar discussion for $Z_3$ models in the beginning of section \ref{identical}.
However, in the case of $Z_3$ models only $E_2$ type events 
give a sharp upper edge and the other two type events, i.e., 
$E_3$ and $E_4$ give 
relatively longer tails (albeit with smaller
endpoints than $E_2$-type events) so that two kinks 
in the combined $M_{ T 2 }$ distribution
from $E_3$ and $E_4$ type events are not clear. 
Therefore, clear sharp kink(s) in the $M_{T2}$ distribution would suggest that the events 
are the result of a $Z_2$ model (as discussed earlier).
%
%

For the case with more than one visible particle in each decay chain, 
the above
idea of using kinks in $M_{ T2 }$ distribution might fail since the edge of 
the sub-distributions is not sharp, even in the case of $Z_2$ models
(as discussed in section \ref{sec:sigfake}).
Instead, 
we can do cross-checks like in the case of non-identical visible 
particle(s) in the two decay chains
(discussed in section \ref{sec:sigfake})
i.e.,
we first measure the masses of 
mother and DM particles by examining the location of the kink present in the maximum $M_{T2}$ as a 
function of the trial DM mass for $E_2$ or $E_2'$-type events, and then 
predict the location of edges in the other types of events.
Of course, 
in order to follow this strategy in the present case, one must first separate the 
events which are mixed, i.e., combination of $E_{2, \; 3, \; 4 }$ in the case of $Z_3$ models 
and $E'_{ 2, \; 3, \; 4 }$ for $Z_2$ models, 
%
%
into each individual type by applying a 
$P_t/H_t$ ratio cut (as explained in detail in Sec.~\ref{sec:pthtcut} for $Z_3$ models). 
Note that even 
mixed events in $Z_2$ models with a second DM-like particle can be separated by a $P_t/H_t$ ratio cut 
because the $E'_3$-type
events also have an imbalance in the energy/momentum of the visible 
on the two decay chains due to the 
difference between $m_{DM}$ and $m_{DM}'$.\footnote{Obviously, one
cannot then use this cut {\em on its own} in order to distinguish $Z_3$ from $Z_2$ models.}
Of course, we can also do a similar separation for the case of {\em one} visible particle
in each decay chain (which was just discussed above) and then 
repeat the strategy which we discussed in section \ref{sec:sigfake} for
the case of visible particle in the two decay chains being non-identical, i.e., consider the 
shape of the separated $M_{ T2 }$ distributions in order to distinguish $Z_3$ from $Z_2$ models.

%
%


\section{Future Considerations}
\label{sec:onshell}

\subsection{On-Shell Decay Processes}

In the previous sections we have focused only on decay chains where the intermediate particles are off-shell.  We have used $H_T$ cuts to reduce the inherent $Z_2$-like background (events with two DM candidates in the final state) to $E_3$-type events (events with three DM candidates in the final state) for the case when each of the pair-produced mothers decay to the \textit{same} SM final state.  As a reminder $E_3$ events do not occur in models with a $Z_2$ stabilization symmetry.  An important question is whether this method can be applied to decays where the intermediate particles are on-shell.  
\newline
\newline
For a quick example on how the decay of the mother particles into intermediate particles that are on-shell can complicate the analysis of the previous section, consider the pair production of mother particles which are charged under a $Z_2$ stabilization symmetry.  We will assume each mother decays via on-shell intermediate particles into a dark matter candidate and identical SM particles.  Now suppose the first mother decays into virtual daughters whose mass differences with each other, the mother and dark matter are large.  The second mother decays into intermediate daughters whose mass differences are all small except for one.  It is clear, in this latter case, there will be events where the transverse momentum of the visible particles can be small.  It is also conceivable, for those events, that phase space available to visible particles in the first mother decay is relatively large.  In these events, on average, the  $H_T$ ratio can be large and therefore can fake the $Z_3$ signal.  
\newline
\newline
To make this more clear, let us go a step further to consider the following explicit example.  The production of two mother particles can be
\begin{equation}
p\, p \to Q\, \overline{Q},
\label{eq:onshellprod}
\end{equation}  
where $Q$ is a new, heavy particle.  $Q$ and $\overline{Q}$ decay to a final state $Q \to q\,\bar{l}\,l\,\chi$ via the following on-shell intermediate particles
\begin{figure}[t]
	\centering
	 \includegraphics[width=7.0truecm,height=5truecm,clip=true]{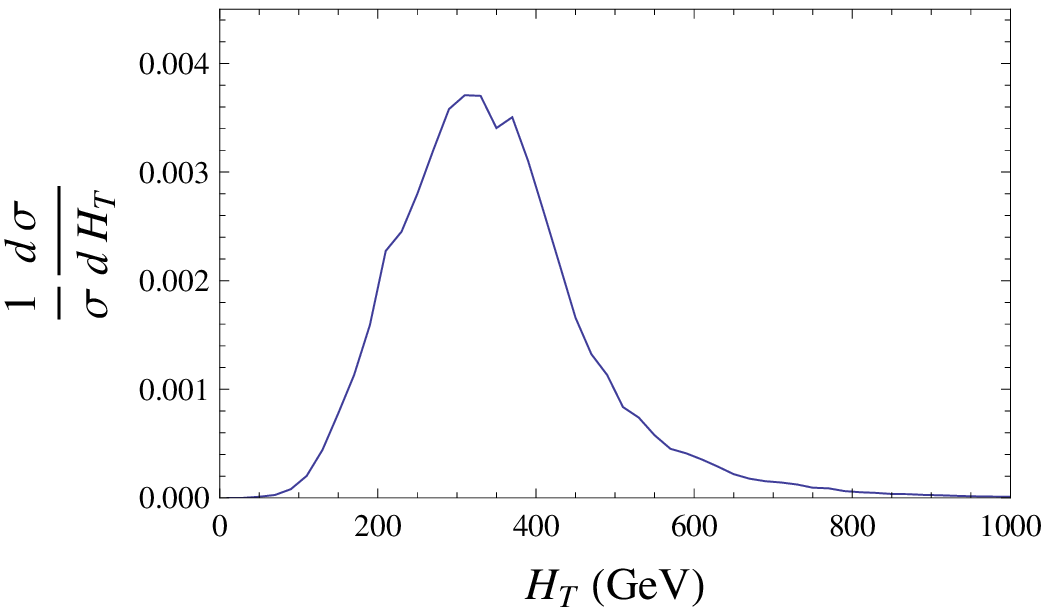}
	\hspace{0.2cm}
	 \includegraphics[width=7.0truecm,height=5truecm,clip=true]{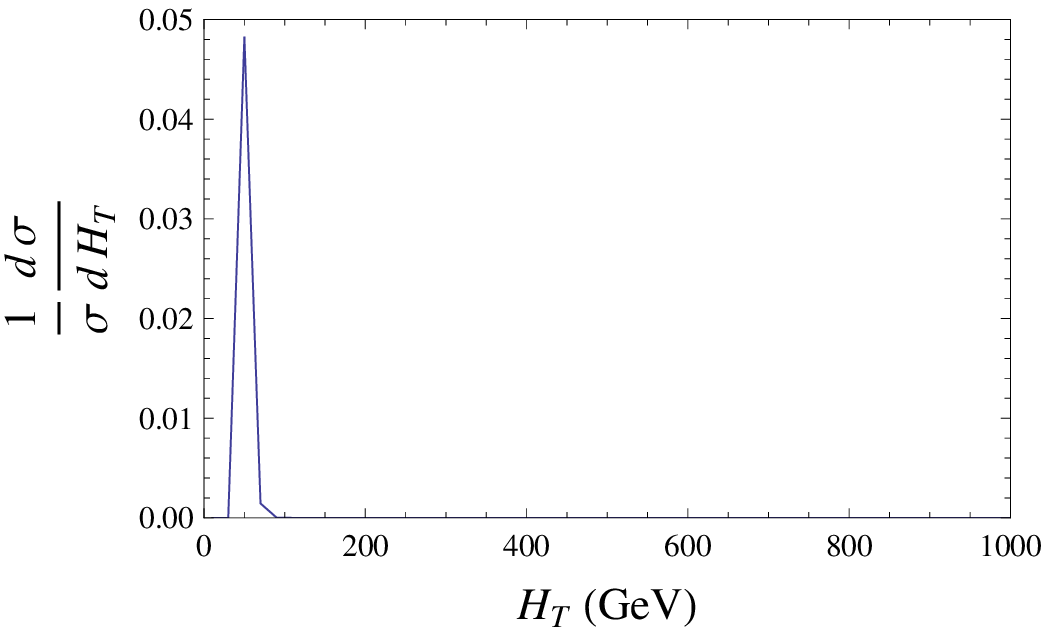}
	\caption{The production and decay of $Q$ in Eqs.~(\ref{eq:onshellprod}) and~(\ref{eq:onshelldecay}).  In the panel on the left the intermediate particles generating the decay have a mass difference of 200 GeV.   On the panel on the right, all of the mass differences are set to 5 GeV with the exception of one which is 590 GeV.  (See Eqs~(\ref{eq:deltam}) through~(\ref{eq:deltatotal}).)  The dark matter mass is the same for both plots and is charged under a $Z_2$ symmetry.}
	\label{fig:onshelll}
\end{figure}
\begin{eqnarray}
Q \to \,A\, \,q &\hspace{0.9cm}&   \overline{Q} \to \,A' \,\,\bar{q}\nonumber \\
A  \to \,B \,\,\bar{l} && A'  \to \,B'\, \,l \label{eq:onshelldecay} \\
B  \to \,l\,\,\chi && B'  \to \,\bar{l}\,\,\chi.   \nonumber
\label{eq:onshelldecay1}
\end{eqnarray} 
Here $\chi$ is the dark matter particle, $A$, $A'$, $B$ and $B'$ are intermediate particles.  The lowercase letters are the other SM particles.  The important mass differences are
\begin{eqnarray}
\delta\,m_{Q A} = m_Q - m_A  &\hspace{0.9cm}& \delta\,m_{\bar{Q} A'} = m_{\bar{Q}} - m_{A'} \\
\delta\,m_{A B} = m_A - m_B & & \delta\,m_{A' B'} = m_{A'} - m_{B'}\\
\delta\,m_{B \chi} = m_B - m_\chi & & \delta\,m_{B' \chi} = m_{B'} - m_\chi. 
\end{eqnarray} 
If the mass differences, $\delta\,m_{Q A}$, $\delta\,m_{A B}$, $\delta\,m_{B \chi}$, are relatively large and the differences, $\delta\,m_{\bar{Q} A'}$, $\delta\,m_{A' B'}$, $\delta\,m_{B' \chi}$, are small then the $H_T$ ratio may be askew.  We show this explicitly in Figure 13 for 
\begin{eqnarray}
&& \delta\,m_{Q A} = \delta\,m_{A B} = \delta\,m_{B \chi} = 200\,\,\mathrm{GeV} \label{eq:deltam} \\
&&\delta\,m_{\bar{Q} A'} = \delta\,m_{A' B'} = 5\,\,\mathrm{GeV} \\
&&\delta\,m_{B' \chi} = 590\,\,\mathrm{GeV}.\label{eq:deltatotal}
\end{eqnarray}
where $m_Q  = 700$ GeV and $m_\chi = 100$ GeV.  $H_T$ is plotted above for the small mass and large mass separation cases.  
This argument is general and can be applied to many on-shell decay scenarios.  For this paper, we considered only off-shell decay scenarios.  We leave these questions to future work.  

\subsection{Invisible $Z$ Decay Background Considerations}

Beyond the on-shell considerations described above, in future work we will also consider acceptance cuts and detector effects consistent with the ATLAS and CMS experiments.  An important irreducible background to our signal with DM in the final state is the emission of $Z$ bosons that decay invisibly.  The standard way to mitigate the effect of this background is to look for events which radiate off a $Z$ boson decaying to electrons or muons.  These $Z$ bosons can be reconstructed with standard invariant mass techniques. Knowing the $Z$ boson branching fractions, one can then reliably estimate the number of invisible $Z$ boson decay events\footnote{This technique is common and used often with analysis of estimating the sensitivity ATLAS and CMS to invisible higgs decay processes.  See for example~\cite{ATDR, CTDR}.  Also, as an additional note, it is conceivable that the techniques introduced in the previous sections can be applied to separate the signal (DM events) from background processes that include invisible $Z$ decays.  However, since estimating the invisible $Z$ branching fraction from data as indicated above is straightforward and reliable, we focus on that method.}.  To be conservative, we will require the signal cross section, after detector and acceptance cuts, to be more than 10\% of this SM background to preclude statistical fluctuations~\cite{conversation}.  

\subsection{Resolving Reconstruction Ambiguities in Multi-Jet Events with Large $\etmiss$}\label{sec:algorithm}

Thus far, we have not applied our analysis to many-particle final states with large $\etmiss$.  Clearly, we want to be able to distinguish $E_3$ versus $E_2$ events for those scenarios.  The problem is reconstruction ambiguities are inherent in, e.g., multi-jet events.  Consider the pair production of mother particles which decay on-shell into a six-jet final state with two SUSY-like dark matter candidates.  In order to reconstruct the decay chain, experimentalists must consider $6$! =  720 combinations of jets!  Such ambiguities can prevent accurate reconstruction of events with dark matter candidates, not to mention distinguishing scenarios with two versus three dark matter particles in the final state.  Thus, the question of searching for dark matter in events with many visible final states is \textit{\textbf{really}} the question of finding methods to enable accurate reconstruction.  
\newline
\newline
Consider the difficult case of decay chains with multi-jet final states and large amounts of missing energy.  Several groups have proposed methods to increase the efficiency of reconstructing such events~\cite{Rajaraman:2010hy,Bai:2010hd}.  These methods require the final state jets to have large $p_T$ cuts.  Because of the large $p_T$ cuts, the regime of phase space where the mother particles are highly boosted is selected.  These methods are therefore essentially variants of the aptly named ``hemisphere method."  By definition, the hemisphere method considers only events where the mother particles are highly boosted.   Thus, final state jets resulting from each mother's decay are correlated in different sides of the detector.  The correlation allows experimentalists to accurately associate the correct final state jets with the right decay chain.  There are, however, inherent problems with this method:
\begin{enumerate}
\item  New particles produced at the LHC are expected to be heavy and therefore produced dominantly at threshold.  Requiring boosted mother particles pushes production into the regime of phase space where the number of signal events are likely to be smaller than that at threshold.

\item Reconstructing multi-jet events requires hard jets with large cone separations ($\Delta R > 0.4$ or $0.5$) between the jets.  When the jets are correlated in different hemispheres of the detector, this requirement will further reduce the number of events which will pass to become signal.  

\end{enumerate}
With the signal events expected to be small (because of the heavy mother production), it is crucial to keep as many of the signal events as possible.  Thus, the \textbf{\textit{true}} goal would then be to accurately reconstruct these events with only the acceptance and kinematical cuts needed to identify the signal and reduce the SM background.  We now briefly outline~\cite{devin} a method to reduce the combinatorics needed to reconstruct events with multi-jet final states with large $\etmiss$.  We also present very preliminary results.  The full results will appear in~\cite{devin}.  To simplify the discussion in the next sections, we assume pair production of mother particles with each decaying into identical dark matter candidates plus jets\footnote{We relax this assumption in~\cite{devin}}.

\subsubsection{Proposed Algorithm}

The reconstruction algorithm is:
\begin{table}[h]
\begin{tabular}{cl}
\hspace{0.3in} Step 0:&Obtain a statistically significant amount of $n$-jet events (with a large amount of\vspace{0.1cm} \\
                                        &$\etmiss$) over the SM background.  \\ \\
\hspace{0.3in} Step 1:&From these events, first consider only the events that remain in the limit of the \vspace{0.1cm} \\
				  &hemisphere method, i.e., the final state jets have a large $p_T$.  The number of \\ 
				  &events retained in this limit must still be statistically significant\footnote{We are evaluating how significant 
				  the background must be and still reconstruct the events properly.} over the re-  \vspace{0.1cm} \\ 
				  &maining background.\\ \\
\hspace{0.3in} Step 2:&In the hemisphere method limit, the topology of the signal events can be ob-\vspace{0.1cm} \\
                                        &tained.\\

\end{tabular}
\end{table}
\newline
To make Step 2 clearer, consider the example of a six-jet final state.  A visual representation of the topologies is in Figure 15.  Again, we assume pair production of mother particles with each decaying into identical dark matter candidates plus the SM; also, in the hemisphere limit, the jets are required to be well separated.   If, in the limit of the hemisphere method, one mother preferentially decays into five jets and the other mother decays to only one, then the ``5-1" topology can be identified.  As described in the previous section, the identification is due to the mother particles being boosted; the resulting decay products are therefore on different hemispheres of the detector and can be readily associated with the right decay chain.  Another example is possible with one mother decaying into two jets and the other decaying into four jets.  This produces a ``4-2" topology.  Thus, Step 2 is crucial to identify which topology (or combination of topologies\footnote{It is possible that multiple signal production processes can generate, e.g., 5-1 and 4-2 topologies.  If both topologies are determined, then reconstructing the full data set requires the assumption of both.  More details are included in~\cite{devin}.  Also, if e.g., a 4-2 topology is produced, this means events with four and eight jet final states are also produced.  By focusing on six-jet final states, we do not consider these events.}) is generated.  
 \begin{table}[h]
\begin{tabular}{cl}
\hspace{0.3in} Step 3:&In the hemisphere limit, reconstruct the $n$-jet invariant masses for the \vspace{0.1cm} \\
                                        &topology(-ies) determined in Step 2.  This is to find kinematic edges.
\end{tabular}
\end{table}
\newline
It is well known that heavy mother decays into the SM and dark matter generate invariant mass ``edges."  Consider again the case of the 4-2 topology.  Reconstructing the four-jet invariant mass associated with the mother decaying to four jets plus dark matter generates an edge.  This edge is roughly at $M_\mathrm{mother} - M_\mathrm{dark\,\,matter}$.  
The measured values of these edges will help to increase the efficiency of reconstructing when going away from the limit of the hemisphere method.  We discuss this more detail in the next section.  Also, please note, if the decays of a mother are on-shell, even more edges are possible than the ones described above.  This will further increase the efficiency of reconstruction.  We show an example of a kinematic edge in the next section.
\newline
\newline
So far the above steps have been done in the limit of the hemisphere method.  The next step is new and is our primary point:
\begin{figure}[t]
	\centering
	 \includegraphics[width=5.0truecm,height=3.2truecm,clip=true]{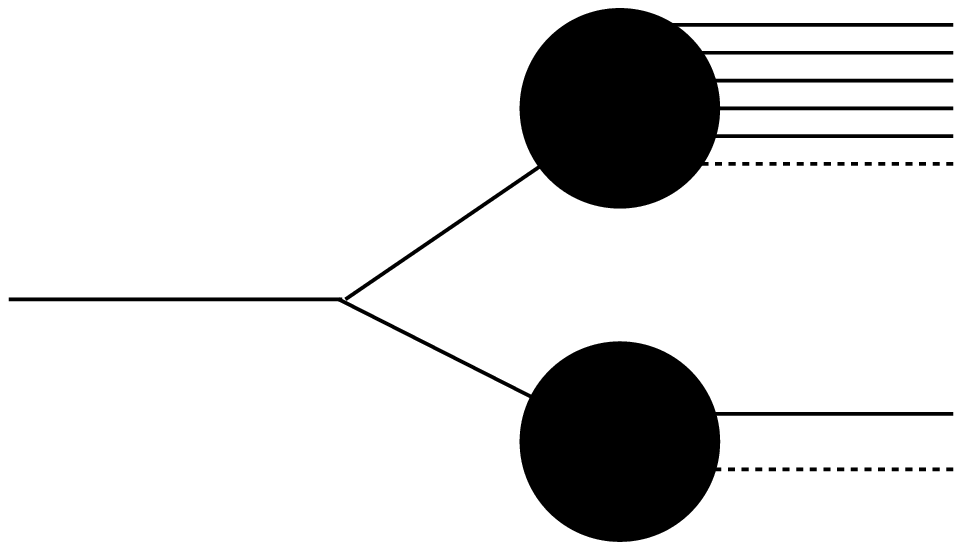}
	\hspace{0.2cm}
	 \includegraphics[width=5.0truecm,height=3.2truecm,clip=true]{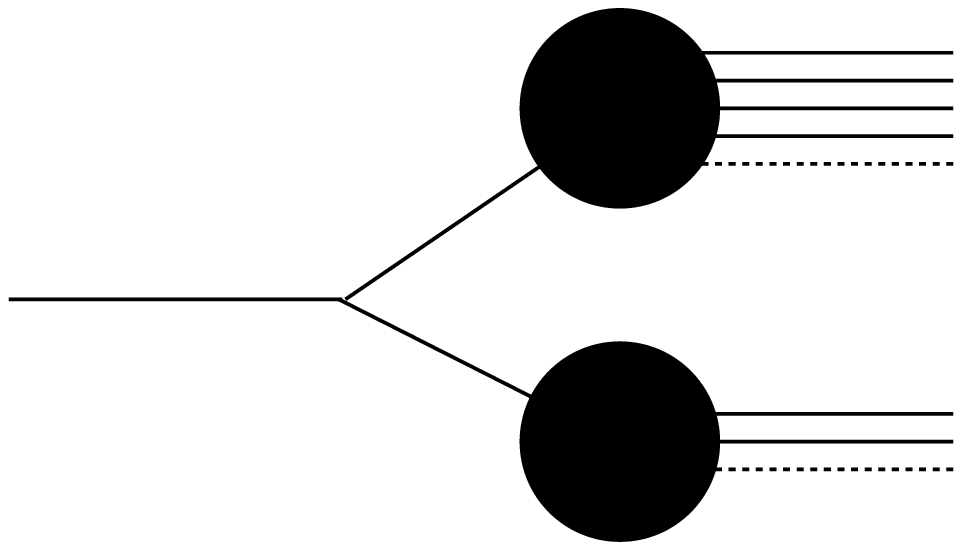}
	 	\hspace{0.2cm}
	 \includegraphics[width=5.0truecm,height=3.2truecm,clip=true]{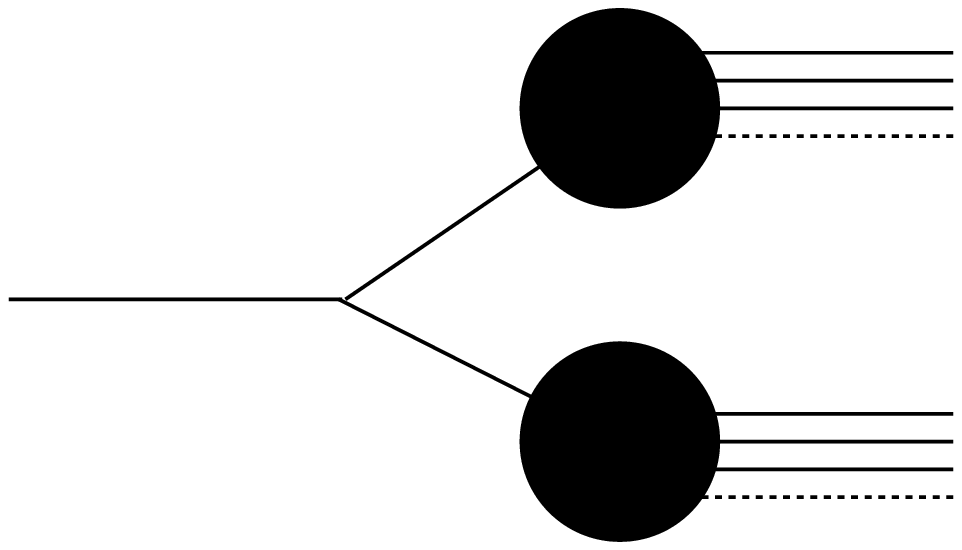}
	\caption{All possible six-jet final state topologies with the assumption of pair production of mother particles which decay to the SM and identical DM candidates.  The 5-1(left), 4-2 (center) and 3-3 (right) topologies are shown. }
	\label{fig:6opo}
\end{figure}
\begin{table}[h]
\begin{tabular}{cl}
\hspace{0.3in} Step 4:&Reconstruct all of the events with the topology(-ies) determined in the limit of   \vspace{0.1cm} \\
                                        &the hemisphere method.  Knowing these topologies serves as a reconstruction \vspace{0.1cm} \\
                                        &template to minimize combinatorics for the events which do not have boosted  \vspace{0.1cm} \\ 
                                        &mother particles.
\end{tabular}
\end{table}
\newline
Thus, our point is the use of the hemisphere method to learn the topologies of the signal event.  This in turn, will \textbf{\textit{reduce the combinatorics}} needed to reconstruct the full set of signal events when only the cuts needed to reduce the SM background are applied.  Consider again the example of a  six-jet final state.  If, in the limit of the hemisphere method, one mother is found to decay into two jets and the other four jets, the number of ways to associate the jets with the correct decay chain is 15.  In the case of the 3-3 and  5-1 topologies, it is 20 and 6, respectively.  The invariant mass edges determined in the hemisphere method limit will further help to reduce the possible ways to associate the various jets with each decay chain even further.  We describe some specifics briefly in the next section.

\subsubsection{Additional Notes on Implementing the Algorithm}

To begin, for each event, the jets are labeled by decreasing $p_T$.   Like in the previous sections, the scalar sum of the $p_T$ of several jets, $H_T$, is defined as
\begin{equation}
H_t = \sum_{a =1}^j |P_t^{\,a}|
\end{equation}
for a number of jets, $j$.  We use $H_t$ to go to the limit of hemisphere method.  To help associate the correct jets with the right decay chain, we also construct the transverse sphericity ($S_T$) for $n$-jets, where $n$ can be less than the total number of final state jets.  The number of jets included in the definition depends on the assumed topology for reconstruction.  $S_T$ is defined as  
\begin{equation}
S_T = \frac{2 \lambda_2 }{ \lambda_1 + \lambda_2}
\end{equation}
where $\lambda_{1,2}$ are the eigenvalues of the $2 \times 2$ sphericity tensor 
\begin{equation}
S_{ij} = \sum_\kappa p_{\kappa i}\, p^{\kappa j}
\end{equation}
where $\kappa$ runs over the number of jets included.  In general, when the jets used to compute $S_T$ are collinear or back-to-back, then $S_T \to 0$.   Because of this property, $S_T$ is used to help determine the topologies of the signal events in the limit of the hemisphere method\footnote{Beyond, $H_T$ and $S_T$ other, more common, kinematic variables also will be used identify the topology(-es) in the hemisphere limit.  We expand on this in detail in~\cite{devin}.}. 
\newline
\newline
In the previous section, we described how reconstructing invariant masses in the hemisphere limit can help reduce combinatoric ambiguities.  
\begin{figure}[t]
	\centering
	 \includegraphics[width=7.2truecm,height=6.0truecm,clip=true]{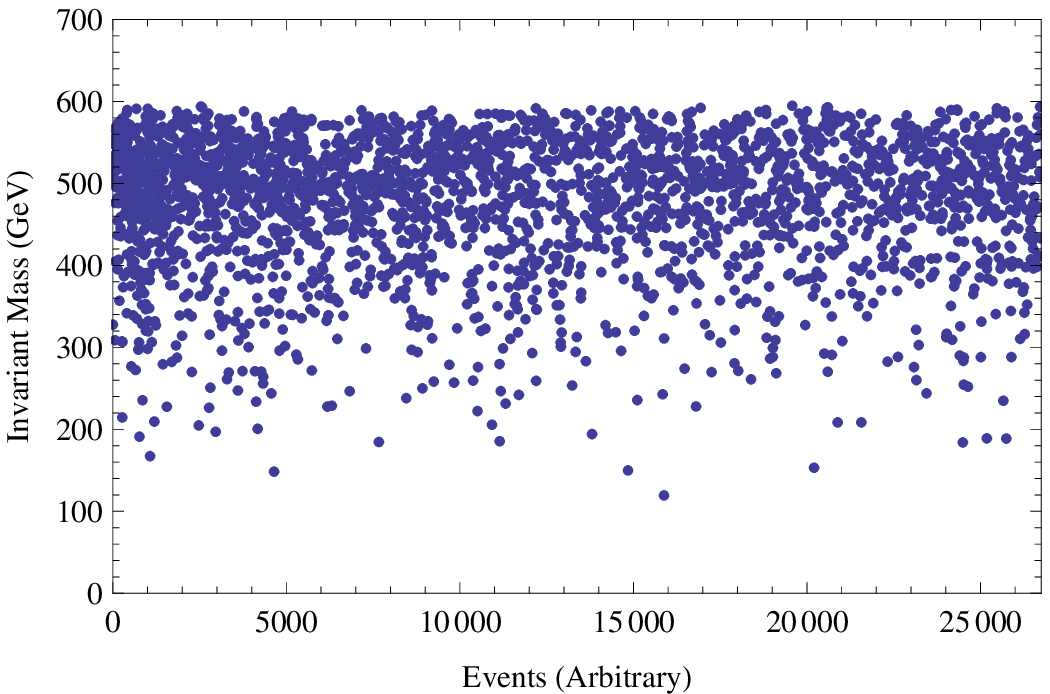}
	\hspace{0.6cm}
	 \includegraphics[width=7.2truecm,height=6.0truecm,clip=true]{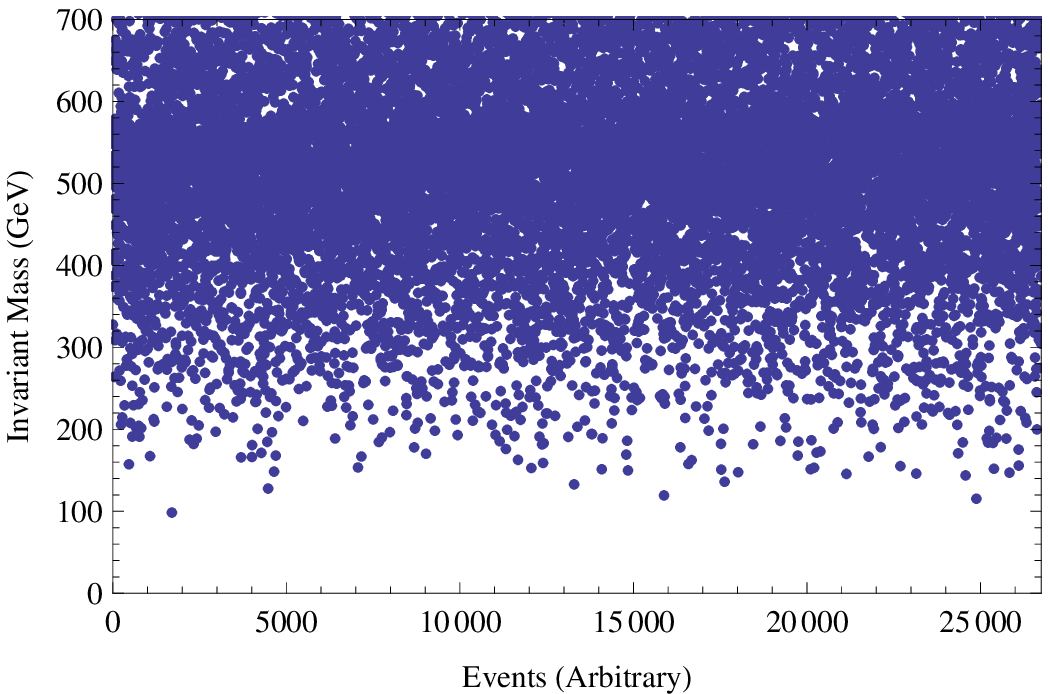}
	\caption{Signal process described in Eq.~(\ref{eq:QhQhsignal}) for a 5-1 topology.  The left panel is the five-jet invariant mass in the limit of the hemisphere method.  Here the hemisphere method is defined as the cuts in Eq.~(\ref{eq:cuts}) with the exception that all of jets have $p_T > 150$ GeV.  The right panel considers the same events with the cuts in Eq.~(\ref{eq:cuts}).  Note:  The reconstruction ambiguities obscure the kinematic edge.  All of the events above the 600 GeV edge are events with poor reconstruction.}
\end{figure}
We want to give a basic example of how this is possible.  Consider again the example of a six-jet final state.  Suppose the signal topology is a 5-1 topology.  In this example, the signal event is generated by pair production of 700 GeV heavy quarks, $Q_H$, one of which decays into five jets and 100 GeV scalar dark matter, $\tilde{\chi}$.  The other mother decays to only one jet and the same dark matter candidate,
\begin{equation}
p\,p \to \overline{Q}_H\,Q_H \to \bar{q}\,q\,\,\bar{q}\,q\,\,\bar{q}\,q\,\,\tilde{\chi}\,\tilde{\chi}^*.
\label{eq:QhQhsignal}
\end{equation}
The virtual particles which contribute to the five-jet decay chain have the masses of 580, 460, 340 and 220 GeV.  The five-jet invariant mass has a kinematic edge at approximately $M_{Q_H} - M_{\tilde{\chi}} = 600$ GeV.  In the left panel of Figure 16, we plot the five-jet invariant mass in the limit of the hemisphere method.  Each dot represents an event.  It is clear an edge has appeared at 600 GeV.  In the right panel, we plot the same five-jet invariant mass with only the cuts needed to reduce the SM backgrounds:
\begin{align}
 p_{T\,\,\mathrm{leading\,\,jet}} > 100\,\mathrm{GeV} && p_{T\,\,\mathrm{non-leading\,\,jets}} > 50\,\mathrm{GeV} \nonumber \\
 |\eta_\mathrm{jets}| < 2.5 && \etmiss > 100\,\mathrm{GeV} \label{eq:cuts} \\
S_{T\,\,\mathrm{all\,\,jets}} > 0.2 && \etmiss > 0.2 \,M_\mathrm{eff}  \nonumber \\
\Delta R_{jj} > 0.4 \nonumber
\end{align}
where effective mass is defined as 
\begin{equation}
M_\mathrm{eff} = \sum p_{T\,\,\mathrm{jets}} + \etmiss.
\end{equation}
This cuts are consistent with the ATLAS and CMS collaborations.  With only these cuts, we have reconstructed the multi-jet events assuming the 5-1 topology determined in the limit of the hemisphere method.  The key point is the edge determined in the hemisphere method limit means all of the invariant masses greater than 600 GeV are mis-reconstructions.  As described above, the amount of mis-reconstructions can be further eliminated if the decay is on-shell and all possible kinematic edges are applied.  At the partonic level, the reconstruction efficiency for on-shell decays for the above example approaches 80\%~\cite{devin}.
\newline
\newline
The next steps to optimize this method is to include QCD backgrounds as well as a variety of signal processes with virtual particles different spins.  A major fault with Figure 16 is that it was generated at the partonic level.  The final study~\cite{devin} should put the signal processes through pythia to get an estimate of hadronization effects.

\section{Conclusions and Outlook}

The LHC experimental program is ramping up. One of its goals is to
test the idea of a WIMP as a DM candidate, especially one which
arises as part of an extension of the SM at the TeV scale. The basic idea
is that there is a new (global) symmetry under which the SM particles are neutral, but some of the new particles are charged. Thus, the lightest of these charged
particles is stable and under certain circumstances -- it certainly has
to be electrically and color neutral -- can be a DM
candidate. 

In general, such an extension of the SM also has
other (heavier, but still TeV mass) particles which are charged under both the new symmetry
{\em and} under the SM gauge symmetries (especially color). So, these
DM ``partners'' (in the sense that they are charged under the symmetry
which stabilizes DM) can be copiously produced at the LHC. In turn, these
DM partners (or ``mother'' particles)
will decay into SM particles and DM, resulting in events with
missing energy and jets/leptons/photons. The goal then would be to piece together
the details of the DM model, including mass of the DM and the mother
particles, by reconstructing the decay chains which lead to such events.

Clearly, the nature of the DM stabilization symmetry plays a crucial
role in the above program, in part because it determines the
number of DM particles in each decay chain of the mother particle:
for example, in the case where a $Z_2$/parity symmetry stabilizes the
DM (typically) only one DM appears in decay of a single mother, whereas
in a $Z_3$ model, both two and one DM are allowed.
Of course, each event must necessarily involve two such mothers being produced.

With the above background, in this paper, we showed how to distinguish
models with $Z_3$ DM stabilization symmetry from $Z_2$. Since
earlier work~\cite{Agashe:2010gt} only studied measurements of the
{\em visible} part of a {\em single} mother decay, here we focussed on
using the information from {\em both} mother decays, including the {\em
missing}
energy in the event which is shared between the two decay chains (or sides).
We found that the variable $M_{ T \; 2 }$ is useful for this purpose.
For simplicity, we 
%
%
studied pair production of the same mother, followed
by decays to SM particles and DM which involve only {\em off}-shell
intermediate particles (i.e., which are heavier than their mother particle). Furthermore, we assumed that we know the division
the SM particles in the full event into two groups corresponding to each
mother [but of course we do not know (a priori) how many DM are in each
decay chain].

Clearly, in a $Z_3$ model, the events can be classified into three types depending on the
total number of DM particles (i.e., two, three or four) vs.~only two DM
particles for $Z_2$ model.
We showed that the edges of the $M_{ T \; 2 }$ distributions
are different in these three types of events in a $Z_3$ model,
again even if the same mother
is produced (vs.~only one edge for $Z_2$ model).
This feature allowed us to distinguish $Z_3$ from
$Z_2$ models. Moreover, we gave predictions for the values of
the edges in the two {\em new} cases, namely, three and four
DM in each event, as functions of mother and DM mass. Thus, we can extract
the mother and DM masses {\em separately} using the measurements of
these different edges for a $Z_3$ model.
This achievement is especially noteworthy
for the case of {\em single} visible particle in each decay chain since
a similar measurement of the mother and DM masses
is not possible in a $Z_2$ model, based solely on using
$M_{ T 2 }$ variable.\footnote{The exception 
to this rule for 
$Z_2$ model is when the total transverse momentum of the two mother particles is \textit{non}-zero (for example, due to initial/final state radiation).}

We emphasized that there are two subcases in the above analysis, namely,
the visible particles in the decay chain with one DM being 
identical or different (respectively) to those in the decay chain with 
two DM (for $Z_3$ models). In the case of the visible particles
{\em not} being identical, it is easy to separate the
events of the three types
mentioned above so that one can then
plot the respective $M_{ T \; 2 }$ distributions. However, in the case of the visible particles being
identical, one obviously has only a {\em single} $M_{ T \; 2 }$ distribution
(i.e., combination of the above three types)
to begin with. Therefore, we developed a new method to separate out
the candidate events with three DM vs. two DM in this case, using the
observation that the visible particles on the side with two DM will have
smaller energy/momentum that the visible particles on the side with one DM
in the same event. This feature is to be compared to the visible energy/momentum being
more ``balanced'' in the case of one DM on each side.

We observed that the above imbalance in the energy/momentum on the two sides
{\em by itself} provides a hint for the appearance of three
DM in the event. However, combining it with edges in $M_{ T \; 2 }$
distributions provides a more powerful discriminator.
Finally, we briefly mentioned the case of intermediate particles
in the decay chain being on-shell. 
%
%

\textbf{Signal Fakes:} We pointed out that models with $Z_2$ as a dark matter stabilization 
symmetry can fake the \textit{three} decay topologies arising in $Z_3$ models.
%
%
An example of such a fake comes from the 
decays in $Z_2$ models which involve 
{\em effectively} a second DM particle (denoted by DM$^{ \prime }$), for example, 
a heavier, neutral $Z_2$-odd, but collider stable particle.
%
%
%
In such a $Z_2$ model, 
three decay topologies are possible if the pair-produced
mother can decay into either DM or
DM$^{ \prime }$, i.e., we can have DM in one side and DM$^{ \prime }$ on the other side, DM on 
both sides or DM$^{ \prime }$ on both sides. 
Nevertheless, 
%
%
we suggested how to discriminate $Z_2$ and $Z_3$ models by using various 
strategies. 
%
%

Another possibility for a fake of $Z_3$-like signals by a $Z_2$ model
comes from pair-production of {\em more than one} type of mother 
particles which decay into visible particles
and, 
for simplicity,
a single DM particle. 
Since the upper edge of $M_{T2}$ distributions depends 
on the mass of the mother particle, the introduction of different types of mother particles can 
easily generate different locations of upper edges so that it could instead be 
mistaken as the multiple decay topologies of a $Z_3$ model (again, with single DM particle). 
This type of degeneracy between $Z_2$ and $Z_3$ models, however, can be easily 
resolved by taking into account the fact that each event in such a $Z_2$ model
is still ``balanced'' in the energy/momentum of visible particles on the two sides 
(again, assuming that there is  
no second
DM-like particle) while events in $Z_3$ models can have imbalance
if they involve one DM on one side and two DM on the other.

Obviously, this and the earlier work ~\cite{Agashe:2010gt} 
are to be viewed as
the first steps in the broad program of distinguishing various
DM stabilization symmetries using collider data. To this end, we outlined some future steps to implement this program.  We leave more
detailed studies along these lines, including relaxing some of the assumptions
outlined above and further studying various experimental
issues, for future work.


\section*{Acknowledgements}

We would like to thank Manuel Toharia for collaboration at an early stage
of this project.
We also thank Hsin-Chia Cheng, Ben Gripaios, Tao Han, Ian Hinchliffe, 
Maxim Perelstein, Marjorie Shapiro, Jesse Thaler, Lian-Tao Wang and Felix Yu
for useful discussions.
K.A. was supported in part by NSF Grant No.~PHY-0652363.  D.W. was supported in part by a University of California Presidential as well as an LHCTI fellowship. L.Z. was supported
by a fellowship from the Maryland Center for Fundamental Physics.


\appendix

\section{The location of $M_{T2}^{\max}$}
\label{app:locofmaxmt2}
In this appendix we will derive the analytic expression for the location of the upper edge in the $M_{T2}$ distribution. We begin with deriving the general expression of the $M_{T2}$ solution for a given set of kinematic configuration, then move on to obtaining the maxima of the balanced/unbalanced $M_{T2}$ solutions, and close with giving the global maximum of the $M_{T2}$ distribution, followed by a simple application.
 
\subsection{The General Expression for the $M_{T2}$ Solution}
The usual $M_{T2}$ variable~\cite{Lester:1999tx, Barr:2002ex} is defined as a generalized transverse mass such that each of pair-produced mother particles decays into visible particles and one dark matter particle of the same type. However, we do not restrict ourselves to such cases, i.e., we extend our consideration to the cases with more than two DM in a full decay chain (e.g., $E_3$ and $E_4$ type events in $Z_3$ models). Nevertheless, in the analysis of $M_{T2}$ variable, we still 
{\em hypothesize} that two dark matter particles with equal mass 
(i.e., one DM per chain) are involved in the full decay process, i.e.,
we employ the ``naive'' $M_{T2}$ method (as mentioned at the beginning of
section \ref{sec:mt2z3}).  

%
%
%
%

\begin{figure}[t]
  \centering
  \includegraphics[height=2.61in,clip]{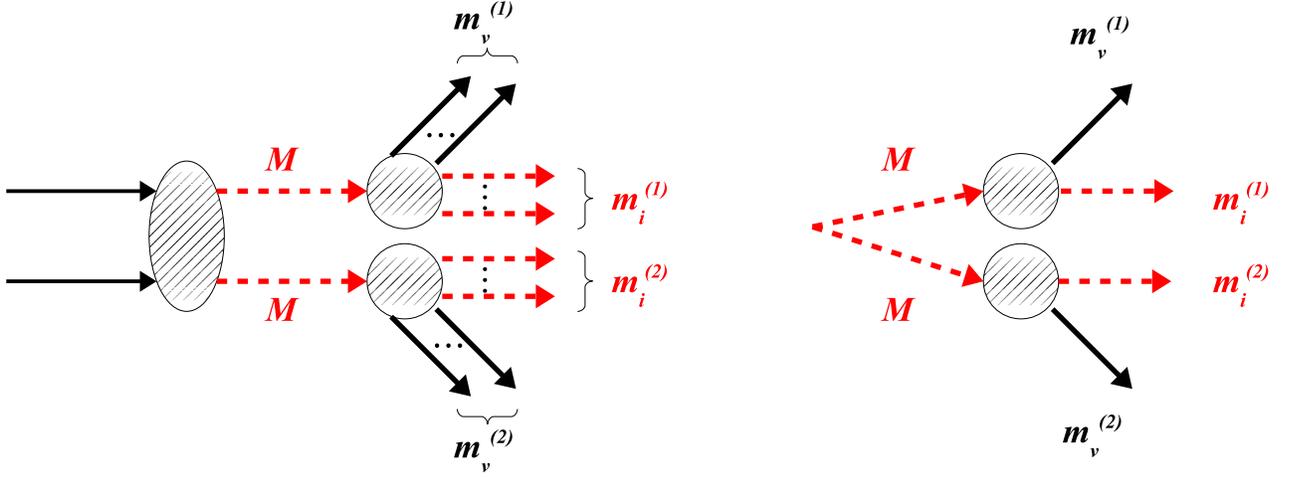}
  \caption{The left panel illustrates the decay process of interest which pair-produced mother particles go through. $M$, $m_{v}^{(a)}$, and $m_{i}^{(a)}$ ($a=1,\;2$) denote the mass of mother particle, total invariant masses of visible particles and invisible particles in the same decay chain, respectively. The right panel illustrates the effective configuration of such a decay process.}
  \label{fig:gendecproc}
\end{figure}

The left diagram of Fig.~\ref{fig:gendecproc} illustrates the decay process of pair-produced mother particles that we are taking into consideration. 
Here the ``blob'' denotes off-shell intermediate particles or an (on-shell) point interaction. The red dashed lines represent any particles charged under dark matter stabilization symmetry whereas the black solid lines represent any visible/SM particles. $M$ is the mass of the mother particle, which must be charged under DM stabilization symmetry. As mentioned above, each mother particle can decay into the multiple number of invisible/DM particles as well as the multiple number of visible/SM particles, and this extended possibility is explicitly depicted by the multiple number of red dashed and black solid arrows behind the two small blobs. Each visible/invisible multi-particle state can be collapsed effectively to a(n) visible/invisible single particle state by introducing invariant (transverse) mass, which will be manifest in the detailed formulas later. In this sense, $m_{v}^{(a)}$ and $m_{i}^{(a)}$ can be understood as the total invariant masses formed by visible or invisible particles belonging to the same decay chain.

For defining the $M_{T2}$ variable one should note that we are not aware of the DM mass in advance. Hence, the best we can do is to introduce trial DM mass. Since we perform the naive $M_{T2}$ analysis as mentioned above, i.e., we assume a single type of DM in each decay chain even if the actual physics could be different, we employ only one type of trial DM mass $\tilde{m}$ and construct the $M_{T2}$ variable as follows~\cite{Lester:1999tx, Barr:2002ex}:
\bea
M_{T2}\left(\textbf{p}_T^{v(1)},m_{T}^{v(1)},\textbf{p}_T^{v(2)},m_{T}^{v(2)};\tilde{m}
\right) \equiv \min_{\textbf{p}_T^{v(1)}+\textbf{p}_T^{v(2)}+\tilde{\textbf{p}}_T^{(1)}+\tilde{\textbf{p}}_T^{(2)}=0} \left[\max \left\{M_T^{(1)},M_T^{(2)}   \right\}    \right]
\eea
Here each transverse mass of the decay product $M_T^{(a)}$ $(a=1,2)$ is given by
\bea
\left(M_T^{(a)}\right)^2=\left(m_T^{v(a)}\right)^2+\tilde{m}^2+2\left(E_T^{v(a)}\tilde{E}_T^{(a)}-
\textbf{p}_T^{v(a)}\cdot \tilde{\textbf{p}}_T^{(a)}               \right) \label{eq:mt}
\eea
where $m_T^{v(a)}$, $\textbf{p}_T^{v(a)}$, and $E_T^{v(a)}$ are the total transverse invariant mass, transverse momentum, and transverse energy of visible particles:
\bea
\left(m_T^{v(a)}\right)^2=\left(t_1^{v(a)}+\cdots+t_n^{v(a)}   \right)^2&=&
\sum_{\alpha}\left(m_{\alpha}^{v(a)}\right)^2+2\sum_{\alpha>\beta}\left(E_{\alpha T }^{v(a)}E_{\beta T }^{v(a)}
-\textbf{p}_{\alpha T }^{v(a)}\cdot \textbf{p}_{\beta T }^{v(a)} \right) \\
\textbf{p}_T^{v(a)} &=& \sum_{\alpha}\textbf{p}_{\alpha T}^{v(a)} \\
E_T^{v(a)} &=& \sum_{\alpha}E_{\alpha T}^{v(a)},
\eea
and $\tilde{\textbf{p}}_T^{(a)}$ and $\tilde{E}_T^{(a)}$ are the transverse momentum and energy of the (assumed-to-be-one) \textit{trial} DM particle in each decay chain. Here $m_{\alpha}^{v(a)}$ indicates the mass of $\alpha$th visible particle in $a$th decay chain ($a=1,\;2$) and $t_l^{v(a)}$ indicates the (1+2) momentum on the transverse plane which is defined as
\bea
t_l^{v(a)} \equiv \left(E_{lT}^{v(a)},\textbf{p}_{lT}^{v(a)}\right)
=\left(\sqrt{\left(\textbf{p}_{lT}^{v(a)}\right)^2+\left(m_l^{v(a)}\right)^2},\textbf{p}_{lT}^{v(a)}\right),
\eea
and the metric for this type of momentum is $\textnormal{diag}(1,-1,-1)$. There arise two noteworthy things: 
\begin{itemize}
\item
As far as the range is concerned, the transverse and the regular invariant masses have the same lower and upper limits. Moreover, since the $M_{T2}$ solutions of interest arise at either of the two limits, one may consider the $M_{T2}$ where $m_T^{v(a)}$ are replaced by the regular invariant masses of visible particles $m_v^{(a)}$:
\bea
\left(m_v^{(a)}\right)^2=\left(p_1^{v(a)}+\cdots+p_n^{v(a)}   \right)^2=
\sum_{\alpha}\left(m_{\alpha}^{v(a)}\right)^2+2\sum_{\alpha>\beta}\left(E_{\alpha  }^{v(a)}E_{\beta  }^{v(a)}
-\textbf{p}_{\alpha  }^{v(a)}\cdot \textbf{p}_{\beta  }^{v(a)} \right).
\eea

\item
As advertised earlier, the entire visible states in the same decay chain can be understood effectively as a single visible particle whose ``effective'' mass is given by $m_v^{(a)}$. On the other hand, the corresponding ``effective'' quantity for invisible particles $m_i^{(a)}$ does not seem to be contained in the $M_{T2}$ variable. In fact, the $M_{T2}$ variable depends \textit{implicitly} on $m_i^{(a)}$, which will be cleared shortly.

\end{itemize}
From these two observations we can reduce the decay of pair-produced mother particles into two multi-particle states (left panel of Fig.~\ref{fig:gendecproc}) to an effective kinematic configuration where there exist two simple 2-body decay chains shown in the right panel of Fig.~\ref{fig:gendecproc}. 

\begin{figure}[t]
	\centering
	 \includegraphics[width=7.5truecm,height=7.5truecm,clip=true]{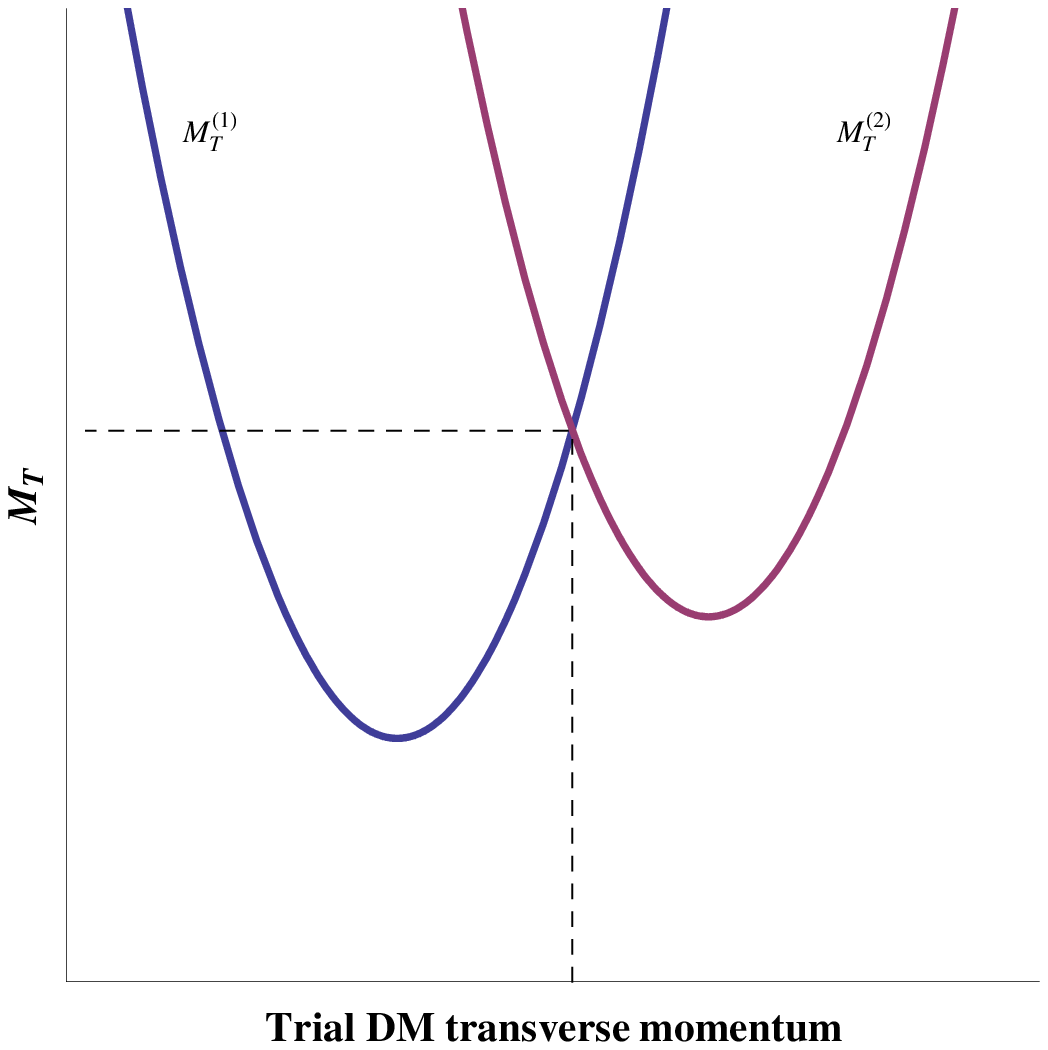}
	\hspace{0.2cm}
	 \includegraphics[width=7.5truecm,height=7.5truecm,clip=true]{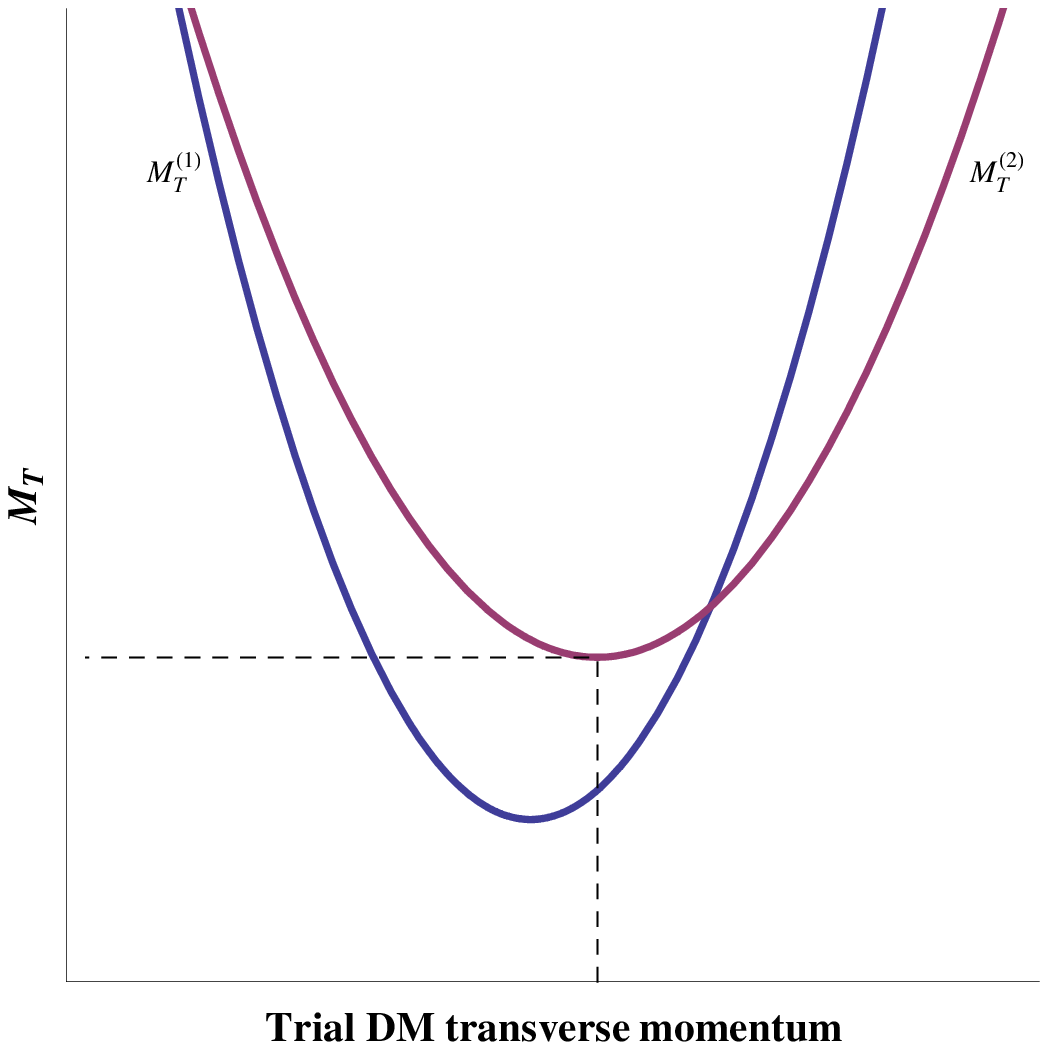}
	\caption{Graphical configurations to give rise to a balanced $M_{T2}$ solution (the left panel) and an unbalanced $M_{T2}$ solution (the right panel). The dashed line indicates the solution for the $M_{T2}$ variable to take.
}
	\label{fig:plotbalunbal}
\end{figure}
For such $M_{T2}$, there are two types of solution which are called the ``balanced'' $M_{T2}$ solution and the ``unbalanced'' $M_{T2}$ solution, and the fact that there always exist some events to give such solutions was proven~\cite{Lester:1999tx, Barr:2002ex, Cho:2007qv}. 
The balanced solution arises when $M_T^{(a)}\geq M_T^{(b)}$ for both $a(=1,\;2)$ with the trial DM momentum having the value to accommodate $M_T^{(b)}$ ($b\neq a$) at their global (or so-called ``unconstrained'') minimum which will be defined shortly (see the left panel of Fig.~\ref{fig:plotbalunbal}), and otherwise, the unbalanced solution arises (see the right panel of Fig.~\ref{fig:plotbalunbal}).\footnote{If the total invariant masses of visible states in both decay chains are the same, only the balanced solutions arise. The reason is because the unconstrained minima for both $M_T$ are identical, there is no possibility that the kinematic configuration like the right panel of Fig.~\ref{fig:plotbalunbal} is made. As an example, if there exists only one massless visible particle in each decay chain, the $M_{T2}$ values are always given by the balanced solution.} The global minima for $M_T^{(1)}$ and $M_T^{(2)}$ are easily evaluated by differentiating Eq.~(\ref{eq:mt}) with respect to the trial DM momentum and finding the stationary point~\cite{Lester:1999tx, Barr:2002ex}:
\bea
\left(M_T^{(1)}\right)_{\min}=m_v^{(1)}+\tilde{m} \label{eq:globalmin1} \\
\left(M_T^{(2)}\right)_{\min}=m_v^{(2)}+\tilde{m} \label{eq:globalmin2}
\eea
The balanced and the unbalanced $M_{T2}$ solutions for a given set of the ``effective'' visible and invisible masses which are shown in the right panel of Fig.~\ref{fig:gendecproc} are as follows~\cite{Lester:1999tx, Barr:2002ex, Cho:2007qv}:
\bea
\left(M_{T2}^{bal}\right)^2
&=&\tilde{m}^2+A+\sqrt{\left(1+\frac{4\tilde{m}^2}{2A-\left(m_v^{(1)}\right)^2-\left(m_v^{(2)}\right)^2}
    \right)\left(A^2-\left(m_v^{(1)}m_v^{(2)}\right)^2\right)} \label{eq:balsol} \\
M_{T2}^{unbal}&=& \tilde{m}+m_v^{(a)} \;\;\; (a=1,\;2)\label{eq:unbalsol}
\eea
for
\bea
A=E^{v(1)}_TE^{v(2)}_T+\vec{\textbf{p}}_T^{v(1)}\cdot\vec{\textbf{p}}_T^{v(2)}.
\eea
Note that the unbalanced solution is simply given by the unconstrained minimum of $M_{T}^{(a)}$, and for a fixed set of visible and invisible masses the balanced solution is bounded above at
\bea
A=E^{v(1)}E^{v(2)}+q^{v(1)}q^{v(2)}. \label{eq:realA}
\eea
In fact, if we take an adequate number of events, we can always find some event which corresponds to such an upper bound~\cite{Cho:2007qv}. Since we are interested in the $M_{T2}^{bal}$ equal to its own upper bound, we henceforth assume that $A$ is understood as Eq.~(\ref{eq:realA}) unless there arises any confusion. 
Here $\left(E^{v(a)}\right)^2=\left(q^{v(a)}\right)^2+\left(m_v^{(a)}\right)^2$ and $q^{v(a)}$ is
the magnitude of the total momentum of visible particles seen in the rest frame of their mother particle. The explicit expression for $q^{v(a)}$ can be easily determined in terms of the masses of
mother, visible, and invisible particles as follows:
\bea
q_v^{(a)}=\frac{1}{2M}\sqrt{\left\{\left(M+m_v^{(a)}\right)^2-\left(m_i^{(a)}\right)^2\right\}
\left\{\left(M-m_v^{(a)}\right)^2-\left(m_i^{(a)}\right)^2\right\}}.
\eea
One thing to be emphasized is that the dependence of the $M_{T2}$ variable on the effective invisible/DM mass $m_i^{(a)}$ first appear in $q_v^{(a)}$. In other words, $M_{T2}$ is an implicit function of $m_i^{(a)}$ via $q_v^{(a)}$ as mentioned before. Furthermore, we include the possibility that multiple (massive) invisible particles are emitted in each decay chain 
unlike the previous studies 
%
%
(which
considered the cases with two invisible particles having non-identical masses).
Hence $m_i^{(a)}$ as well as $m_v^{(a)}$ have their own range once multiple visible 
and invisible particles are involved in the given decay process. For off-shell intermediate particles the respective ranges are given by (See, for example,~\cite{Byckling:1973bk})
\bea
\sum_{\alpha}m_{\alpha}^{v(a)}=m_{v,\textnormal{ min}}^{(a)}\leq m_v^{(a)} \leq
m_{v,\textnormal{ max}}^{(a)}=M-\sum_{\beta}m_{\beta}^{i(a)} \label{eq:rangevis} \\
\sum_{\beta}m_{\beta}^{i(a)}=m_{i,\textnormal{ min}}^{(a)}\leq m_i^{(a)} \leq
m_{i,\textnormal{ max}}^{(a)}=M-\sum_{\alpha}m_{\alpha}^{v(a)}. \label{eq:rangeinvis}
\eea
The lower limit corresponds to the situation in which particles described by $p_{\alpha}^{v(a)}\left(p_{\beta}^{i(a)}\right)$ are at rest in their center of mass frame so that they move with the same velocity in any frame. The upper limit corresponds to the situation in which particles described by $p_{\beta}^{i(a)}\left(p_{\alpha}^{v(a)}\right)$ are at rest in the overall center of mass
frame of the final state described by $p_{\alpha}^{v(a)}$ and $p_{\beta}^{i(a)}$.

\subsection{The Maximum Balanced and Unbalanced $M_{T2}$ Solutions}
For the decay with visible/invisible multi-particle final states, it is obvious that balanced/unbalanced $M_{T2}$ solutions have their own range due to the existence of the range in either $m_v^{(a)}$ or $m_i^{(a)}$ or both of them. As far as the upper edge in the $M_{T2}$ distribution is concerned, either the maximum balanced or the maximum unbalanced solution appears as the global maximum. For the unbalanced solution, one can easily derive the following relationship from Eqs.~(\ref{eq:unbalsol}) and~(\ref{eq:rangevis}): 
\bea
M_{T2}^{\max, unbal}=\tilde{m}+\max \left[m_{v,\max}^{(1)},\;m_{v,\max}^{(2)}\right]. \label{eq:unbalgen}
\eea

For the balanced solution, however, it is not easily seen which values of $m_v^{(a)}$ and $m_i^{(a)}$ $(a=1,\;2)$ will form the maximum balanced solution because of the complication in the corresponding expression $\left(M_{T2}^{bal}\right)^2$ given in Eq.~(\ref{eq:balsol}). In order to identify those values, we are required to carefully investigate the functional behavior of $\left(M_{T2}^{bal}\right)^2$ according to the changes in $m_i^{(a)}$ and $m_v^{(a)}$, which will be considered in order. 

\subsubsection{The Change in $m_i^{(a)}$}
To see the dependence of $\left(M_{T2}^{bal}\right)^2$ on $m_i^{(1)}$, we simply take the partial derivative:
\bea
\frac{\partial \left(M_{T2}^{bal}\right)^2}{\partial \left(m_i^{(1)}\right)^2} = \frac{D}{2X\sqrt{B^3 C}}
\eea
where
\bea
B\equiv 2A-\left(m_v^{(1)}\right)^2-\left(m_v^{(2)}\right)^2, \;\;\; C\equiv A^2-\left(m_v^{(1)}\right)^2 \left(m_v^{(2)}\right)^2, \;\;\; X\equiv \sqrt{B+4\tilde{m}^2} \\
D \equiv (BC'-B'C)X^2+2A'\sqrt{B^3 C}X+B'BC \hspace{3.5cm}
\eea
with the following notations:
\bea
A' \equiv \frac{\partial A}{\partial \left(m_i^{(1)}\right)^2}, \;\;\; B' \equiv \frac{\partial B}{\partial \left(m_i^{(1)}\right)^2}=2A', \;\;\; C' \equiv \frac{\partial C}{\partial \left(m_i^{(1)}\right)^2}=2AA'.
\eea
One can easily see that $A$, $B$, and $C$ are always positive for any set of $m_v^{(a)}$ and $m_i^{(a)}$, and that only positive $X$ is allowed by construction. Also, one can easily prove that $A'$ is negative. 

The solutions to $D=0$ are given as follows:
\bea
X_1=-\frac{\sqrt{BC}}{A-\left(m_v^{(1)}\right)^2} \\
X_2=-\frac{\sqrt{BC}}{A-\left(m_v^{(2)}\right)^2}.
\eea
For $m_v^{(1)} \neq m_v^{(2)}$ it can be proven that either $A-\left(m_v^{(1)}\right)^2$ or $A-\left(m_v^{(2)}\right)^2$ must be positive and the other is positive or negative depending on the parameter space formed by $m_v^{(1)}$ and $m_v^{(2)}$~\cite{Cho:2007qv}. Hence, one of the two solutions given above must be negative, which is unphysical, the other is either physically allowed or not. Actually, it turns out that the signs of $A-\left(m_v^{(1)}\right)^2$ and $A-\left(m_v^{(2)}\right)^2$ are connected to the coefficient of $X^2$ in $D$ in the following way:
\bea
BC'-B'C=2A'\left(A-\left(m_v^{(1)}\right)^2\right)\left( A-\left(m_v^{(2)}\right)^2\right).
\eea
Let us assume that $m_v^{(1)}$ is larger than $ m_v^{(2)}$. In this case, $A-\left(m_v^{(2)}\right)^2$ is always positive, i.e., $X_2$ is always unphysical. Since $A'<0$ as mentioned above, if $A-\left(m_v^{(1)}\right)^2$ is positive as well, then $D$, which is a quadratic function in $X$, becomes a parabola bounded above, and the two solutions $X_1$ and $X_2$ all are negative, i.e., unphysical. Therefore, $D<0$ for arbitrary (physically-allowed) $X$ or $\tilde{m}$. On the other hand, if $A-\left(m_v^{(1)}\right)^2$ is negative, then $D$ turns into a parabola bounded below, and $X_1$ becomes a physically allowed solution. Therefore, $D<0$ for $0<X<X_1$ and $D>0$ for $X>X_1$. However, in~\cite{Cho:2007qv} it was shown that $X_1$ gives rise to 
\bea
M_{T2}^{bal}(X=X_1)=\tilde{m}+m_v^{(1)},
\eea   
which is simply the unbalanced solution for $m_v^{(1)}> m_v^{(2)}$. Moreover, they showed that this implies that $X_1$ corresponds to the boundary between the balanced domain and the unbalanced domain. In other words, with $X$ being larger than $X_1$ the balanced solution is reduced to the unbalanced solution. One can make the same argument and lead to the same conclusion for the opposite configuration, i.e., $m_v^{(1)}< m_v^{(2)}$. Also, the dependence on $m_i^{(2)}$ can be easily checked by following similar arguments. Based on hitherto observations, we have
\bea
\frac{\partial \left(M_{T2}^{bal}\right)^2}{\partial \left(m_i^{(a)}\right)^2} &<& 0 \\
M_{T2}^{\max, bal}&=&M_{T2}^{bal}\left(m_i^{(1)}= m_{i,\min}^{(1)},\;m_i^{(2)}= m_{i,\min}^{(2)}\right)
\eea
for any set of $m_v^{(a)}\;(a=1,\;2)$. 

\subsubsection{The Change in $m_v^{(a)}$}
The early work on the dependence of $\left(M_{T2}^{bal}\right)^2$ on $m_v^{(a)}$ was made in~\cite{Cho:2007qv}. Here we simply provide the final results and mention some modification from the original expression. 
\bea
\frac{\partial \left(M_{T2}^{bal}\right)^2}{\partial \left(m_v^{(a)}\right)^2}&&
\left\{
\begin{array}{l}
\leq 0 \hspace{2cm} \hbox{for $\tilde{m} < m'$} \cr
\cr
\geq 0 \hspace{2cm} \hbox{for $\tilde{m} \geq m'$}
\end{array}\right. \\
M_{T2}^{\max, bal}&=&
\left\{
\begin{array}{l}
M_{T2}^{bal}\left(m_v^{(1)}= m_{v,\min}^{(1)},\;m_v^{(2)}= m_{v,\min}^{(2)} \right) \hbox{ for $\tilde{m} < m'$} \cr
\cr
M_{T2}^{bal}\left(m_v^{(1)}= m_{v,\max}^{(1)},\;m_v^{(2)}= m_{v,\max}^{(2)} \right)\hbox{ for $\tilde{m} \geq m'$}. \label{eq:endresult}
\end{array}\right.
\eea
Here the ``kink'' location $m'$ can be identified as the true dark matter mass $m_{DM}$ if only a single type of DM is involved~\cite{Cho:2007qv}. 
However, in general, it differs from $m_{DM}$ because we do not restrict our consideration to the case with one single-typed DM emitted in each decay chain. Therefore, its expression is written in terms of all parameters (i.e., $M$, $m_v^{(a)}$, and $m_i^{(a)}$), and it can be calculated by solving the following equation~\cite{Cho:2007qv}: 
\bea
\sqrt{B+4m'^2}=\frac{\sqrt{BC}(1-2\bar{A})}{2\bar{A}\left(A-\left(m_v^{>}\right)^2\right)+A-\left(m_v^{<}\right)^2} \label{eq:kinklocgencase}
\eea
where $m_v^>$ and $m_v^<$ denote the heavier and the lighter (invariant) visible masses between the two decay sides, respectively, and $\bar{A}$ is defined as
\bea
\bar{A}=\frac{\partial A}{\partial \left(m_v^>\right)^2}.
\eea 

\subsection{Discussions and Application} 
It is a well-known fact that there arises a ``kink'' in the $M_{T2}^{\max}$ as a function of the trial mass once there exists more than one visible particle in each decay chain and its location is at $\tilde{m}=m_{DM}$ for the cases with a single identical DM particle per decay chain. For more extended case, i.e., $m_i^{(1)} \neq m_i^{(2)}$, one can simply solve Eq.~(\ref{eq:kinklocgencase}). In turns out, however, that this is not the only way of obtaining the kink location. An alternative and simpler way is to find the intersecting point between the maximum balanced and unbalanced solutions. In other words, the solution to satisfy Eq.~(\ref{eq:kinklocgencase}) also satisfies the relation $M_{T2}^{\max, bal}=M_{T2}^{\max, unbal}$. For simplicity, let us assume that $m_{v, \max}^{(1)}>m_{v,\max}^{(2)}$. We then have 
\bea
\left(M_{T2}^{\max, unbal}\right)^2=\left(\tilde{m}+m_{v,\max}^{(1)}\right)^2 \label{eq:max1}
\eea
and
\bea
\left(M_{T2}^{\max,bal}\right)^2=\tilde{m}^2+A+\sqrt{\frac{C}{B}\left(B+4\tilde{m}^2\right)} \label{eq:max2}
\eea
where $A$ is evaluated at $m_i^{(a)}=m_{i,\min}^{(a)}$ as discussed before. Letting Eqs.~(\ref{eq:max1}) and~(\ref{eq:max2}) be equated and doing some tedious algebra, one can end up with 
\bea
A^2=\left(m_v^{(1)}\right)^2\left(m_v^{(2)}\right)^2,
\eea
which is valid only with $m_v^{(a)}$ being their maximum. Note that $M_{T2}^{\max, bal}$ at $\tilde{m}=m'$ arises when $m_v^{(a)}=m_{v,\max}^{(a)}$ from Eq.~(\ref{eq:endresult}). Hence, the above-given relationship holds, and the location of the kink can be evaluated by finding the intersection between the maximum balanced and unbalanced solutions. 

This observation, actually, leads us to the expressions for $M_{T2}^{\max, bal}$ and $M_{T2}^{\max, unbal}$. Note that it was proven that the balanced solution contributes to the upper edge of the $M_{T2}$ distribution at $\tilde{m}<m'$ in~\cite{Cho:2007qv}. Also, it is straightforward to prove that the maximum unbalanced solution is larger than the maximum balanced solution at $\tilde{m} \geq m'$. Therefore, as long as the values to give the upper edge of the $M_{T2}$ distribution are concerned, it can be (effectively) understood that the maximum balanced solutions occur at $m_v^{(a)}=m_{v,\min}^{(a)}$ and $m_i^{(a)}=m_{i,\min}^{(a)}$ and the maximum unbalanced solutions do at the maximum of the two $m_{v,\max}^{(a)}\; (a=1,\;2)$ for any $\tilde{m}$. As an example, if all visible particles are assumed massless, the maximum balanced and unbalanced solutions are given as follows:
\bea
M_{T2}^{\max , bal}=\sqrt{\frac{\left(M^2-m_{i,\min}^{(1)2}\right)\left(M^2-m_{i,\min}^{(2)2}\right)}{4M^2}}
+\sqrt{\frac{\left(M^2-m_{i,\min}^{(1)2}\right)\left(M^2-m_{i,\min}^{(2)2}\right)}{4M^2}+\tilde{m}^2} \label{eq:mt2balgen} \\
M_{T2}^{\max, unbal}=\tilde{m}+\max\left[m_{v,\max}^{(1)},\;m_{v,\max}^{(2)}\right] =\tilde{m}+M-\min \left[m_{i,\min}^{(1)},\;m_{i,\min}^{(2)}\right] \hspace{2.6cm} \label{eq:mt2unbalgen}
\eea

Obviously, the upper edge in the $M_{T2}$ distribution is determined by the maximum value among many events for a given trial DM mass.
\bea
M_{T2}^{\max}(\tilde{m})=\max_{\hbox{many events}}[M_{T2}(\tilde{m})] \label{eq:manyevent}
\eea
Based on the above-discussed understanding, one could expect that taking the maximum between $M_{T2}^{\max , bal}$ and $M_{T2}^{\max , unbal}$ will result in the same value as the above-given Eq.~(\ref{eq:manyevent}).
\bea
M_{T2}^{\max}=\max\left[M_{T2}^{\max , bal},\;M_{T2}^{\max ,unbal} \right] \label{eq:comp}
\eea
It turns out, however, it is true only for the case where there exists more than one visible particle on each decay chain. In the case where there is only one visible particle per decay chain, one can prove that the maximum unbalanced solution is less than the maximum balanced solution for any $\tilde{m}$ so that the $M_{T2}^{\max}$ is simply governed by the $M_{T2}^{\max, bal}\left(m_i^{(a)}=m_{i,\min}^{(a)}\right)$. 

\section{The Existence of a Kink in $M_{T2}^{\max}$ Versus $\tilde{m}$}
\label{app:kink}

As discussed in App.~\ref{app:locofmaxmt2}, it is obvious that for the cases where there is only a single visible particle in each decay chain, the $M_{T2}^{\max}$ as a function of the trial DM mass behaves like a smoothly increasing curve because the upper edge is solely governed by the ``balanced'' solution in Eq.~(\ref{eq:balsol}). However, if there exists more than one visible particle per decay chain, the competition between the ``balanced'' and the ``unbalanced'' solutions, which is explicitly given in Eq.~(\ref{eq:comp}), gives rise to the possibility of a kink (i.e., no longer smooth) in the plot of $M_{T2}^{\max}$ versus $\tilde{m}$. In fact, this approach, the competition between the two types of solutions, enables us to examine easily whether or not there exists a ``kink'' on the function of the location of $M_{T2}^{\max}$.

\begin{figure}[t]
  \centering
  \includegraphics[height=3.0in,clip]{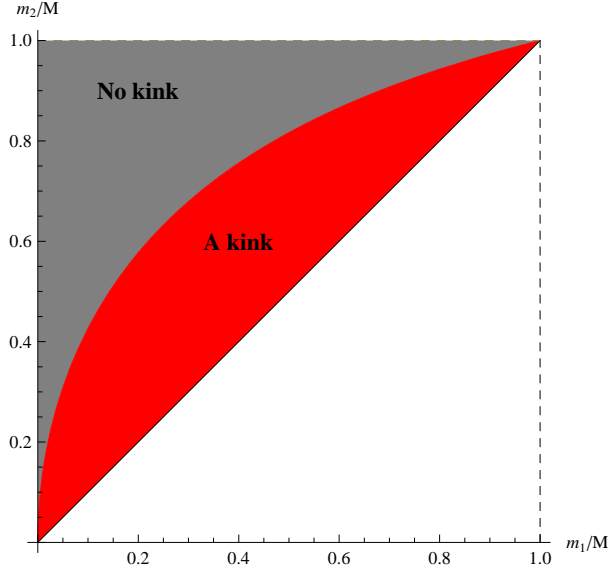}
  \caption{The kinematic regions to have a kink or no kink in the graph of $\frac{m_2}{M}$ versus $\frac{m_1}{M}$.}
  \label{fig:kinkregion}
\end{figure}
In order to have a kink in $M_{T2}^{\max}$ as a function of $\tilde{m}$, the two functions of the maximum balanced and unbalanced solutions over the trial DM mass, i.e., $M_{T2}^{\max ,bal}(\tilde{m})$ and $M_{T2}^{\max ,unbal}(\tilde{m})$, must cross each other. From Eqs.~(\ref{eq:balsol}) and~(\ref{eq:unbalsol}) they are monotonic functions in $\tilde{m}$, and the slope of Eq.~(\ref{eq:balsol}) is not greater than that of Eq.~(\ref{eq:unbalsol})($=1$) over the entire range. These two observations tell us that once a crossover is made, no additional crossovers are made. Therefore, it is sufficient to check whether or not the relative sizes of their corresponding functional values at $\tilde{m}=0$ and $\tilde{m}\rightarrow\infty$ are flipped for ensuring such a crossover. Let us assume that the visible particles are massless for simplicity.\footnote{One can easily apply the same argument for the case of massive visible particles.} From Eqs.~(\ref{eq:mt2balgen}) and~(\ref{eq:mt2unbalgen}) one can easily prove that $M_{T2}^{\max , unbal}$ is larger than $M_{T2}^{\max , bal}$ at $\tilde{m}\rightarrow \infty$, and thus $M_{T2}^{\max , bal}$ should be larger than $M_{T2}^{\max , unbal}$ at $\tilde{m}=0$ to obtain a kink. Their functional values at $\tilde{m}=0$ are expressed as follows:
\bea
M_{T2}^{\max ,bal}(\tilde{m}=0)&=&\sqrt{\frac{(M^2-m_1^2)(M^2-m_2^2)}{M^2}} \\
M_{T2}^{\max ,unbal}(\tilde{m}=0)&=&M-m_1
\eea
where $m_{i,\min}^{(1)}\equiv m_1$ and $m_{i,\min}^{(2)}\equiv m_2$, and we assumed $m_1\leq m_2$ without loss of generality.
Therefore, the condition to have a kink is simply given by
\bea
\sqrt{\left(1-\frac{m_1^2}{M^2}\right)\left(1-\frac{m_2^2}{M^2}\right)}>1-\frac{m_1}{M},
\eea
which can be further simplified to
\bea
\frac{m_1}{M}<\frac{m_2}{M}<\sqrt{\frac{2m_1}{M+m_1}}.
\label{eq:condkink}
\eea
Likewise, one can easily find the condition to have no kink as follows:
\bea
\frac{m_2}{M}>\sqrt{\frac{2m_1}{M+m_1}}.
\label{eq:condnokink}
\eea
Fig.~\ref{fig:kinkregion} shows the regions for a kink or no kink pictorially; the red area represents all possible kinematic configurations for the existence of a kink while the gray one all possible kinematic configurations for no kink. Elsewhere is not physically allowed due to the assumption that $m_1 \leq m_2$.

There is a special case where each decay chain emits only one \textit{single}-typed DM particle, i.e., $m_1=m_2$. The $Z_2$ models or the $E_2$ type events of $Z_3$ models belong to this case. The range to satisfy Eq.~(\ref{eq:condnokink}), i.e., the condition to have no kink, is $\frac{m_1}{M}>1$ or $\frac{m_1}{M}<-2$ both of which are not physically allowed. Therefore, we always obtain a kink in $M_{T2}^{\max}$ as a function of the trial DM mass as 
expected~\cite{Cho:2007qv}.

As another concrete example, let us take $E_3$ type events of $Z_3$ models, where one of the two decay chains emits a single dark matter particle whereas the other one emits two dark matter particles with intermediate particles \textit{off}-shell. Like before, we assume that all DM particles to be emitted in the full decay process have the same mass so that the \textit{minimum} of the effective dark matter mass of the two DM side to give the maximum balanced solution is $m^{eff}_{DM} = 2m_{DM}$, i.e., $m_1=m_{DM}$ and $m_2=2m_{DM}$. From Eqs.~(\ref{eq:condkink}) and~(\ref{eq:condnokink}) the conditions to have a kink or no kink are more simplified to
\bea
\hbox{A kink:}&&0<\frac{m_{DM}}{M}<\frac{\sqrt{3}-1}{2} \\
\hbox{No kink:}&&\frac{\sqrt{3}-1}{2}<\frac{m_{DM}}{M}<1,
\label{eq:condz3}
\eea
which was mentioned in Sec.~\ref{sec:z3morethanone} and demonstrated in Fig.~\ref{fig:mt2max}.

\section{Algorithm to Find the Upper Edge of $M_{T2}$ Distribution}
\label{app:algorithm}

In this appendix, we describe an algorithm to identify the $M_{T2}^\text{max}$ for events after the $R_{P_t}$ cut. As we discussed in section \ref{sec:onevisibleidentical}, in $Z_3$ models where there is only one visible particles per decay chain, and the visible particles in the decay chains with one DM and two DM are identical, the total $M_{T2}$ distribution becomes a combination of the distributions of $E_2$ and $E_3$
events. So the idea is to apply an $R_{P_t}$ cut (a cut on the ratio of $P_t$'s of visible particles on the two decay chains in the same event) to ``remove'' the $E_2$ events. This in principle can give us a relatively pure sample of $E_3$ events, which has a smaller $M_{T2}^\text{max}$. But in practice/reality, there is still a small number of $E_2$ events that survive the $R_{P_t}$ cut. Therefore, the upper edge of $M_{T2}$ distribution for events after the $R_{P_t}$ cut is hard to be determined due to the ``contamination'' of $E_2$ type events. This is shown in the right panel of Fig. \ref{fig:combinedE2E3}, which shows clearly that there is a small number of events which has $M_{T2}$ beyond the $M_{T2}^\text{max}$ of $E_3$ type events. Here we propose an algorithm to identify/extract the ``would-be'' $M_{T2}^\text{max}$ for $E_3$ events by removing $E_2$ contamination events and then do a fitting to the resulting distribution.

First, we need to ``subtract'' the contaminating events. To do this, we calculate the moving average of the number of events per bin including the last $n$ bins in the $M_{T2}$ distribution: $A_n$. The choice of moving average makes this quantity rather stable under statistical fluctuations as we increase $n$. However, as we increase $n$ to the point below $M_{T2}^\text{max}$ for $E_3$ type events, we start to get a sharp rise on $A_n$. Based on this, we define $n_\text{max}$ to be the bin such that $A_{n_\text{max} + 1} \ge 2.5 A_{n_\text{max}}$. This bin is considered as a rough separation point between ``contaminating'' $E_2$ type events and the start of $E_3$ type events. And we treat $A_{n_\text{max}}$ as a rough estimate for the $E_2$ type events contribution to the number of events per bin.

Next, we pick events with $n > n_\text{max}$ and subtract $A_{n_\text{max}}$ from the number of events in each bin. This gives us an approximate $M_{T2}$ distribution for pure $E_3$ type events. Since we do not have an analytical formula for the $M_{T2}$ distribution for $E_3$ type events, we can only do a fitting for events near and to the left of the bin $n_\text{max}$ to find the upper edge of the $M_{T2}$ distribution. We choose two fitting functions, one linear function and one quadratic function, and did the fitting separately.  Our final answer for the $M_{T2}^\text{max}$ is given by the average of the values obtained by these two fitting methods, and their difference is regarded as the error due to fitting.

\end{document}